\documentclass[aps,prd,showpacs,twocolumn,superscriptaddress,nofootinbib,preprintnumbers]{revtex4-1} 

\makeatletter
\def\p@subsection{}
\makeatother

\usepackage{graphicx,amsmath,amssymb,lipsum}

\usepackage[usenames,dvipsnames]{xcolor}
\definecolor{xlinkcolor}{rgb}{0.7752941176470588, 0.22078431372549023, 0.2262745098039215}

\usepackage[colorlinks=true,citecolor=xlinkcolor,linkcolor=xlinkcolor,urlcolor=xlinkcolor, backref=false,pdfborder={0 0 0}]{hyperref}

\usepackage[normalem]{ulem}
\usepackage[utf8]{inputenc} 
\usepackage{mathtools}
\usepackage{aas_macros}
\usepackage{float}

\DeclareMathOperator*{\argmax}{arg\,max}

\usepackage{ulem}
\usepackage{diagbox}

\usepackage[dvipsnames]{xcolor}
\usepackage{mathrsfs}

\newcommand{\be}{\begin{equation}}
\newcommand{\ee}{\end{equation}}
\newcommand{\beqa}{\begin{eqnarray}}
\newcommand{\eeqa}{\end{eqnarray}}

\renewcommand\k{{\bf k}}

\renewcommand\a{\alpha}

\def\d{\partial}
\newcommand{\bseq}{\begin{subequations}}
\newcommand{\eseq}{\end{subequations}}

\def\gsim{\raise0.3ex\hbox{$\;>$\kern-0.75em\raise-1.1ex\hbox{$\sim\;$}}}
\def\lsim{\raise0.3ex\hbox{$\;<$\kern-0.75em\raise-1.1ex\hbox{$\sim\;$}}}

\def\beqn#1{\begin{equation}\label{#1}}
\def\eeqn{\end{equation}}

\def\beqa#1{\begin{eqnarray}\label{#1}}
\def\eeqa{\end{eqnarray}}

\def\kmax{{k_\text{max}}}
\def\hMpc{h{\text{Mpc}}^{-1}}
\def\Mpch{h^{-1}{\text{Mpc}}}

\def\Z2{$\mathcal{Z_2}$}

\newcommand {\ignore}[1]{}

\begin{document}

\preprint{MIT-CTP/5684}

\title{Full-shape analysis 
with simulation-based priors:\\
constraints on single field inflation 
from BOSS
}

\author{Mikhail M. Ivanov}
\email{ivanov99@mit.edu}
\affiliation{Center for Theoretical Physics, Massachusetts Institute of Technology, Cambridge, MA 02139, USA}

\author{Carolina Cuesta-Lazaro}
\email{cuestalz@mit.edu}
\affiliation{The NSF AI Institute for Artificial Intelligence and Fundamental Interactions, Cambridge, MA 02139, USA}
\affiliation{Department of Physics, Massachusetts Institute of Technology, Cambridge, MA 02139, USA}
\affiliation{Center for Astrophysics | Harvard \& Smithsonian, 60 Garden Street, MS-16, Cambridge, MA 02138, USA}

\author{Siddharth Mishra-Sharma}
\email{smsharma@mit.edu}
\affiliation{The NSF AI Institute for Artificial Intelligence and Fundamental Interactions}
\affiliation{Center for Theoretical Physics, Massachusetts Institute of Technology, Cambridge, MA 02139, USA}
\affiliation{Department of Physics, Harvard University, Cambridge, MA 02138, USA}

\author{Andrej Obuljen}
\email{andrej.obuljen@uzh.ch}
\affiliation{Department of Astrophysics, University of Zurich, Winterthurerstrasse 190, 8057 Zurich, Switzerland}

\author{Michael W. Toomey}
\email{mtoomey@mit.edu}
\affiliation{Center for Theoretical Physics, Massachusetts Institute of Technology, 
Cambridge, MA 02139, USA}

\begin{abstract} 
Perturbative, or effective field theory (EFT)-based, full-shape analyses of 
galaxy clustering data involve ``nuisance parameters'' to capture various observational effects such as the galaxy-dark matter
connection (galaxy bias).
We present an efficient approach to set informative physically motivated 
priors on these parameters. 
We extract these priors from 
simulated galaxy catalogs based on 
halo occupation distribution (HOD) models.
First, we build a joint distribution 
between EFT galaxy bias and HOD parameters
from a set of 10,500 HOD mock catalogs. 
We use the field level EFT technique that allows 
for cosmic variance cancellation, enabling a precision calibration of EFT parameters
from computationally inexpensive small-volume simulations. 
Second, we use neural density estimators -- normalizing flows --
to model the marginal probability density of the EFT parameters,
which can be used as a prior distribution
in full shape analyses. 
As a first application, we use our HOD-based priors 
in a new analysis of  
galaxy power spectra and bispectra from the BOSS survey 
in the context of 
single field 
primordial 
non-Gaussianity. 
We find that our priors lead to 
a reduction of the posterior volume
of bias parameters by an order of magnitude.
We also find 
$f_{\rm NL}^{\rm equil} = 320\pm 300$ and 
$f_{\rm NL}^{\rm ortho} = 100\pm 130$
(at 68\% CL) in a combined two-template analysis,
representing a $\approx 40\%$ improvement
in constraints on single field primordial non-Gaussianity,
equivalent to doubling the survey volume.
\end{abstract}

\maketitle

\section{Introduction}

The last three decades of precision cosmological observations have 
yielded significant evidence 
for new physical phenomena in the form of dark matter, 
dark energy, and primordial accelerated 
expansion of the Universe, called cosmic inflation. 
Future progress in understanding of these phenomena will depend on our ability 
to extract cosmological information from upcoming large-scale structure surveys, 
such as DESI~\cite{Aghamousa:2016zmz}, Euclid~\cite{Laureijs:2011gra}, LSST~\cite{LSST:2008ijt}, and Roman Space Telescope~\cite{Akeson:2019biv}.

On scales larger than $\sim 10$ Mpc, 
large scale structure 
formation 
occurs in the mildly non-linear regime. 
In this regime the data can be described with perturbation theory 
(its consistent formulation is 
called effective field theory (EFT) of large-scale structure~\cite{Baumann:2010tm,Carrasco:2012cv,Ivanov:2022mrd}),
which provides a systematic framework for modeling
galaxy clustering based only on 
symmetries and scale separation.
Currently, this is the only approach to galaxy clustering 
that offers a sub-percent accuracy for 
galaxy separations larger than 10 Mpc~\cite{Nishimichi:2020tvu,Ivanov:2021zmi}.
Another key advantages of EFT is its efficiency.  
Namely, EFT model templates can be computed in less than 1 second~\cite{Chudaykin:2020aoj,Chen:2020zjt,DAmico:2020kxu}.
EFT thus offers an unmatched flexibility
in analyzing cosmological models beyond the vanilla $\Lambda$CDM model (see e.g.~\cite{Ivanov:2020ril,Chudaykin:2020ghx,DAmico:2020kxu}).
All these virtues made EFT a useful tool for 
the analysis of the galaxy clustering data 
on large scales, see 
e.g.~\cite{Ivanov:2019pdj,DAmico:2019fhj,Philcox:2021kcw,Chen:2021wdi}
for a sample of analyses of BOSS data~\cite{Alam:2016hwk}.

The main disadvantages of 
EFT techniques are: (a) the breakdown
on small scales (where the statistical power of data 
can be significant), and (b) the proliferation of free ``nuisance'' parameters. 
These ``nuisance'' parameters
include, e.g. the classical 
perturbative 
bias parameters such as linear, quadratic, and tidal biases $b_1,b_2,b_{\mathcal{G}_2}$ etc~\cite{Desjacques:2016bnm}.
These parameters encapsulate 
effects of small-scale
galaxy formation physics and 
typically are 
determined from data in actual analyses.
However, marginalization over EFT parameters within  
uninformative priors
leads to a significant loss of information
in current EFT-based full-shape analyses~\cite{Wadekar:2020hax,Cabass:2022epm,Philcox:2022frc}.\footnote{One may also hope that simulation based priors could reduce Bayesian parameter projection effects that may complicate the interpretation of full-shape results, see~\cite{Ivanov:2019pdj,Chudaykin:2020ghx,Philcox:2021kcw,Ivanov:2023qzb} for more detail. }

The loss of information due to free EFT parameters can be 
avoided by using informative priors. 
These priors can be 
extracted from phenomenological or empirical galaxy formation models.
In the EFT context, they are UV-complete models 
available for matching calculations. 

Phenomenological approaches
such as 
the local Lagrangian approximation, 
co-evolution of galaxies and 
dark matter, the peak-background split etc. (see~\cite{Desjacques:2016bnm,Eggemeier:2021cam} and references therein), 
predict certain relationships between galaxy bias 
parameters, which have been often used in actual data analyses, see e.g.~\cite{Beutler:2016arn}.
However, accurate determination of these parameters from numerical simulations
have shown that commonly used phenomenological 
analytic relations are not very accurate in practice, 
see e.g.~\cite{Abidi:2018eyd,Eggemeier:2021cam,Ivanov:2021kcd,Barreira:2021ukk}.
This suggests that their use in data analysis can bias
cosmological parameter recovery. 
An alternative is to match the EFT parameters from 
empirical galaxy formation models based on
hydrodynamical simulations or the halo occupation 
distribution (HOD) approach~\cite{Berlind:2001xk,Zheng:2004id,Zheng:2007zg,Wechsler:2018pic}.

In what follows we will focus on the HOD framework.
This approach is motivated by the fact 
that galaxies fundamentally 
reside inside dark matter halos.
Based on that, the HOD framework naturally assumes that key 
galaxies'
properties are derived from those
of the halos. Emulators based on HOD models have been 
successfully applied to data, see e.g.~\cite{Kobayashi:2021oud,Paillas:2023cpk,Cuesta-Lazaro:2023gbv,Hahn:2023udg,Valogiannis:2023mxf},
which proves that they can reproduce the 
observed galaxy clustering with sufficient accuracy even though their robustness on small scales remains to be determined. 

There have been notable efforts to determine 
EFT parameters from hydrodynamical 
simulations and HOD catalogs~\cite{Barreira:2020kvh,Eggemeier:2021cam,Barreira:2021ukk,Kokron:2021faa,Zennaro:2021pbe}. In particular, Ref.~\cite{Eggemeier:2021cam}
measured non-linear bias parameters of 3 BOSS-like HOD galaxy samples
and 4 halo catalogs using a combination of the power spectrum and bispectrum.
Ref.~\cite{Barreira:2021ukk} extracted second order galaxy 
bias parameters from $\approx 30$
different samples of galaxies from IllustrisTNG hydrodynamical simulations. 
The relatively small size of data points in both of these analyses 
does not allow one to robustly explore the EFT parameter distribution,
which is expected to have a complicated correlation structure.
In particular, one can estimate that building 
an accurate distribution for 14 dimensional parameter space of HOD 
and EFT parameters requires a total of $10^4$ samples.\footnote{This estimate is based on 
a multi-variate normal distribution with random covariance and means, whose 
density is modeled with a normalizing flow. 
}
Previously,
\cite{Zennaro:2021pbe}
measured Lagrangian bias parameters
of a hybrid EFT model
for $\sim$8000 
galaxy and halo samples.
Their results, however, 
cannot be 
directly applied to 
pipelines based on 
the traditional EFT models.

In this paper, we present a new approach for precise 
determination of priors for EFT-based full shape analyses.
The key object of our study is a precise map between 
EFT parameters and HOD parameters for BOSS-like 
galaxy catalogs. We build this map from 10,500 mocks, 
which represents a significant improvement in size over 
previous EFT measurements. 
This map can be used in multiple ways. 
For instance, we consider joint and conditional distributions of 
EFT and HOD parameters, 
$p(\theta_{\rm EFT},\theta_{\rm HOD})$ and $p(\theta_{\rm EFT}\mid\theta_{\rm HOD})$,
respectively,
which clearly display the response of galaxy bias parameters to variations of HOD 
parameters. These distributions give us new insights into the physical meaning 
of galaxy bias parameters and help us connect galaxy bias models on small and large scales. 

Finally, we propose to use the marginal density of the 
EFT parameters from our samples $p(\theta_{\rm EFT})$
as a prior distribution in EFT-based full-shape analyses. 
Similar ideas have been previously presented in \cite{Sullivan:2021sof} in the context of the Halo Zel'dovich perturbaiton theory model.
We demonstrate the power of this approach in 
a new analysis of the BOSS data in the context of inflationary models
with non-local 
single-field 
primordial non-Gaussianity (PNG)
captured by the equilateral 
and orthogonal templates.
This type of non-Gaussianity probes inflaton self-interactions
and the propagation speed~\cite{Arkani-Hamed:2003juy,Alishahiha:2004eh,Senatore:2004rj,Chen:2006nt,Creminelli:2006xe,Cheung:2007st,Cheung:2007sv,Senatore:2009gt}.  
In the past, it has been shown
that constraints 
on single-field PNG 
from large-scale structure
depend significantly
on the assumptions about
non-linear bias parameters
\cite{Cabass:2022wjy,Cabass:2022ymb,Cabass:2022epm,DAmico:2022gki,Chen:2024bdg} (see also \cite{Castorina:2019wmr,MoradinezhadDizgah:2020whw,Cabass:2022ymb,Lazeyras:2022koc,Barreira:2022sey,Barreira:2023rxn,Green:2023uyz} for recent related work and discussions in the context of multi-field inflation).
We find that with our
HOD-based priors 
the constraints improve by $\approx 40\%$,
analogous to a twofold increase
of the survey volume. 
The volume of the 
posterior distribution of non-linear galaxy bias parameters for each 
independent chunk of BOSS data 
shrinks by an order of magnitude.

Our paper is structured as follows.
We describe the technical aspects of 
how to produce a map of EFT and HOD 
parameters in Sec.~\ref{sec:method}.
Sec.~\ref{sec:maps} presents the map
and selected conditional distributions. 
Our PNG constraints on BOSS with HOD-informed 
priors are presented in Sec.~\ref{sec:png}.
Sec.~\ref{sec:hod} compares 
the optimal values of the HOD 
parameters implied by our EFT full shape
analysis to those based on 
other techniques involving 
small scale data. 
Finally, Sec.~\ref{sec:disc} draws conclusions
and discusses directions of further work. 
Additional plots are presented
in Appendix~\ref{app:add}.
Appendix~\ref{app:flows}
summarizes details of our 
normalizing flow
training, while Appendix~\ref{sec:uvsens}
studies the residual UV-dependence of our results.

\section{The Method} 
\label{sec:method}

The main idea of our method is to create a detailed mapping 
(joint probability density)
between EFT and HOD parameters. 
From the physical point of view, it will be interesting to consider a conditional 
version of this mapping, 
$p(\theta_{\rm EFT}|\theta_{\rm HOD})$,
which will describe the 
dependence of the EFT parameters
on the underlying halo model physics.  
In the context of EFT-based full shape analyses, one is interested in the marginal distribution of the EFT parameters ``informed'' by the HOD models,
\be
\label{eq:margdist}
p(\theta_{\rm EFT})=\int p(\theta_{\rm EFT}\mid\theta_{\rm HOD})p(\theta_{\rm HOD})d\theta_{\rm HOD}\,,
\ee
where $p(\theta_{\rm HOD})$ is the prior distribution of the HOD 
parameters. Creating $p(\theta_{\rm EFT})$
is the main practical goal of our paper.

To build a joint EFT/HOD distribution, we measure EFT parameters from 
a large set of simulations with varying HOD parameters. 
This poses several challenges. The first one 
is cosmic variance. A traditional way to 
determine the EFT parameters is to extract them by fitting a
combination of simulated 
correlation functions, typically the power
spectrum and bispectrum, see e.g.~\cite{Saito:2014qha}. 
In this approach, the fits have to be 
performed on large scales where EFT is applicable. But the large scales 
are also most affected by cosmic variance. As a consequence, 
a high precision measurement of the EFT parameters requires
computationally expensive simulations with large volume. 
For example, a measurement
of the quadratic bias parameters with absolute errorbars $\sigma\simeq 0.02$~\cite{Ivanov:2021kcd,Philcox:2022frc}
requires a cumulative simulation volume of 566 $h^{-3}$Gpc$^3$. 
We solve this problem 
by employing a field-level EFT~\cite{Schmittfull:2014tca,Lazeyras:2017hxw,Abidi:2018eyd,Schmidt:2018bkr,Schmittfull:2018yuk,
Elsner:2019rql,
Cabass:2019lqx,Modi:2019qbt,
Schmidt:2020tao,
Schmidt:2020viy,Schmittfull:2020trd,Lazeyras:2021dar,Stadler:2023hea,Rubira:2023vzw,Nguyen:2024yth}
that allows for cosmic variance 
cancellation, and thus enables precision 
calibration of EFT parameters from computationally cheap 
small-volume simulations. 
The field-level EFT 
also provides a 
computationally 
inexpensive way to 
take into account 
information beyond 
the two-point function,
which is important
in order to break
degeneracies 
between EFT parameters. 

In what follows we describe in detail the creation of 
HOD catalogs and  field level EFT fits.

\subsection{HOD mocks}

We build a large set of HOD mocks for BOSS-like galaxies based on the \textsc{AbacusSummit} \citep{Maksimova:2021ynf} suite of simulations.  
For the purposes of this work, we use a set of mocks with an underlying 
\textit{Planck} 2018 baseline cosmological model~\cite{Aghanim:2018eyx}. 
In principle, the EFT parameters depend on cosmology,
so a full analysis of the BOSS data 
with variations of cosmological
parameters 
will require a distribution that samples cosmological 
parameters. This will be presented in future work.
 
We produce a suite of 10,500 mock catalogs 
of galaxies similar to luminous red galaxies of BOSS.
Specifically, we generate galaxies whose linear bias is similar to that of BOSS galaxies (around $\approx 2$), and number density is less or equal to that of the CMASS sample
of BOSS, $\bar n\approx 3.6\cdot 10^{-4}~h^3\text{Mpc}^{-3}$.
Note that the HOD-to-EFT mapping 
will have to be calibrated anew 
for galaxy samples different 
from that of BOSS, e.g. for 
emission line galaxies of DESI.

We 
use available $N$-body simulation boxes with periodic boundary 
conditions from the small boxes \textsc{AbacusSummit}  covariance suite. 
This suite has a total of 1883 independent realizations of 
dark matter initial conditions.
Each box has a site length 500 $h^{-1}$Mpc. 
Our mocks are 
produced from snapshots at $z=0.5$, which matches the typical
redshift of the CMASS-type galaxies of BOSS.
The mocks are fitted with a resolution of $256$ pixels and have been corrected for the cloud-in-cell (CIC) window.
Thanks to the small volume of our simulations, 
we can generate galaxy catalogs and perform EFT fits 
in a very short time, $\sim 20$ and $\sim 40$ sec., 
respectively.

Each $N$-body simulation is populated with HOD galaxies~\cite{Zheng:2007zg}.
We sample  
HOD parameters $\{\log M_{\rm cut},\log M_{1},\log\sigma$,
$\alpha$, $\kappa$, $B_{\rm cen}$, $B_{\rm sat}\}$
from uniform uninformative priors specified in~\cite{Paillas:2023cpk}.
These are: 
\be 
\begin{split}
& \log M_{\rm cut}\in [12.4,13.3]\,,\quad 
 \log  M_1\in [13.2,14.4]\,,\\
& \log \sigma \in [-3.0,0.0]\,,\quad  \alpha \in [0.7,1.5]\,,\quad \kappa \in [0.0,1.5]\,,\\
& B_{\rm cen} \in [-0.5,0.5]\,,\quad 
B_{\rm sat} \in  [-1.0,1.0]\,.
\end{split}
\ee 
In the HOD model, average numbers of central and satellite galaxies are given by 
\be
\begin{split}
& \langle N_c\rangle(M)=\frac{1}{2}\left[1+\text{Erf}\left(\frac{\log M-\log M_{\rm cut}}{\sqrt{2}\sigma}\right)\right]\,,\\
& \langle N_s\rangle(M)=\langle N_c\rangle(M)\left(\frac{M-\kappa M_{\rm cut}}{M_1}\right)^{\alpha}\,,
\end{split} 
\ee
where $\text{Erf}(x)$ is the Gauss error function, 
$M_{\rm cut}$ is the minimum mass of a halo that can host a galaxy, 
$M_1$ is
 the typical (i.e. most probable) halo mass that hosts one satellite galaxy,\footnote{ \cite{Zheng:2007zg} uses $M_1'$ to denote the same quantity.}
$\kappa M_{\rm cut}$ is the minimum mass of a halo that can host a satellite,
$\alpha$ is the slope of the satellite probability distribution function.

We focus on PNG 
analysis in this work.
As mentioned in the Introduction, 
the only relevant parameters
in these case are non-linear bias
parameters $b_2,b_{\mathcal{G}_2}$ which can be 
extracted from real space mocks. 
Hence, for our purpose, 
it is sufficient to 
consider real space clustering only, 
and hence we ignore the velocity bias parameters that are relevant 
in redshift space.

In addition, 
we use  $B_{\rm cen}$, $B_{\rm sat}$ to capture assembly bias
for satellites and centrals~\cite{Yuan:2021izi,2020MNRAS.493.5506H,2021MNRAS.502.3242X}.
Specifically, the 
galaxy
occupation enhancement
is captured by promoting 
$M_{\rm cut}$
and $M_1$ to functions of
$\delta_R$ via
\be 
\begin{split}
& \log M_{\rm cut}\to \log M_{\rm cut} + B_{\rm cen}(\delta_R -0.5)\,,\\
& \log M_{1}\to \log M_{1} + B_{\rm sat}(\delta_R -0.5)\,,
\end{split}
\ee 
where $\delta_R$ is the (smoothed) dark matter overdensity 
around halo centers
ranked within
halo mass bin and normalized to range from 0 to 1,
and we use
a top-hat filter with 
radius 
$R=5~\Mpch$ for smoothing, 
following \cite{Yuan:2021izi}.
Note that the main halo itself is excluded when computing 
$\delta_R$, i.e 
only neighbor halo 
contributions are taken into
account. 
Positive (negative) values of $B_{\rm sat/cen}$ mean 
that galaxies of the relevant type preferably form in 
more (less) dense environments. 

\begin{figure*}
\centering
\includegraphics[width=0.99\textwidth]{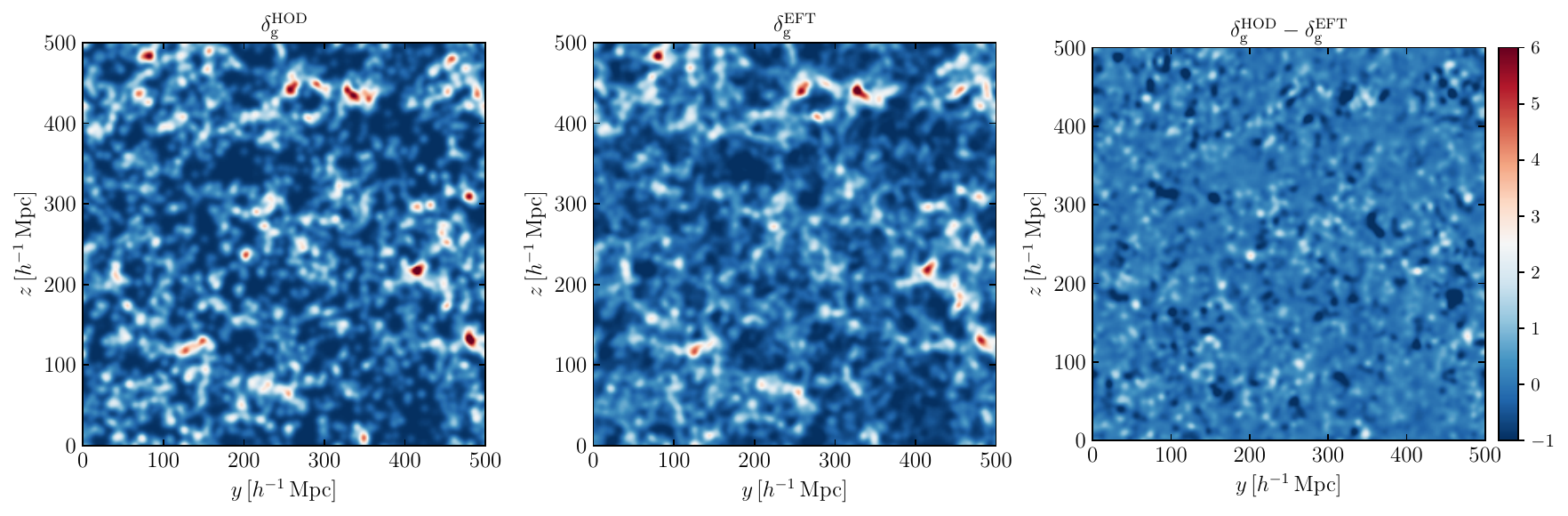}
   \caption{A typical HOD mock galaxy distribution from our set (left), field-level EFT fit to it (center), 
   and the residuals (right). The overdensity field has been smoothed with a $R=4\,\Mpch$ 3D Gaussian filter and the depth of each panel is $\approx60\,\Mpch$.
    } \label{fig:field}
\end{figure*}

\subsection{Field-level EFT}

The standard Eulerian EFT bias model relevant for the one-loop EFT-based
full shape analyses is given by~\cite{Ivanov:2019pdj}
\be 
\label{eq:Eulb}
\delta_{\rm g} = b_1\delta +\frac{b_2}{2}(\delta^2 -\sigma^2)
+b_{\mathcal{G}_2} \mathcal{G}_2 +b_{\Gamma_3}\Gamma_3
-b_{\nabla^2\delta}\nabla^2\delta
+\varepsilon\,,
\ee 
where and $\varepsilon $ is the stochastic density component, 
$\delta$ is the nonlinear dark matter field, $\nabla\equiv \d^i \d_i$, the ``non-local''
bias operators are given by~\cite{Assassi:2014fva}
\be 
\label{eq:Euleft}
\begin{split}
&\mathcal{G}_2 = \left(\frac{\d_i\d_j}{\nabla^2}\delta\right)^2-\delta^2\,,\\
&\Gamma_3 = \frac{4}{7}\delta\mathcal{G}_2 -\frac{4}{7}
\frac{\d_i\d_j\delta}{\nabla^2}\frac{\d_i\d_j\mathcal{G}_2}{\nabla^2} 
\,,\\
\end{split}
\ee 
and $\sigma^2\equiv \langle \delta^2 \rangle$, enforcing 
non-renormalization of the background density.
The direct use of the above model in field-level EFT, 
however, leads to a poor match to simulations because 
the fully non-linear density field $\delta$
above has a strong UV 
sensitivity~\cite{Abidi:2018eyd,Schmittfull:2018yuk}. 
This sensitivity also makes it hard to compare field level results with those based on traditional 
correlation functions because the latter feature 
renormalized biased parameters as opposed to ``bare'' ones in the field-level calculations. In order to compare field-level 
and correlation function EFT calculations in a more consistent manner, 
one would want to evaluate the Eulerian bias model~\eqref{eq:Eulb}
with perturbative matter fields, as is done in the Eulerian EFT loop expansion~\cite{Bernardeau:2001qr,Ivanov:2022mrd}.
The perturbative fields, however, miss large contributions from 
enhanced Zel'dovich displacements~\cite{Zeldovich:1969sb,Bernardeau:2001qr}, which again, leads to a failure 
of naive field-level EFT models to predict the galaxy density field.
A way around is to use shifted operators
proposed in~\cite{Schmittfull:2018yuk,Schmittfull:2020trd,Obuljen:2022cjo},
which have a well controlled small-scale behavior
and retain large displacements at the same time.
A shifted operator $\tilde{\mathcal{O}}$ 
is obtained by shifting the Lagrangian operator $\mathcal{O}({\bf q})$ 
by the Zel'dovich displacement ${\boldsymbol \psi}_1$:
\be
\tilde{\mathcal{O}}(\k)=\int d^3q~\mathcal{O}({\bf q}) 
e^{-i\k\cdot({\bf q}+{\boldsymbol \psi}_1({\bf q}))}\,,
\ee
where ${\bf q}$ are Lagrangian coordinates. Note that in this approach 
higher order Lagrangian displacements are treated perturbatively, 
which is appropriate given that they are suppressed just like 
other operators
in perturbation theory~\cite{Desjacques:2016bnm}.
The statistics of shifted operators
can be shown to be equivalent to IR-resummed Eulerian EFT~\cite{Schmittfull:2018yuk} 
(see~\cite{Senatore:2014via,Baldauf:2015xfa,Blas:2015qsi,Blas:2016sfa,Ivanov:2018gjr} for details of IR resummation).

In the context of real-space clustering that we study here, the 
field level EFT forward
model based on shifted operators takes the form 
\be 
\label{eq:eftmod}
\begin{split}
\delta_{\rm g} \big|_{\rm EFT}= & \beta_1 \tilde\delta_1 + \beta_2(\tilde\delta^2_1)^{\perp} + 
\beta_{\mathcal{G}_2}\tilde{\mathcal{G}}^{\perp}_2   + \beta_3(\tilde\delta^3_1)^{\perp} 
  \,,
\end{split}
\ee 
where $\tilde\delta_1$ is the shifted linear density field $\delta_1$,
$\beta_{n}$ are scale-dependent 
transfer functions.
Importantly, the forward
model~\eqref{eq:eftmod}
is built from orthogonalized
operators that satisfy
\be
\langle \tilde{\mathcal{O}}^{\perp}_m 
\tilde{\mathcal{O}}^{\perp}_n \rangle =0\,,\quad \text{for}\quad m\neq n\,,
\ee 
which 
allows us to 
remove UV-sensitive 
two-loop corrections 
to the transfer 
functions~\cite{Abidi:2018eyd}.

Note that our model contains an extra cubic operator $\delta^3$
that has a first non-trivial contribution 
in the bispectrum at the one loop order~\cite{Eggemeier:2018qae,Philcox:2022frc,DAmico:2022ukl}.
We extract the $k$-dependent 
transfer functions from each mock and then fit them 
with the appropriate EFT templates on mildly-nonlinear scales. 
The 
transfer functions are computed by taking expectation 
values of the simulated galaxy density field multipled by appropriate shifted operators.  
Specifically, the transfer 
function $\beta_i$  of a Fourier space bias operator $\tilde{\mathcal{O}}_i(\k)$ is computed as 
\be
\label{eq:betai}
\beta_i(\k)=\frac{\langle \delta^{\text{HOD}}_g(\k)\tilde{\mathcal{O}}^{\perp}_i{}^*(\k)\rangle}{
	\langle |\tilde{\mathcal{O}}^{\perp}_i(\k)|^2\rangle
}\,,
\ee
where $\delta^{\text{HOD}}_g$ is the galaxy density field from 
simulations.
The use of expectation values
allows us to 
connect the transfer functions 
with renormalized EFT parameters
that appear, e.g. 
in separate universe 
simulations~\cite{Lazeyras:2015lgp} or in correlation functions. 
In particular, at the formal one-loop order, 
we have the following 
expressions 
in the $k\to 0$ limit
\footnote{Note that we
have absorbed the dark matter sound speed parameter into $b_{\nabla^2\delta}$ here. } 
\be 
\label{eq:tflowk}
\begin{split}
 \beta_1 & = b_1  + b_{\nabla^2\delta} k^2 
+ \left( b_{\Gamma_3}+\frac{b_1}{6}+\frac{5}{2}b_{\mathcal{G}_2}\right)\frac{\langle \tilde\delta_1 \tilde \Gamma_3\rangle}{\langle \tilde\delta_1 \tilde\delta_1 \rangle }
\\&+\frac{b_{2}}{2}
\frac{\langle \tilde\delta_1 \tilde \delta_1^2\rangle}{\langle \tilde\delta_1 \tilde\delta_1 \rangle } 
+ \left(b_{\mathcal{G}_2}+\frac{2b_1}{7}\right)
\frac{\langle \tilde\delta_1 \tilde{\mathcal{G}}_2\rangle}{\langle \tilde\delta_1 \tilde\delta_1 \rangle } -b_1
\frac{\langle \tilde\delta_1 \tilde{\mathcal{S}}_3\rangle}{\langle \tilde\delta_1 \tilde\delta_1 \rangle }\,, \\
 \beta_2 & = \frac{b_2}{2}\,,\quad \beta_{\mathcal{G}_2}=b_{\mathcal{G}_2}+\frac{2}{7}b_1, \quad \beta_3 = \frac{b_3}{6}\,.
\end{split}
\ee 
Note that the constant coefficients
here are different from those of~\cite{Schmittfull:2018yuk} because we use the Eulerian
bias parameters from Eq.~\eqref{eq:Euleft} (matched at the cubic order),
while the parameters used in~\cite{Schmittfull:2018yuk} are  more closely related 
to the Lagrangian bias parameters. 
The cubic operator $\tilde{\mathcal{S}}_3$ is the shifted version of
\be 
\mathcal{S}_3={\boldsymbol \psi}_2({\bf q})\cdot \nabla \delta_1({\bf q})\,,
\ee
which is produced by the second order 
displacement ${\boldsymbol \psi}_2$. 
Note that the presence of this operator 
is enforced by the equivalence principle.

In practice, 
we use public \texttt{Hi-Fi mocks}\footnote{\url{https://github.com/andrejobuljen/Hi-Fi_mocks}} to produce the EFT forward model.
A typical snapshot generated with the field-level EFT forward model and its residual with the original simulated galaxy field are shown in Fig.~\ref{fig:field}. 

\begin{figure*}
\centering
\includegraphics[width=1.00\textwidth]{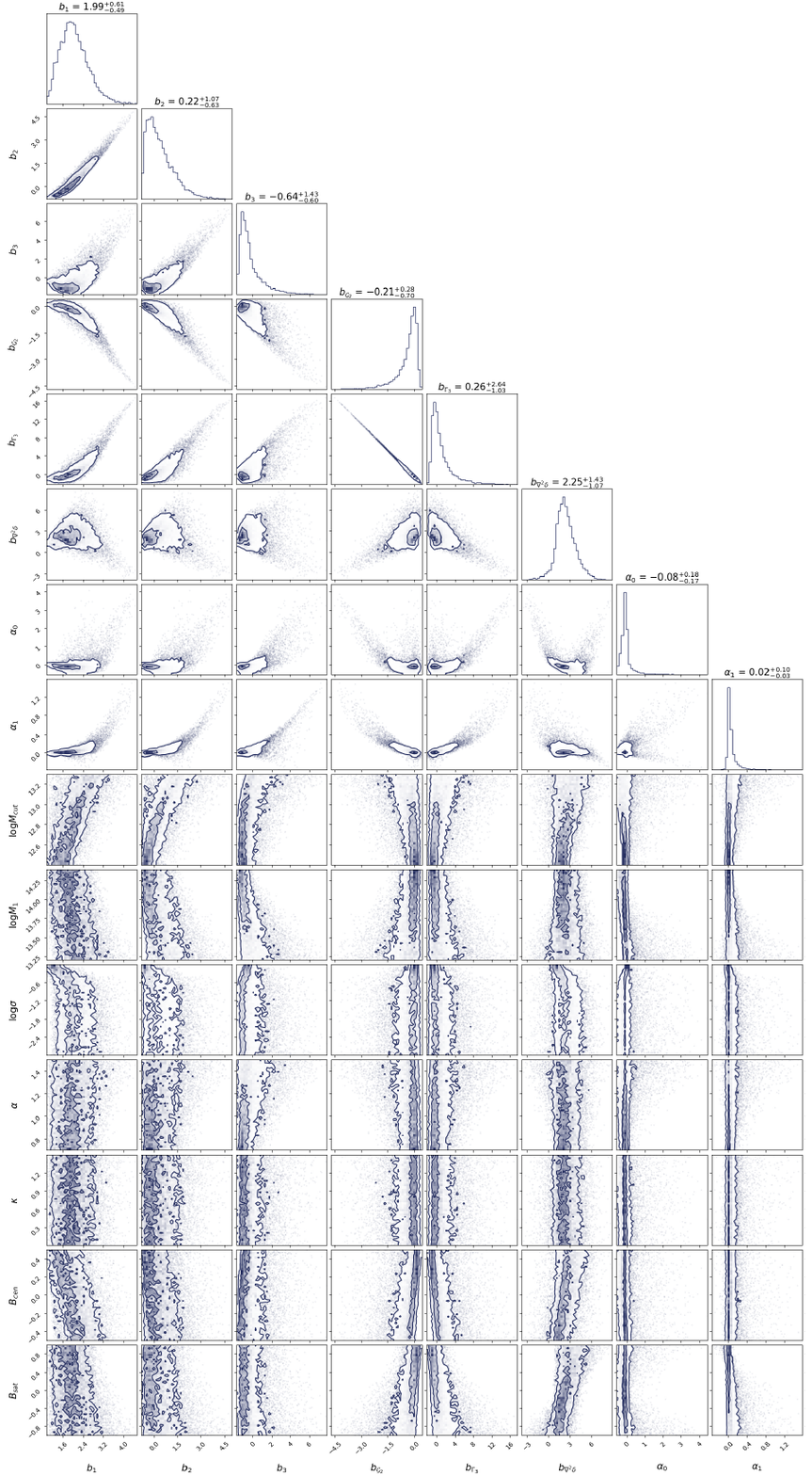}
   \caption{The joint distribution of EFT and HOD parameters
   extracted from 10,500 HOD mocks for BOSS-like galaxies.
Density levels correspond to two-dimensional $1$-$\sigma$
  and $2$-$\sigma$ intervals (i.e. 39.3\% and 86.5\% of samples). Individual 
  samples are also shown as dots. They are
   especially pronounced in the 
  tails.
    } \label{fig:dist}
\end{figure*}

We extract the bias parameters from the $k\to 0$ limit of the transfer functions
using Eq.~\eqref{eq:tflowk}. In practice we use $\kmax=0.4~\hMpc$, 
for which the one-loop EFT models are reliable~\cite{Chudaykin:2020hbf,Ivanov:2021fbu}.
We have checked that a
more conservative 
choice of $\kmax=0.3~\hMpc$
gives results consistent
with those of $\kmax=0.4~\hMpc$, but
with a somewhat larger scatter
in their distribution. This is 
the reason we adopt 
$\kmax=0.4~\hMpc$ as a baseline choice. 

To account for the scatter 
in transfer functions 
on large scales, we adopt error weights
for $k$-bins based on the number of  Fourier modes.
For $b_2,b_{\mathcal{G}_2},b_3$, we fit the transfer functions with 
a polynomial $c_0 + c_2 k^2  + c_4 k^4$, and then match $c_0$
to the constant values of bias parameters in Eq.~\eqref{eq:tflowk}.
As far as $\beta_1$ is concerned, we  
calculate the power spectra and cross-spectra of shifted
operators in Eq.~\eqref{eq:tflowk}, and use them, along with the best-fit
values of $b_2$ and $b_{\mathcal{G}_2}$ from the previous step, to 
fit the transfer function at low $k$,
which yields $b_{\nabla^2\delta}$
and $b_{\Gamma_3}$. Plots with typical 
transfer functions fits can be found in 
Appendix~\ref{app:add}.

\begin{figure*}[htb!]
\centering
\includegraphics[width=0.95\textwidth]{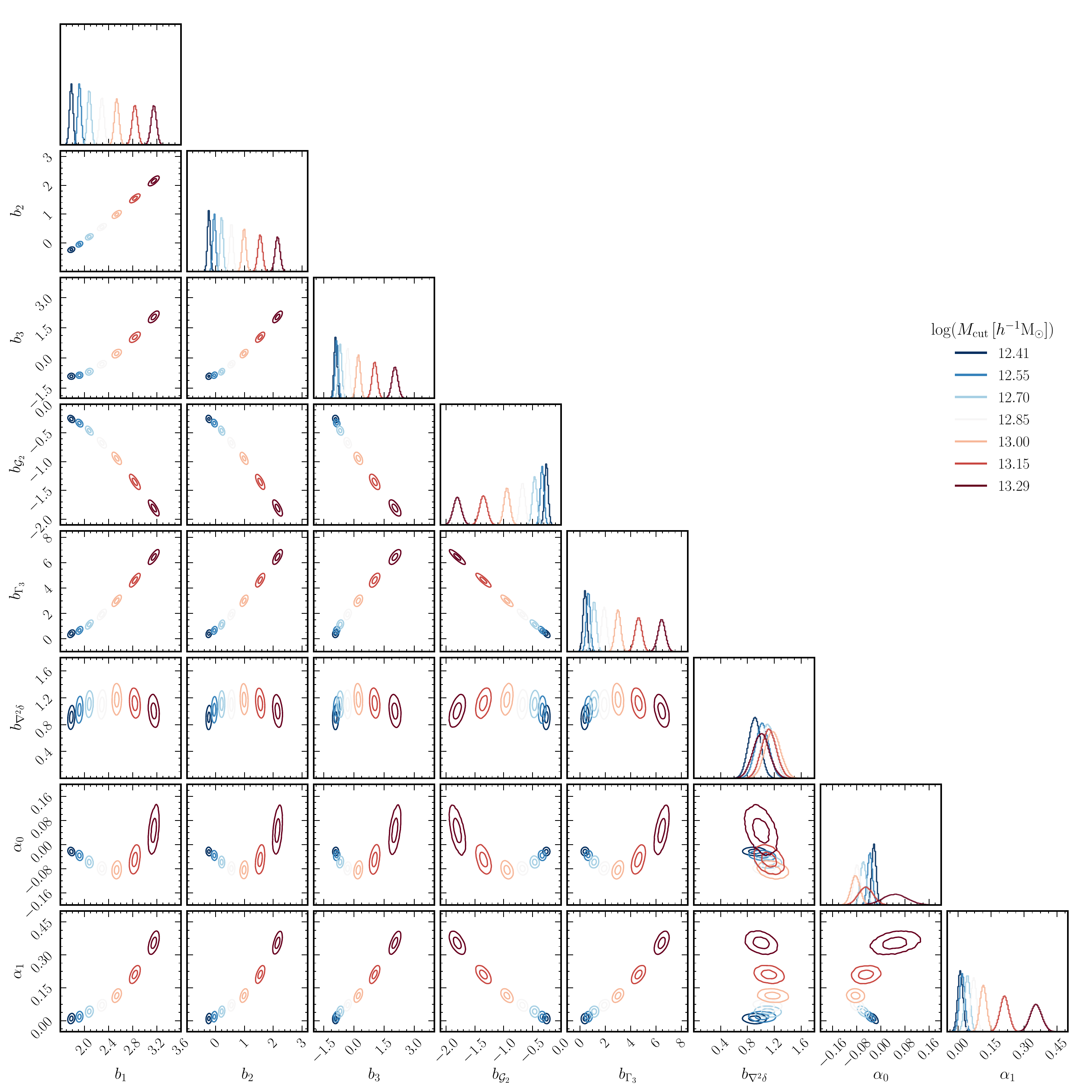}
   \caption{\small Conditional distributions of EFT parameters for different values of 
   $\log M_{\rm cut}$ (see legend). Other HOD parameters are kept fixed. Density levels correspond to two-dimensional $1$-$\sigma$
  and $2$-$\sigma$ intervals.
    } \label{fig:cond}
\end{figure*}
\begin{figure*}[htb!]
\centering
\includegraphics[width=0.95\textwidth]{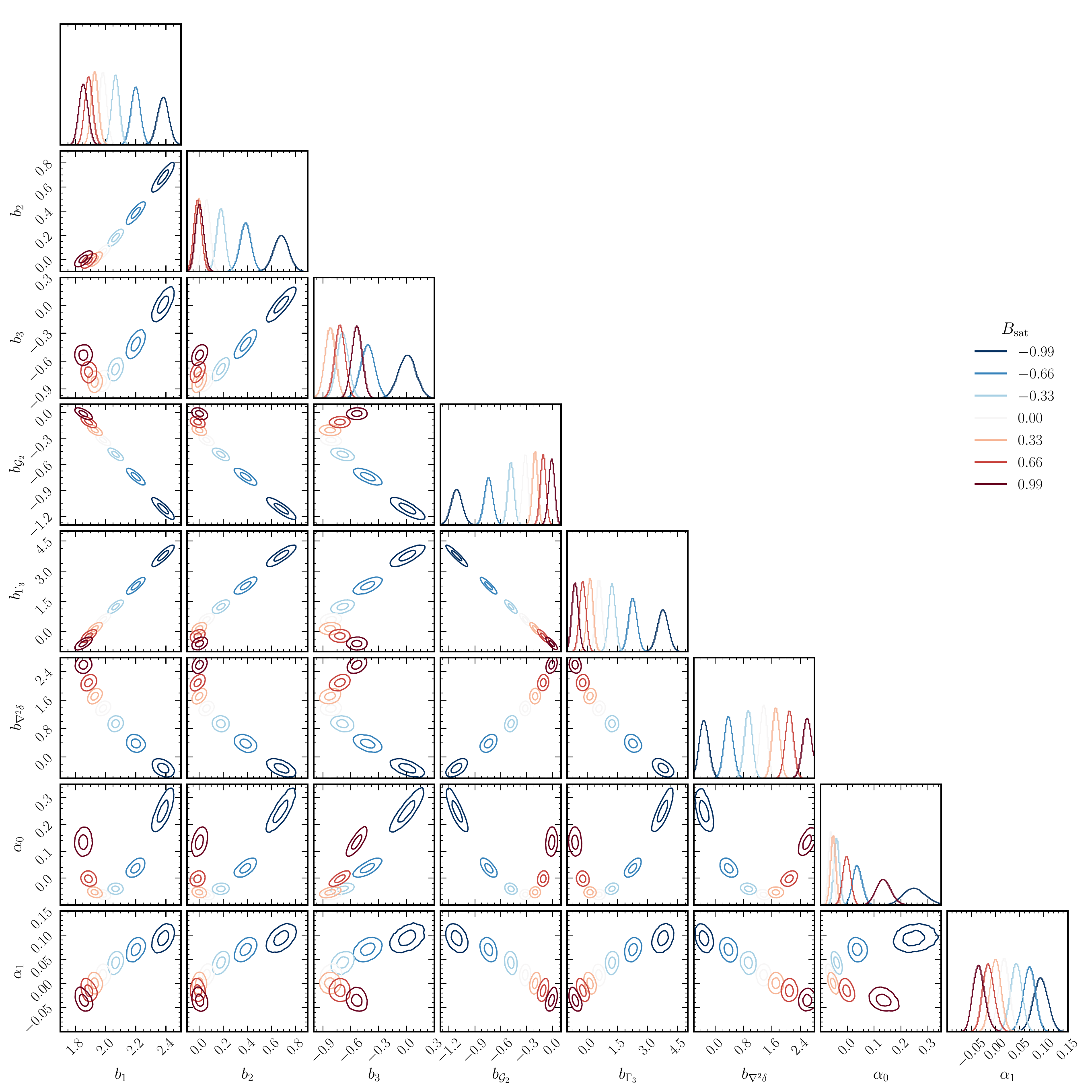}
   \caption{\small Same as Fig.~\ref{fig:cond}, but 
   $B_{\rm sat}$ is varied (see legend) while other parameters are kept fixed. 
    } \label{fig:cond2}
\end{figure*}

The final ingredient that we need 
is the distribution of stochasticity parameters, characterizing 
the power spectrum of the $\varepsilon$ field. 
In practice, we calculate the error power spectrum as 
\be
P_{\rm err}(k)=\langle |\delta_g^{\rm HOD}(\k)-\delta^{\rm EFT}_g(\k)|^2 \rangle \,.
\ee
Theoretical consistency dictates that on 
large scales it should match the EFT prediction 
for the stochastic contribution to the galaxy power spectrum
~\cite{Schmittfull:2018yuk,Ivanov:2021zmi,Ivanov:2021kcd} 
\be 
\label{eq:perreft}
\begin{split}
& P_{\rm err}(k) = \frac{1}{\bar n}\left(1+\a_{0} + \a_{1} \left(\frac{k}{k_{\rm NL}}\right)^2\right)\,,
\end{split}
\ee
where $\bar n=V/N_{\rm gal}$ is the number density of mock galaxies ($V$ being the simulation box volume), 
and we chose $k_{\rm NL}=0.45~\hMpc$ following~\cite{Ivanov:2021kcd}.
Note that
the EFT fitting pipelines~(e.g.~\cite{Ivanov:2021zmi,Ivanov:2021kcd})
use parameters 
$P_{\rm shot}$ and $a_0$ 
that are similar to our
$\alpha_{0}$
and $\alpha_{1}$, respectively. 
However, as we discuss in detail later, these ``standard''
EFT models
also absorb additional 
contributions into the stochastic 
parameters, which make 
them somewhat different from
our $\alpha_0$ and $\alpha_1$.

Before closing this part, 
let us comment on
errors in our measurements
of bias parameters.
There are two main sources of errors: 
the residual UV dependence 
of the transfer functions
on large scales
and the numerical noise due to 
the use of a single realization
per mock in our fits. 
The use of orthogonalized 
shifted operators
and $k$-dependent transfer functions
allowed us to 
remove the bulk of 
the UV sensitivity of our 
bias parameters. 
Within the original field-level method 
for cosmic variance cancellation, see e.g.~\cite{Abidi:2018eyd},
the bias parameters stem 
from a global fit of a constant to many $k$-bins, which is sensitive to small scale nonlinearity and higher-order loop effects. 
In order to reduce this sensitivity, 
one may introduce a relatively aggressive cutoff for the linear fields (\cite{Abidi:2018eyd} used
a Gaussian filter with $R_s=20~\Mpch$). This cutoff generates a mismatch between the 
field-level and $n$-point functions results, as the latter are formally extracted in the $R_s\to 0$ limit.\footnote{In principle, this mismatch can be compensated with the ``renormalization group'' technique of~\cite{Rubira:2023vzw}.} This is because  many EFT codes, e.g.~\texttt{CLASS-PT}~\cite{Chudaykin:2020aoj} use dimensional regularization for loop integrals that compute correlation functions, in which all convergent integrals are done without 
any cutoff, while the divergent pieces 
are set to zero. In the field level language this means the matching should be done after the orthogonalization (which removes divergent corrections proportional to the mass variance integrals),
and without smoothing of the linear fields (i.e. at $R_s\to 0$).

In this work, however, we 
follow the method~\cite{Schmittfull:2018yuk}
and extract the bias parameters from
the low-$k$ limits of the transfer
functions, in which case the UV-sensitive higher-order corrections are absorbed into 
the $k^2$ and $k^4$ polynomials
of our fit, leaving the constant part 
unaffected provided one uses a reasonably high cutoff. As an explicit check, we demonstrate that the low-$k$ transfer functions used in our fits
are stable w.r.t. variations 
of the grid resolution, which provides
an effective cutoff for all fields, see Appendix~\ref{sec:uvsens}
for more detail. This test shows that our results have converged w.r.t. the $R_s\to 0$ limit relevant for the matching to the $n$-point functions in dimensional regularization.

As far as numerical noise is concerned, 
results of Ref.~\cite{Schmittfull:2018yuk}
obtained with a similar box size
suggest that the corresponding error
is below the level 
of the scatter induced by variations 
of HOD models, implying 
that these effects are
negligible for parameter constraints.\footnote{If needed, the numerical 
noise can be reduced 
by averaging over multiple independent 
realizations.}

\subsection{Normalizing flow for density estimation and sampling}
\label{sec:flows}

The crux of our approach is the ability to effectively model the marginal and HOD-conditional EFT parameter distributions, $p\left(\theta_\mathrm{EFT}\right)$ and $p\left(\theta_\mathrm{EFT}\mid\theta_\mathrm{HOD}\right)$ respectively, given a set of samples $\{\theta_\mathrm{EFT},\theta_\mathrm{HOD}\} \sim p\left(\theta_\mathrm{EFT},\theta_\mathrm{HOD}\right)$ as described above. Ideally, we would like to be able to generate new samples from these distributions (i.e., use them as priors), and evaluate the likelihood of a new set of samples under the modeled distributions.

Generative models are a class of statistical models that aim to encode potentially complex target distributions. While high-dimensional distribution modeling is an inherently challenging task, machine learning has brought this into the realm of tractability. Normalizing flows~\cite{rezende2015variational}, a class of \emph{deep} generative models, are especially suited for our purposes, as they allow for seamless density estimation as well as sampling, including in the conditional regime. Briefly, normalizing flows model the target density $\hat p(\theta)$ as a bijective (invertible) learnable function $\theta = f_\varphi(u)$ from a simple distribution $\pi(u)$, e.g. a multivariate Gaussian,
\begin{equation}
\hat{p}(\theta)=\pi(u)\left|\operatorname{det}\left(\frac{\partial u}{\partial \theta}\right)\right|=\pi\left(f_\varphi^{-1}(\theta)\right)\left|\operatorname{det} J_{f_\varphi^{-1}}(\theta)\right|
\label{eq:flow}
\end{equation}
where $\left|\operatorname{det} J_{f_\varphi^{-1}}(\theta)\right|$ is the Jacobian determinant of the inverse transformation and is by construction easy to compute. $\varphi$ are the parameters of the learnable transformation, usually modeled through an appropriate invertible neural network.

Given a set of samples $\{\theta\}\sim p(\theta)$ (e.g, HOD and EFT parameters), we can maximize Eq.~\eqref{eq:flow} over those samples to train the flow, and thus build an approximation for our target density. The optimal parameters $\varphi^*$ of the transformation are obtained, in practice through stochastic gradient descent or a variant thereof, as
\begin{equation}
    \varphi^* = \argmax_\varphi\,\langle\log \hat p(\theta)\rangle_{\theta\sim p(\theta)}.
\end{equation}
The target density can then easily be sampled from, by drawing $u\sim \pi(u)$ from the simple (Gaussian) base density and running the forward transformation $f_\varphi(u)$, as well as evaluated for a new sample $\theta'$. The flows are implemented using the \texttt{nflows}\footnote{\url{https://github.com/bayesiains/nflows}} library, with training and evaluation performed using \texttt{PyTorch}~\cite{paszke2019pytorch}. Further details of the normalizing flow implementation, training, and validation are given in Appendix~\ref{app:flows}.

\section{Distributions of galaxy bias and HOD parameters}
\label{sec:maps}

The joint samples of EFT parameters for given HOD 
models are displayed in Fig.~\ref{fig:dist}. 
Numerical values of 
$b_{\nabla^2\delta}$ are 
given in $[\Mpch]^2$ units.
We do not show the 
samples of the HOD parameters as they simply scan over 
their uniform priors.
We see that the 
EFT parameters are strongly correlated among each other.
This is the behavior expected on the basis that the (infinite) entire 
set of EFT parameters must be produced by only 7 HOD parameters.
(In the particle physics context, the situation here is analogous to the textbook 
matching of chiral perturbation theory 
to the linear sigma model, where 
low energy constants are either zero, or obey correlations set by
a few parameters of the linear sigma model.)
The tightest correlation is between $b_{\Gamma_3}$ and $b_{\mathcal{G}_2}$,
which follows the linear law
\be
 b_{\Gamma_3}\approx -3.8 b_{\mathcal{G}_2}-0.5\,.
\ee
Note that this law is steeper 
than the local Lagrangian
model prediction $b_{\Gamma_3}=-\frac{23}{12}b_{\mathcal{G}_2}$,
and the co-evolution model relation $b_{\Gamma_3}\sim -1.5b_{\mathcal{G}_2}$~\cite{Eggemeier:2021cam}.
The strong correlation 
between $b_{\mathcal{G}_2}$
and $b_{\Gamma_3}$ may be an artifact 
of our forward model, which absorbs 
the contribution of the $\Gamma_3$ operator
into $\beta_1$, thus making 
our measurement of $\Gamma_3$ sensitive 
to details of the fitting procedure. 
It will be interesting 
to compare our results
with independent measurements 
of $b_{\Gamma_3}$ from a forward 
model with the shifted 
$\Gamma_3$ operator.

Another important observation is that the higher-derivative stochastic counterterm
$\alpha_{1}$
is very close to zero for most of the samples. 
A similar pattern was 
discovered before 
for the clustering of dark matter 
halos~\cite{Schmittfull:2018yuk}.
It would be interesting 
to see if this pattern is specific to HOD 
models, 
i.e. if other galaxy formation prescriptions can produce larger $\alpha_{1}$ matching the natural EFT 
expectation that this parameter should be proportional 
to the Lagrangian radius of the host halo. 
This expectation is supported by the positive correlation
between $\alpha_{1}$ and $b_1$.
Note that $\alpha_{1}$
is mostly correlated with $M_{\rm cut},M_1$ and the satellite environmental assembly bias 
$B_{\rm sat}$. 
These correlations can be understood as follows: non-locality in the galaxy stochasticity 
is sourced by the stochastic non-locality of the underlying halo 
(set by $M_{\rm cut}$)
plus the stochasticity from 
the satellites, whose spatial
distribution increases the  
non-locality scale, thus generating 
additional 
$M_1$ and $B_{\rm sat}$ dependencies.

Let us discuss now the correlation between $P_{\rm shot}$ (deviation of white noise
amplitude from $\bar n^{-1}$) and $b_1$. Due to halo exclusion effects~\cite{Casas-Miranda:2001dwz,Cooray:2002dia,Baldauf:2013hka,Baldauf:2015fbu}, 
we expect that samples with large enough $b_1$ should feature sub-Poisson 
stochastic shot noise, 
i.e. have negative $\alpha_{0}$.
This is indeed the case, 
on average. However, massive halos also tend to have more satellites, 
which would increase $\bar n$, and hence compensate for the sub-Poisson trend 
produced by the halos. 
This is also is supported by correlations between 
$\alpha_{0}$ and satellite properties such as $B_{\rm sat}$.
Indeed, the observation 
that satellite galaxy
assembly bias 
produces super-Poissonian
sampling was
pointed out before in \cite{Hearin:2015jnf}.
This is the reason why we do not see a strong sub-Poisson constant shot 
noise contribution for galaxies that reside in massive halos. 
A similar trend
was pointed out in a previous work on the field-level calibration 
of $\alpha_{0}$ from HOD~\cite{Kokron:2021faa}, with which we find 
excellent agreement. 

Note that other prominent correlations are $b_1-B_{\rm cen}$
and $b_{\nabla^2\delta}-B_{\rm sat}$, which suggest that the measurements
of these parameters can be used to diagnose the presence of environmental 
assembly bias.

In order to visualize the response of the bias parameters
to variations of the HOD parameters, we study the 
conditional distribution $p(\theta_{\rm EFT}\mid\theta_{\rm HOD})$. 
It indicates what are the most likely EFT parameters
for a given fixed set of HOD parameters.
This distribution is extracted by modeling the original samples 
with normalizing flows as described in Sec.~\ref{sec:flows}.

The fiducial conditional distribution
of EFT parameters that we study is calculated for the set of HOD parameters fixed to 
the best-fit values of the CMASS BOSS sample~\cite{Paillas:2023cpk},
\be 
\begin{split}
& \log M_{\rm cut}=12.66\,,\quad \log M_{1}=13.66\,,\quad \alpha = 1.34\,,\\
& \log\sigma = -0.5\,,\quad \kappa = 0.03\,,\\
& B_{\rm cen}=-0.43\,,\quad B_{\rm sat}=-0.22\,.
\end{split}
\ee
As a first example, we plot $p(\theta_{\rm EFT}\mid\theta_{\rm HOD})$
for 7 HOD models with different $\log M_{\rm cut}$ and all other
parameters fixed. The result is shown in Fig.~\ref{fig:cond}.
As anticipated from the simple thresholding picture~\cite{Desjacques:2016bnm}, 
the higher cutoff mass of the halo 
implies a lower probability of halo formation and 
hence higher bias of the host halos and galaxies.
Note that the higher-derivative bias 
$b_{\nabla^2\delta}$ is not very sensitive to the cutoff mass,
which is at odds with the naive expectation that this parameter 
should be proportional to the Lagrangian radius of the host halo.
It would be interesting to understand the origin of this behavior. 
This intuition is, however, confirmed by the higher derivative 
stochastic counterterm $\alpha_{1}$ which grows with cutoff mass.

Having confirmed that our conditional distribution 
reflects the basic intuition about the galaxy bias, 
let us consider now the response of 
EFT coefficients to a less intuitive parameter, the environment-based
assembly (secondary)
bias of satellites $B_{\rm sat}$. 
The corresponding 
conditional distributions for 7 samples of galaxies
with different $B_{\rm sat}$ (other HOD parameters are fixed) 
is shown in Fig.~\ref{fig:cond2}.
The first relevant observation is that 
local-in-matter-density bias parameters $b_1,b_2,b_3$
display only a weak dependence on $B_{\rm sat}$. This is consistent
with the expectation that these parameters are mostly determined 
by halo mass. The small residual dependence on $B_{\rm sat}$
can be explained as the tendency to have more/less galaxies 
for positive/negative $B_{\rm sat}$, which leads to lower/higher 
values of local biases. The presence of satellite assembly bias is best reflected by 
non-local bias parameters $b_{\mathcal{G}_2},b_{\Gamma_3},b_{\nabla^2\delta}$
and higher-derivative stochastic counterterm $\alpha_{1}$, which is consistent with 
the fact that assembly bias is an intrinsically non-local property
of galaxies. Another interesting observation is that a strong satellite assembly bias 
leads to super-Poisson shot noise (positive $P_{\rm shot}$) regardless 
of the sign of $B_{\rm sat}$.

The distributions of EFT parameters conditioned to other 
HOD models have a similar behavior that can be qualitatively 
understood on the basis of simple physical arguments
such as thresholding.

\section{ PNG constraints from BOSS with HOD-informed priors}
\label{sec:png}

In this section, we re-analyze the BOSS data within initial 
conditions with non-local primordial non-Gaussianity. 
In this analysis, following the standard practice~\cite{Planck:2019kim,Cabass:2022wjy,Cabass:2022ymb}, 
we fix the ``standard''
cosmological parameters to the Planck best-fit values,
and vary only the EFT and PNG parameters. Single-field PNG is captured by the equilateral and orthogonal templates identical to the ones used in~\cite{Cabass:2022wjy}. 
Note that 
we use the orthogonal template that has the correct 
physical scaling in the squeezed limit.

\begin{figure*}[htb]
\centering
\includegraphics[width=0.9\textwidth]{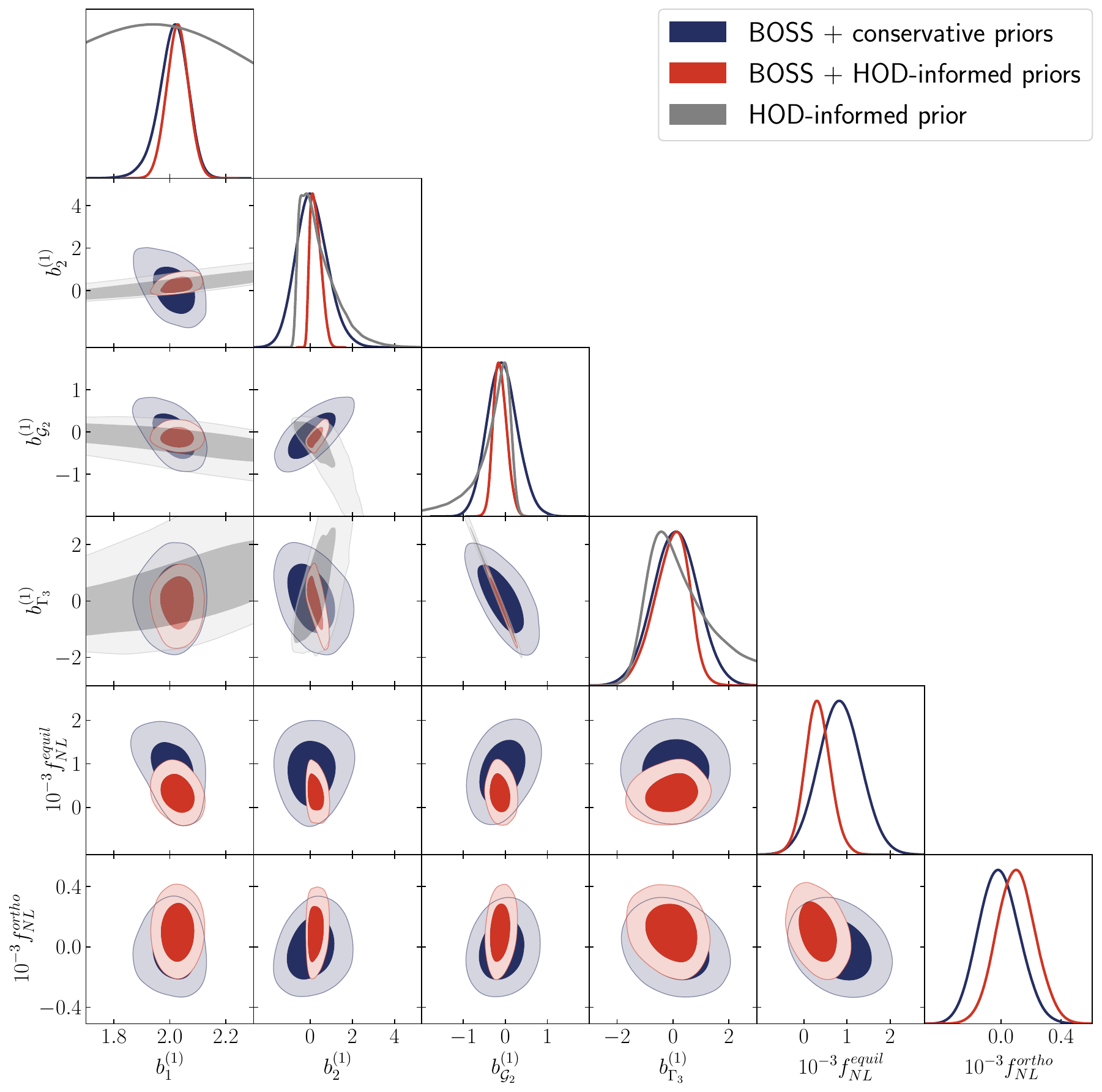}
   \caption{\small Corner plots with 2d and 1d marginalized posterior
   distribution of the PNG parameters from the full BOSS 
   DR12 dataset and the 
   galaxy bias parameters from BOSS NGC high-z sample.
   HOD priors on bias parameters
   are shown in gray.
   Bias parameters for other samples
   are shown in Appendix.
   Density levels correspond to $68\%$
   and $95\%$ CL.
    } \label{fig:triangle_small}
\end{figure*}

\begin{table*}
\begin{tabular}{|l|c|c|c|c|}
 \hline
 \multicolumn{5}{|c|}{BOSS PNG with HOD-informed priors} \\
\hline
PParam & best-fit & mean$\pm\sigma$ & 95\% lower & 95\% upper \\ \hline
$b^{(1)}_{1 }$ &$2.04$ & $2.028_{-0.04}^{+0.04}$ & $1.949$ & $2.106$ \\
$b^{(1)}_{2 }$ &$-0.02737$ & $0.2632_{-0.33}^{+0.18}$ & $-0.164$ & $0.7636$ \\
$b^{(1)}_{{\mathcal{G}_2 }}$ &$-0.2879$ & $-0.1257_{-0.18}^{+0.14}$ & $-0.4172$ & $0.21$ \\
$b^{(1)}_{{\Gamma_3} }$ &$0.5578$ & $-0.04866_{-0.51}^{+0.72}$ & $-1.357$ & $1.062$ \\
$b^{(2)}_{1 }$ &$2.189$ & $2.164_{-0.054}^{+0.055}$ & $2.058$ & $2.272$ \\
$b^{(2)}_{2 }$ &$0.149$ & $0.4281_{-0.32}^{+0.18}$ & $-0.01538$ & $0.9462$ \\
$b^{(2)}_{{\mathcal{G}_2 }}$ &$-0.4428$ & $-0.2394_{-0.18}^{+0.15}$ & $-0.564$ & $0.1174$ \\
$b^{(2)}_{{\Gamma_3} }$ &$1.166$ & $0.3628_{-0.59}^{+0.71}$ & $-1.022$ & $1.624$ \\
$b^{(3)}_{1 }$ &$1.899$ & $1.886_{-0.041}^{+0.041}$ & $1.806$ & $1.966$ \\
$b^{(3)}_{2 }$ &$-0.1266$ & $0.06905_{-0.29}^{+0.17}$ & $-0.3182$ & $0.5086$ \\
$b^{(1)}_{{\mathcal{G}_2 }}$ &$-0.173$ & $-0.03168_{-0.17}^{+0.12}$ & $-0.3022$ & $0.2751$ \\
$b^{(3)}_{{\Gamma_3} }$ &$0.1139$ & $-0.3867_{-0.44}^{+0.65}$ & $-1.562$ & $0.6154$ \\
$b^{(4)}_{1 }$ &$1.974$ & $1.942_{-0.054}^{+0.055}$ & $1.834$ & $2.049$ \\
$b^{(4)}_{2 }$ &$-0.1232$ & $-0.00013_{-0.21}^{+0.13}$ & $-0.3064$ & $0.3515$ \\
$b^{(1)}_{{\mathcal{G}_2 }}$ &$-0.2592$ & $-0.2181_{-0.12}^{+0.13}$ & $-0.4913$ & $0.04831$ \\
$b^{(4)}_{{\Gamma_3} }$ &$0.4536$ & $0.3303_{-0.48}^{+0.44}$ & $-0.6572$ & $1.373$ \\
$10^{-3}f^{\rm equil}_{{\rm NL} }$ &$0.4511$ & $0.3233_{-0.31}^{+0.29}$ & $-0.2824$ & $0.9446$ \\
$10^{-3}f^{\rm ortho}_{{\rm NL} }$ &$-0.03236$ & $0.09964_{-0.14}^{+0.13}$ & $-0.1556$ & $0.3596$ \\
\hline
 \end{tabular} 
 \begin{tabular}{|l|c|c|c|c|}
 \hline
  \multicolumn{5}{|c|}{BOSS PNG with conservative priors} \\
\hline
Param & best-fit & mean$\pm\sigma$ & 95\% lower & 95\% upper \\ \hline
$b^{(1)}_{1 }$ &$2.003$ & $2.013_{-0.048}^{+0.058}$ & $1.903$ & $2.119$ \\
$b^{(1)}_{2 }$ &$0.1239$ & $0.07676_{-0.8}^{+0.7}$ & $-1.426$ & $1.598$ \\
$b^{(1)}_{{\mathcal{G}_2 } }$ &$-0.01367$ & $-0.08101_{-0.38}^{+0.34}$ & $-0.782$ & $0.6405$ \\
$b^{(1)}_{{\Gamma_3} }$ &$0.08801$ & $0.09261_{-0.81}^{+0.82}$ & $-1.505$ & $1.693$ \\
$b^{(2)}_{1 }$ &$2.146$ & $2.147_{-0.061}^{+0.068}$ & $2.018$ & $2.274$ \\
$b^{(2)}_{2 }$ &$0.3097$ & $0.1222_{-0.81}^{+0.74}$ & $-1.418$ & $1.694$ \\
$b^{(2)}_{{\mathcal{G}_2 } }$ &$0.1676$ & $0.0117_{-0.4}^{+0.38}$ & $-0.7551$ & $0.7937$ \\
$b^{(2)}_{{\Gamma_3} }$ &$0.7921$ & $0.2227_{-0.89}^{+0.87}$ & $-1.512$ & $1.955$ \\
$b^{(3)}_{1 }$ &$1.895$ & $1.888_{-0.049}^{+0.052}$ & $1.785$ & $1.99$ \\
$b^{(3)}_{2 }$ &$-0.2482$ & $-0.3937_{-0.64}^{+0.66}$ & $-1.681$ & $0.8862$ \\
$b^{(3)}_{{\mathcal{G}_2 } }$ &$-0.2097$ & $-0.1933_{-0.32}^{+0.3}$ & $-0.805$ & $0.4273$ \\
$b^{(3)}_{{\Gamma_3} }$ &$0.1215$ & $-0.006442_{-0.78}^{+0.77}$ & $-1.523$ & $1.531$ \\
$b^{(4)}_{1 }$ &$1.955$ & $1.948_{-0.063}^{+0.062}$ & $1.823$ & $2.072$ \\
$b^{(4)}_{2 }$ &$-0.835$ & $-0.6685_{-0.72}^{+0.73}$ & $-2.073$ & $0.7085$ \\
$b^{(4)}_{{\mathcal{G}_2 } }$ &$-0.5038$ & $-0.3575_{-0.36}^{+0.35}$ & $-1.049$ & $0.326$ \\
$b^{(4)}_{{\Gamma_3} }$ &$1.588$ & $1.086_{-0.84}^{+0.86}$ & $-0.6011$ & $2.746$ \\
$10^{-3}f^{\rm equil}_{{\rm NL} }$ &$0.5977$ & $0.8062_{-0.51}^{+0.49}$ & $-0.1857$ & $1.793$ \\
$10^{-3}f^{\rm ortho}_{{\rm NL} }$ &$-0.02401$ & $-0.008201_{-0.15}^{+0.13}$ & $-0.2793$ & $0.2653$ \\
\hline
 \end{tabular} 
 \caption{Best-fits and 1d marginalized limits for PNG and galaxy bias parameters 
 from BOSS with HOD-informed priors (left panel)
 and the usual conservative priors (right panel).
 Upper scripts $(1,2,3,4)$ refer to 
 NGCz3, SGCz3, NGCz1, and SGCz1 samples (z1$=0.38$, z3$=0.61$).}
 \label{eq:tab1}
 \end{table*}

Our dataset is BOSS DR12 galaxy clustering in
redshift space~\cite{BOSS:2016wmc}. 
The BOSS DR12 data is split in four chunks: 
low-z ($z_{\rm eff}=0.38$) and high-z ($z_{\rm eff}=0.61$), 
South and North Galactic Caps. 
For each chunk, our datavector is 
$\{P_0,P_2,P_4,Q_0,B_0,B_2,B_4,\alpha_\parallel,\alpha_\perp\}$,
where $P_{\ell}, B_\ell$ are window-free power spectrum and bispectrum 
multipoles, $\ell=0,2,4$~\cite{Chudaykin:2020ghx,Philcox:2021kcw,Ivanov:2023qzb}, 
$Q_0$ is the real-space power spectrum proxy~\cite{Ivanov:2021fbu}, 
while $\alpha_\parallel$,$\alpha_\perp$ are the post-reconstructed 
BAO parameters~\cite{Philcox:2020vvt}. 
We use scale cuts $\kmax^{P_\ell}=0.2~\hMpc$,
$\kmax^{B_\ell}=0.08~\hMpc$, 
$\kmax^{Q_0}=0.4~\hMpc$ and $k_{\rm min}=0.01~\hMpc$.
The covariance matrix is estimated from Multidark Patchy mocks~\cite{Kitaura:2015uqa,Rodriguez-Torres:2015vqa}. 
We assume that EFT parameters in four chunks are independent
of each other. Other aspects 
of our analysis are identical to those of~\cite{Cabass:2022epm,Ivanov:2023qzb}. 

We first run the usual analysis with the 
conservative priors on galaxy bias parameters similar to~\cite{Cabass:2022epm}.
For reference, the priors
on the galaxy bias 
parameter are 
\be 
\begin{split}
& b_1\in [1,4]\,,\quad 
b_2\sim \mathcal{N}(0,1)\,,\\
& b_{\mathcal{G}_2}\sim \mathcal{N}(0,1)\,,\quad 
b_{\Gamma_3}\sim 
\mathcal{N}\left(\frac{23}{42}(b_1-1),1\right)\,,
\end{split}
\ee
where $\mathcal{N}(a,\sigma)$
denotes a normal distribution with
mean $a$ and variance $\sigma^2$.

As a second step, we replace the conservative priors from the previous 
analysis with the marginal likelihood $p(\theta_{\rm EFT})$, see Eq.~\eqref{eq:margdist}. We use $p(\theta_{\rm EFT})$ 
to set priors for parameters $\theta_{\rm EFT}=\{b_1,b_2,b_{\mathcal{G}_2},b_{\Gamma_3}\}$ of the BOSS EFT likelihood. 
The other
EFT 
parameters from our samples, 
i.e. $\{b_3, b_{\nabla^2\delta},\alpha_0,\alpha_1\}$
are not included 
in the distribution, i.e. are
effectively marginalized over,
because of the 
following reasons.

Parameters $P_{\rm shot}$ and $a_0$ in the  BOSS EFT likelihood are quite different 
from $\alpha_0,\alpha_1$ that we have extracted from the HOD mocks. 
First, $P_{\rm shot}$ in 
the EFT models currently implemented in 
\texttt{CLASS-PT}~\cite{Chudaykin:2020aoj} absorbs the constant 
deterministic 
contribution from the one-loop auto-spectrum of
$\delta^2$, see 
\cite{Ivanov:2023yla} for a recent discussion. 
This is the first 
reason why $P_{\rm shot}$ is
different from our $\alpha_{\rm shot}$, which corresponds
to a truly stochastic component. Second, our
$\alpha_{\rm shot}$ captures departures 
from the Poissonian shot noise in a periodic box geometry,
which is different from the number density 
of actual galaxies in a survey that is
subject to additional weights, e.g. FKP (see e.g. a discussion in~\cite{Ivanov:2021zmi}).
Third, stochastic counterterms in
the BOSS EFT pipeline additionally 
absorb contributions
from fiber collisions~\cite{Hahn:2016kiy,Ivanov:2019hqk,Ivanov:2021zmi}. 
Given these reasons, 
we keep using 
uninformative priors on $P_{\rm shot}$
and $a_0$ in our analysis. 
We also do not use a prior on $b_3$
as this parameter does not appear 
in our current analysis based on the tree-level
bispectrum. Finally, we do not use
the prior on $b_{\nabla^2\delta}$
in the current 
analysis 
because the contribution of 
the 
$\nabla^2\delta$
operator to the 
galaxy power spectrum 
in redshift space is degenerate
with other 
redshift space 
counterterms that we have not studied yet. 
Their precision measurement with the redshift-space
field-level EFT of~\cite{Schmittfull:2020trd}
is left for future work. Since the 
PNG parameters are primarily degenerate with the quadratic
bias parameters $b_2$ and $b_{\mathcal{G}_2}$,
we do not expect our results 
to depend significantly on the RSD 
counterterms and parameters $\{P_{\rm shot},a_0\}$
discussed above. For this reason, 
we adopt standard conservative priors
for them in all analyses presented here.  

Additionally, we have marginalized
over the non-Gaussian bias 
parameter $b_\zeta$ 
within conservative 
priors of~\cite{Cabass:2022wjy}
in both analyses.
In principle, one should extract priors
on these parameters from
mocks too. However, 
marginalization over $b_\zeta$ 
does not have a noticeable 
impact on the single-field
PNG constraints presented here.
We leave a calibration 
of these parameters from 
simulations for future work. 

A comment is in order on the redshift-dependence 
of the EFT parameters. 
We have calibrated priors at $z=0.5$,
which is somewhat different from 
$z_{\rm eff}\approx 0.4$ and $0.6$ of 
BOSS chunks. 
Within the HOD models, the galaxy
distribution is determined by 
local-in-time properties of halos. 
Therefore, a redshift dependence of bias
parameters 
can be fully compensated by 
a change of HOD parameters such as $M_{\rm cut}$, which are varied
in our samples. Physically, we expect some residual sensitivity to the past evolution, e.g.
to merger and assembly histories, which is captured 
within EFT by the expansion along the past fluid 
trajectory~\cite{Mirbabayi:2014zca,Desjacques:2016bnm}. 
These effects 
appear only at the two-loop order in the EFT, 
and hence are irrelevant for our analysis
based on the one-loop calculation. 
All in all, the redshift dependence 
of the EFT parameters is adequately 
captured by variations of HOD parameters
at the current precision level.

Our main results are displayed in figure~\ref{fig:triangle_small} and 
in table~\ref{eq:tab1}. 
A corner plot with 
EFT parameters for each chunk 
can be found 
in Appendix~\ref{app:add}. 
The first relevant observation is 
that the posteriors for \textit{non-linear} galaxy bias parameters 
shrink significantly after applying the HOD priors. 
In particular, the 
1d marginalized 
errorbars on $b_2$, $b_{\mathcal{G}_2}$
and  $b_{\Gamma_3}$
shrink by a factor of few. To quantify the improvement in a more 
rigorous manner, 
we introduce 
a figure of merit (FoM)
for bias parameters,
defined along 
the lines of the FoM 
for the dark energy equation of state. 
Namely, for 
each individual BOSS data 
slice we write 
\be 
\text{FoM}_{\rm bias}=\left[\det{C(b_1,b_2,b_{\mathcal{G}_2},b_{\Gamma_3})}\right]^{-1/2}\,,
\ee 
where $C$ is the covariance
matrix of the bias parameters
after marginalizing 
over other parameters
in the chain. 
In the Gaussian case, our FoM is proportional to 
the inverse volume of the 4-dimensional ellipsoid
enclosing the 
posterior distribution of 
the bias coefficients.
A relevant parameter
in our comparison is 
the FoM ratio 
between the old 
and new bias 
parameter measurements,
e.g. for the 
NGCz3 we have
\be 
\frac{\text{FoM}_{\rm bias}|_{\rm HOD priors}}{\text{FoM}_{\rm bias}\big|_{\rm cons. priors}}\Bigg|_{\rm NGCz3}=60.3~\,,
\ee 
which can 
be interpreted 
as a factor of 
$\approx 60$ 
reduction in cumulative inverse variance
of the posterior 
bias parameter distribution.
The cumulative 
figure of merit 
of bias parameters from all four BOSS 
data chunks is $1.2\cdot 10^7$.

The second important observation is 
that 
our HOD-based measurements 
of bias parameters 
are in perfect agreement 
with the EFT results based on the non-informative priors:
the HOD-based posteriors are  
located almost at the
centers of posterior 
densities 
from the standard EFT analysis. 
This is a non-trivial consistency check 
of both our prior calibration
technique 
and the BOSS EFT pipeline since
our measurements
of non-linear bias 
parameters 
are dominated by the 
galaxy bispectrum data\footnote{
Fitted with the EFT pipeline of~\cite{Ivanov:2021kcd} and 
estimated
with the window-free 
method of~\cite{Philcox:2020vbm,Philcox:2021ukg}.}, while 
the HOD-informed priors
were extracted from 
the field-level fits
to HOD mocks.
One can also note that 
the constraints on 
the linear bias parameter $b_1$
do not improve. 
This is because the data 
itself is much more informative  
than the prior for this parameter,
cf. the standard BOSS posterior 
and the HOD prior in Fig.~\ref{fig:triangle_small}.  
Indeed, our HOD samples are consistent 
with a wide range of $b_1$, while 
the actual measurements 
from BOSS 
are at the level of few percent. 

The third important observation is that the 
1d marginalized errorbar on $f_{\rm NL}^{\rm equil}$
has narrowed by $\approx 40\%$, from 
$f_{\rm NL}^{\rm equil}=806_{-510}^{+490}$ (conservative priors) 
to $f_{\rm NL}^{\rm equil}=0.323_{-0.31}^{+0.29}$ (HOD-informed).
The main channel of improvement here is the breaking 
of degeneracy between $f_{\rm NL}^{\rm equil}$ and $b_{\mathcal{G}_2}$
in the galaxy bispectrum. Since $f_{\rm NL}^{\rm ortho}$ was not 
significantly 
correlated with bias parameters to begin with, the 
improvement for this parameter is only at $\approx 10\%$, 
$f_{\rm NL}^{\rm equil}=-8.2_{-150}^{+130}$ (conservative priors) 
vs. $f_{\rm NL}^{\rm equil}=100_{-140}^{+130}$ (HOD-informed).
In line with the previous
discussion, let us introduce 
the FoM for non-Gaussian
amplitudes
\be 
\text{FoM}_{\rm PNG}=\left[\det{C(f_{\rm NL}^{\rm equil},f_{\rm NL}^{\rm ortho})}\right]^{-1/2}\,,
\ee 
which calculates the inverse
area (up to a factor of $1/\pi$) of the error ellipse in
the $(f_{\rm NL}^{\rm equil},f_{\rm NL}^{\rm ortho})$ plane. 
Then we have 
\be 
\frac{\text{FoM}_{\rm PNG}|_{\rm HOD priors}}{\text{FoM}_{\rm PNG}\big|_{\rm cons. priors}}=1.76~\,,
\ee 
or a $\approx 40\%$ reduction
of the total
2-dimensional 
posterior 
variance
of PNG parameters.

\begin{figure*}
    \centering
    \includegraphics[width=0.7\textwidth]{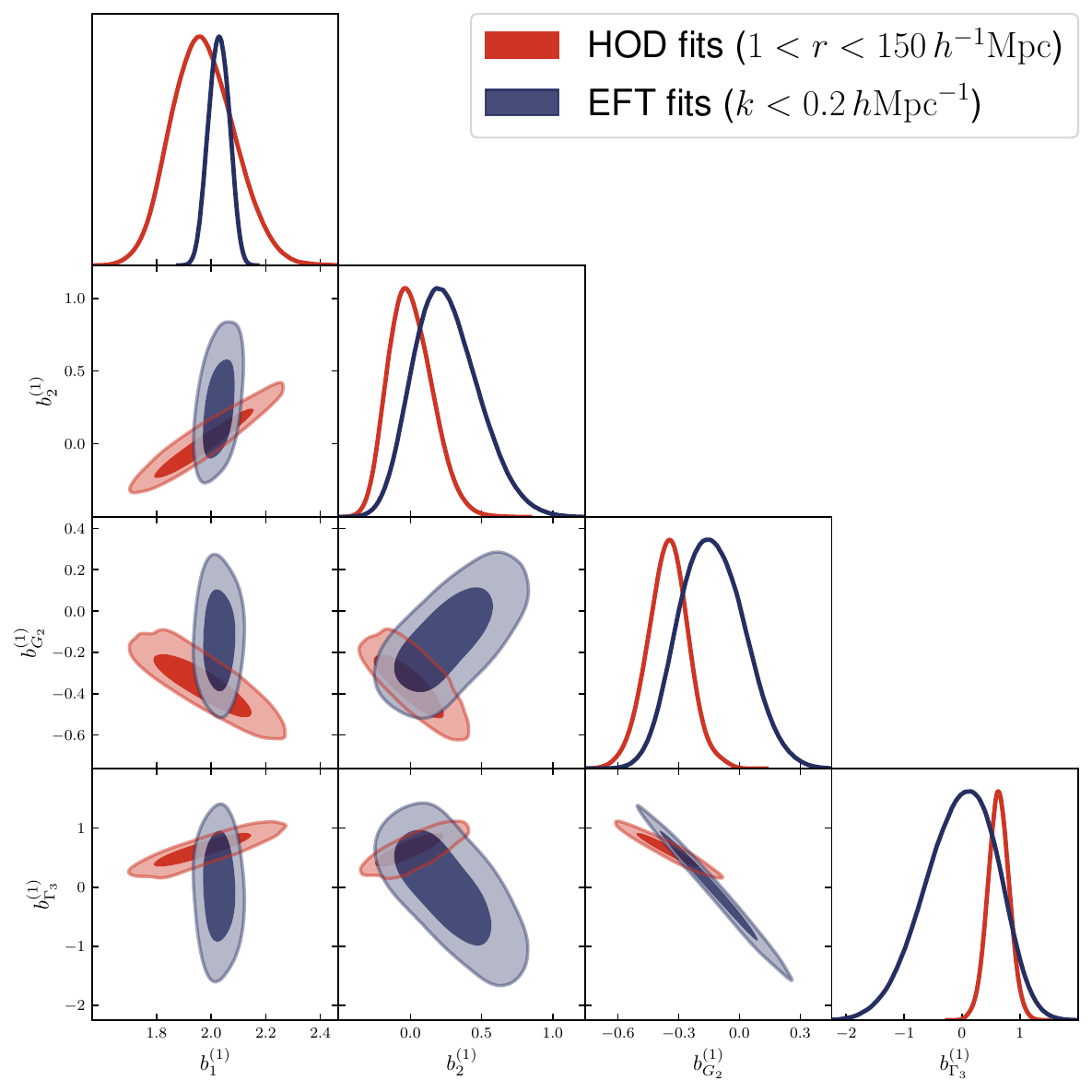}
    \caption{Comparison of the EFT biased parameters found by small scale simulation-based fits to density split and two-point statistics based on HOD models (HOD fits), and those found through EFT fits to the power spectrum and the bispectrum on large scales (EFT fits).}
    \label{fig:hod_vs_eft}
\end{figure*}

Finally, let us mention that the analysis with the conservative priors implies a marginal $2\sigma$ preference for non-zero $f_{\rm NL}^{\rm equil}$, which 
cannot be physical as the central value $f_{\rm NL}^{\rm equil}=800$ 
is in strong tension with the CMB measurements~\cite{Planck:2019kim}.
The preference 
for non-zero $f_{\rm NL}^{\rm equil}$ 
reduces in
our new analysis,
which highlights the 
importance of physically-motivated 
priors for the full-shape 
analysis.

\section{Comparison with small-scale measurements } 
\label{sec:hod}

An interesting application of the method presented here is the comparison between simulation-based models fitting clustering data on small scales, see e.g.~\cite{Paillas:2023cpk,Cuesta-Lazaro:2023gbv,Hahn:2023udg,Valogiannis:2023mxf}, and perturbation theory based models constrained by large scales. In particular, we use the conditional model $p(\theta_{\rm EFT}\mid\theta_{\rm HOD})$ to convert the HOD posterior chains from \cite{Paillas:2023cpk} into EFT bias parameters. These posteriors were obtained by fitting the two-point correlation function and density split statistics of the CMASS galaxy sample in the scale range $1 < r < 150 \, \Mpch$. The model used for inference is a neural network trained to reproduce the clustering of HOD-based galaxy mocks. 

In Figure~\ref{fig:hod_vs_eft}, we compare the estimated bias parameters, denoted as HOD fits, with those found in this paper, denoted as EFT fits. Interestingly, although the range of scales used is widely different, both posteriors agree remarkably. Note that some important differences beyond scale ranges are the parameters that are being fitted, since \citep{Paillas:2023cpk} also fits the cosmological parameters, and the statistics used.

\section{Discussion and Outlook} 
\label{sec:disc}

We have presented a new framework for deriving informative priors
on EFT ``nuisance'' parameters from galaxy formation simulations. 
For concreteness, we focused here on the HOD models, although our
approach can be straightforwardly applied to other
galaxy formation models. 
The central object of our method is the conditional distribution
between parameters of galaxy formation models and EFT parameters.
We build this distribution by generating 
a large sample of EFT parameters extracted from 
HOD mocks with the field-level technique. 
In this work, we have used the
resulting distribution of the perturbative galaxy bias parameters 
as a prior 
in the analysis of the BOSS galaxy clustering data in the 
context of primordial non-Gaussianity. 
We have found a significant improvement 
in the PNG constraints, as well as in the measurement 
of the galaxy bias parameters. Importantly, 
the values of bias parameters that we have extracted 
are fully consistent with previous analyses of BOSS
based on conservative priors, as well as with the 
measurements based on the short-scale density-split
statistic. This impressive consistency 
can be interpreted as a sign of convergence of 
various independent techniques to model 
the large-scale structure of the Universe. 

We note that in this work we have focused 
on the real space analysis of galaxy clustering that allows us to study  perturbative galaxy bias parameters. 
The informative priors on these parameters are sufficient for our main goal here: to improve limits on single-field primordial non-Gaussianity. A more general analysis, including full redshift-space nuisance parameters and HOD models, is currently in progress.

A next step in our 
program will be to carry out a full re-analysis of the BOSS
data including cosmological parameters such as $\sigma_8,\Omega_m$ etc.
Another important research direction is to extend our
calibration of EFT parameters to 
more general HOD models relevant, e.g. 
for eBOSS quasars~\cite{Ata:2017dya,Hou:2020rse,Neveux:2020voa,Chudaykin:2022nru} and emission
line galaxies~\cite{deMattia:2020fkb,Ivanov:2021zmi}, as well as to 
other galaxy formation models,
such as abundance matching and hydrodynamical simulations. 

Note that the approach presented here can be used for a new 
type of emulator where small scale data are modeled 
with simulations, and then large scale clustering is 
reconstructed from the small simulation box using EFT.
Similar ideas have been put forward before (see e.g.~\cite{Obuljen:2022cjo,Modi:2023drt}).
A central tool in this emulator is the conditional 
distribution between EFT and simulation parameters,
identical to the one that we derived here for HOD 
models. The work on this emulator is currently underway. 

Finally, it would also be interesting 
to extend our approach to imaging clustering data and intensity mapping, see~\cite{Obuljen:2022cjo} for recent  
work in this direction. 

Code to reproduce and use the simulation-based priors $p(\theta_{\rm EFT})$ and $p(\theta_{\rm{EFT}} \mid \theta_{\rm HOD})$ is available at \url{https://github.com/smsharma/eft-hod}.

\section*{Acknowledgments}

We thank Kazuyuki Akitsu, 
Stephen Chen, 
Marko Simonovi\'c,
Fabian Schmidt, 
and Marcel Schmittfull 
for useful discussions.
The authors also thank the IAIFI Astro-ML Hackathon where the first
draft of this work was completed.
Monte Carlo Markov Chains with PNG 
parameters were generated 
with the \texttt{Montepython} code~\cite{Audren:2012wb,Brinckmann:2018cvx}.
This work is supported by the National Science Foundation under Cooperative Agreement PHY-2019786 (The NSF AI Institute for Artificial Intelligence and Fundamental Interactions, \url{http://iaifi.org/}). This material is based upon work supported by the U.S. Department of Energy, Office of Science, Office of High Energy Physics of U.S. Department of Energy under grant Contract Number  DE-SC0012567. AO acknowledges financial support from the Swiss National Science Foundation (grant no CRSII5{\_}193826). MWT  acknowledges financial support from the Simons Foundation (Grant Number 929255).
\appendix

\section{Additional plots}
\label{app:add}

\begin{figure*}
\centering
\includegraphics[width=0.47\textwidth]{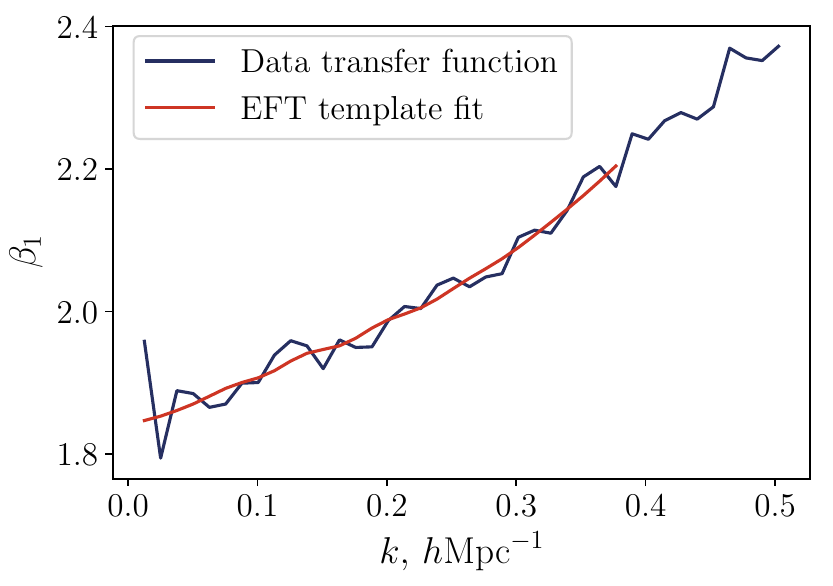}
\includegraphics[width=0.49\textwidth]{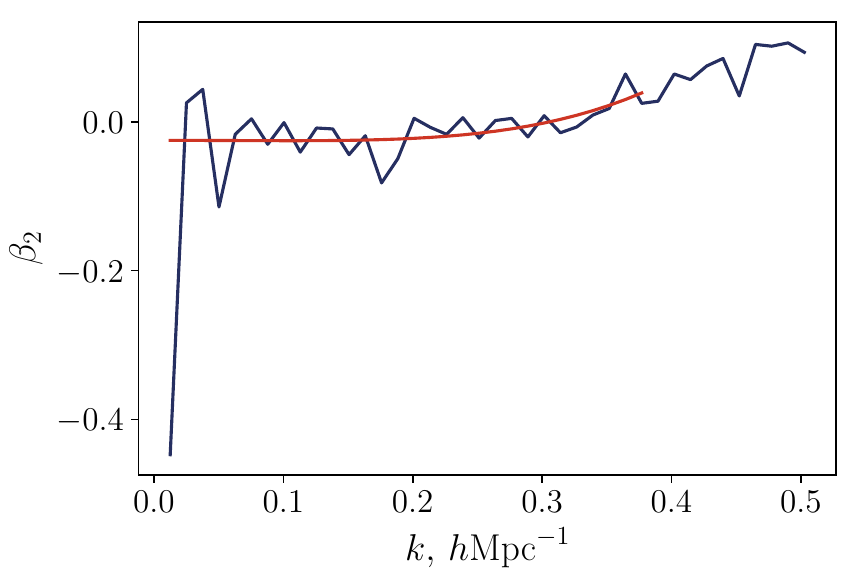}
\includegraphics[width=0.49\textwidth]{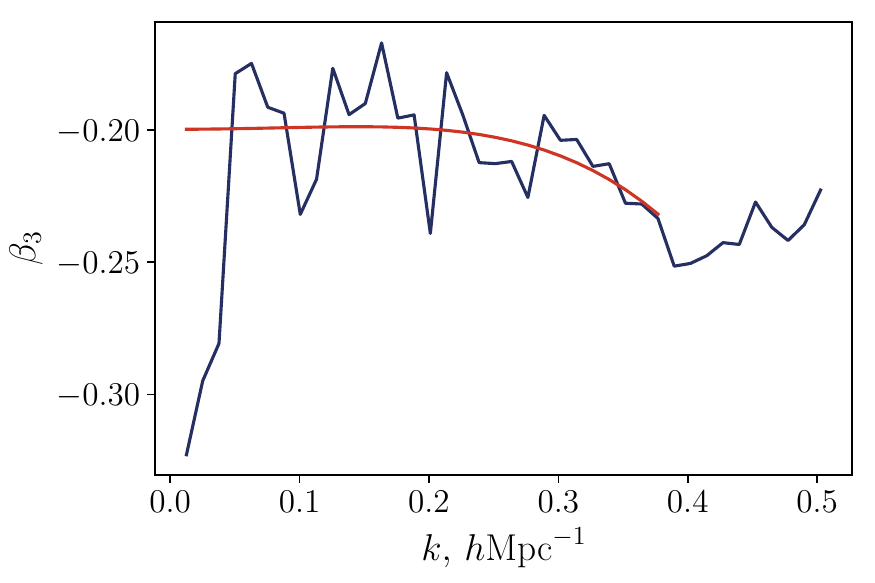} 
\includegraphics[width=0.45\textwidth]{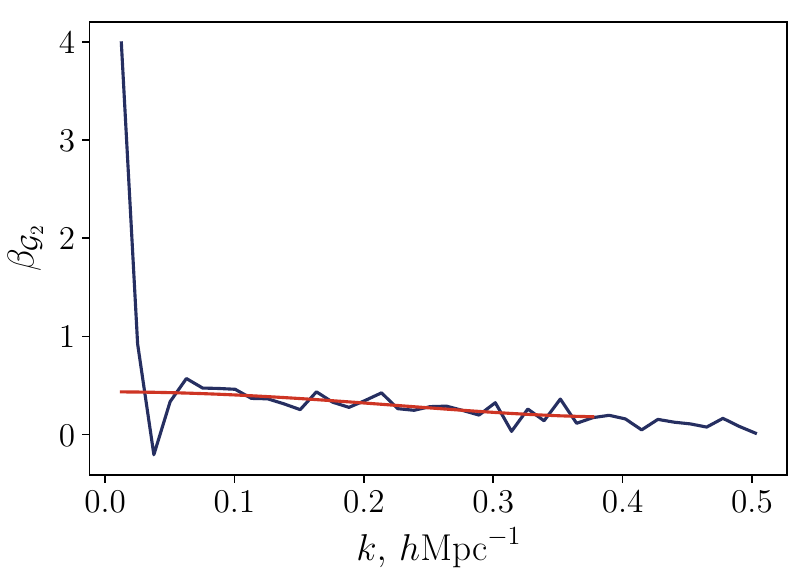}
   \caption{Transfer functions from our forward model and their EFT template fits
    for a typical HOD mock. 
    } \label{fig:betas}
\end{figure*}

\begin{figure*} 
\centering
\includegraphics[width=0.51\textwidth]{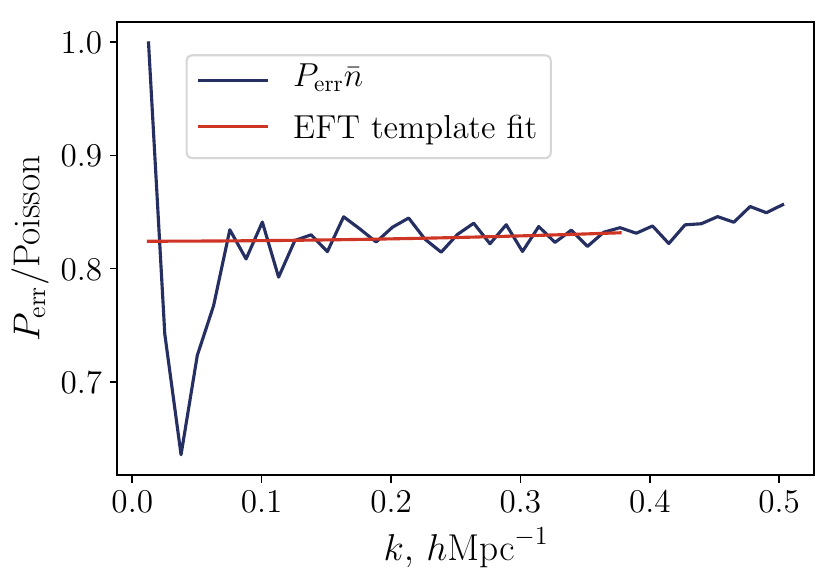}
\includegraphics[width=0.45\textwidth]{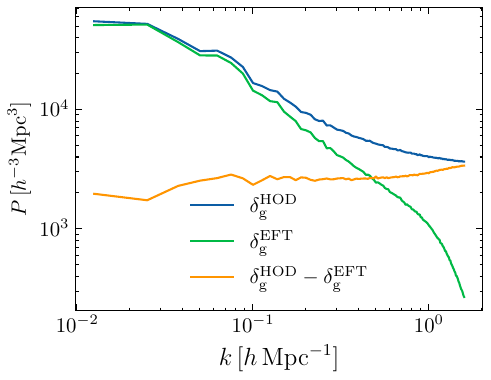}
   \caption{\textit{Left panel:} Error power spectrum from the forward model and its EFT template fit for a typical HOD mock in our sample. The number density of this mock is $\bar{n}=3.5\cdot 10^{-4}~[\hMpc]^3$.
   \textit{Right panel}: Power spectra of 
a typical HOD galaxy field, 
 the deterministic part of the EFT forward model, and the residual between the two (the error power spectrum $P_{\rm err}$).
    } \label{fig:perr}
\end{figure*}

Fitting the EFT parameters from the field-level
model is visualized 
in Fig.~\ref{fig:betas} where we show the data transfer functions 
and their EFT template fits for a typical mock from our sample.
The left panel of 
Fig.~\ref{fig:perr} shows 
the extraction of the 
stochastic $\alpha_{0}$ and $\alpha_1$
parameters by fitting the error power spectrum
from the data to the EFT template~\eqref{eq:perreft}.
Typical power spectra of the 
HOD galaxies and the EFT forward model
are shown 
in the right panel of Fig.~\ref{fig:perr}. 
The difference between the EFT and the 
HOD density fields by construction
is the error field density, 
whose power spectrum is also shown
in Fig.~\ref{fig:perr}. 

\begin{figure*}[htb]
\centering\includegraphics[width=0.99\textwidth]{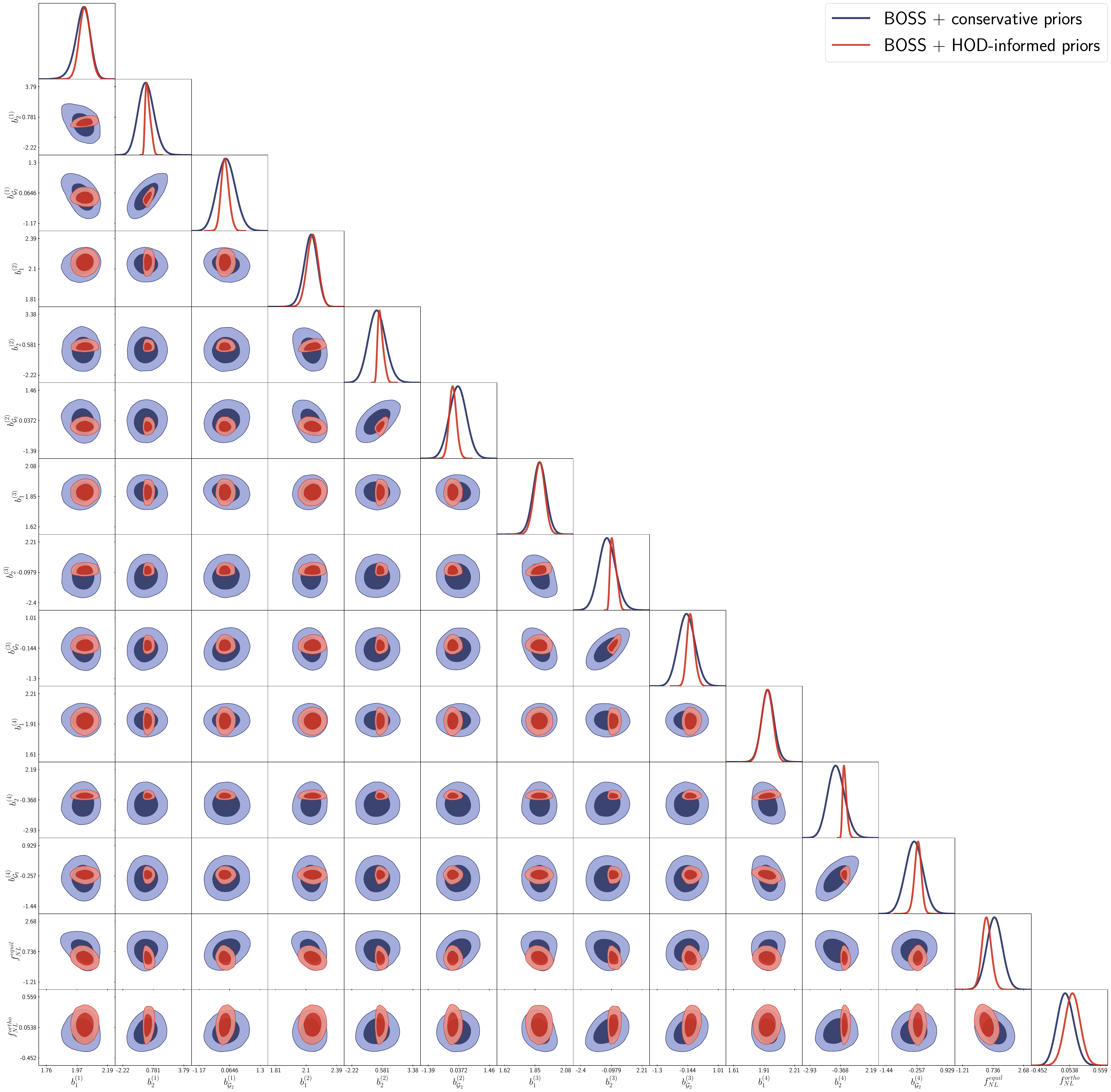}
   \caption{\small Corner plots with 2d and 1d marginalized posterior
   distribution of PNG and galaxy bias parameters $b_1$, $b_2$, $b_{\mathcal{G}_2}$ from BOSS DR12 power
   spectra and bispectra for  four independent data chunks: NGCz3 (1), SGCz3 (2), NGCz1 (3), 
   SGCz3 (4).
    } \label{fig:triangle}
\end{figure*}

In Fig.~\ref{fig:triangle}
we show 
a corner plot with marginalized 
posterior distribution 
for PNG parameters 
plus linear and quadratic 
bias parameters for all four BOSS data chunks.

\section{Normalizing flow training and validation}
\label{app:flows}

\begin{figure*} 
\centering
\includegraphics[width=0.45\textwidth]{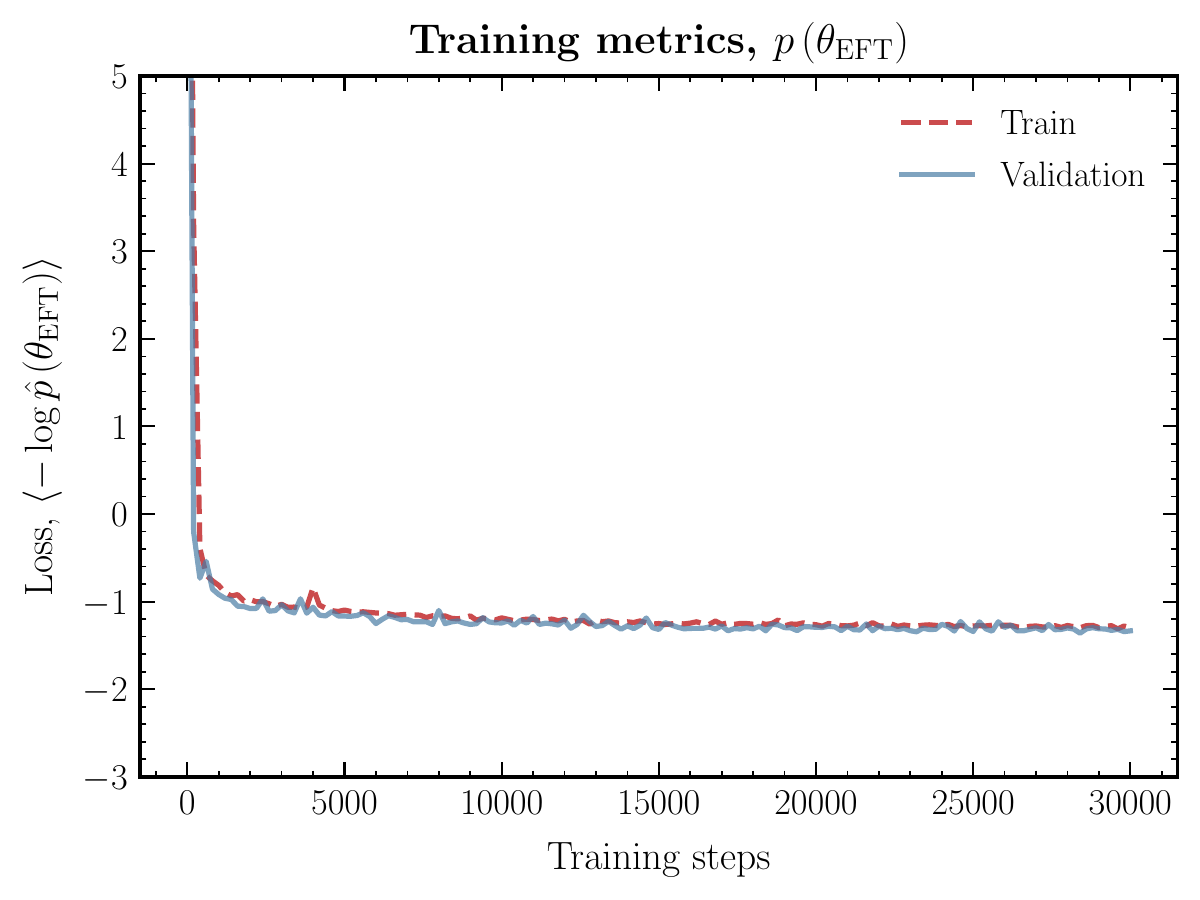}
\includegraphics[width=0.45\textwidth]{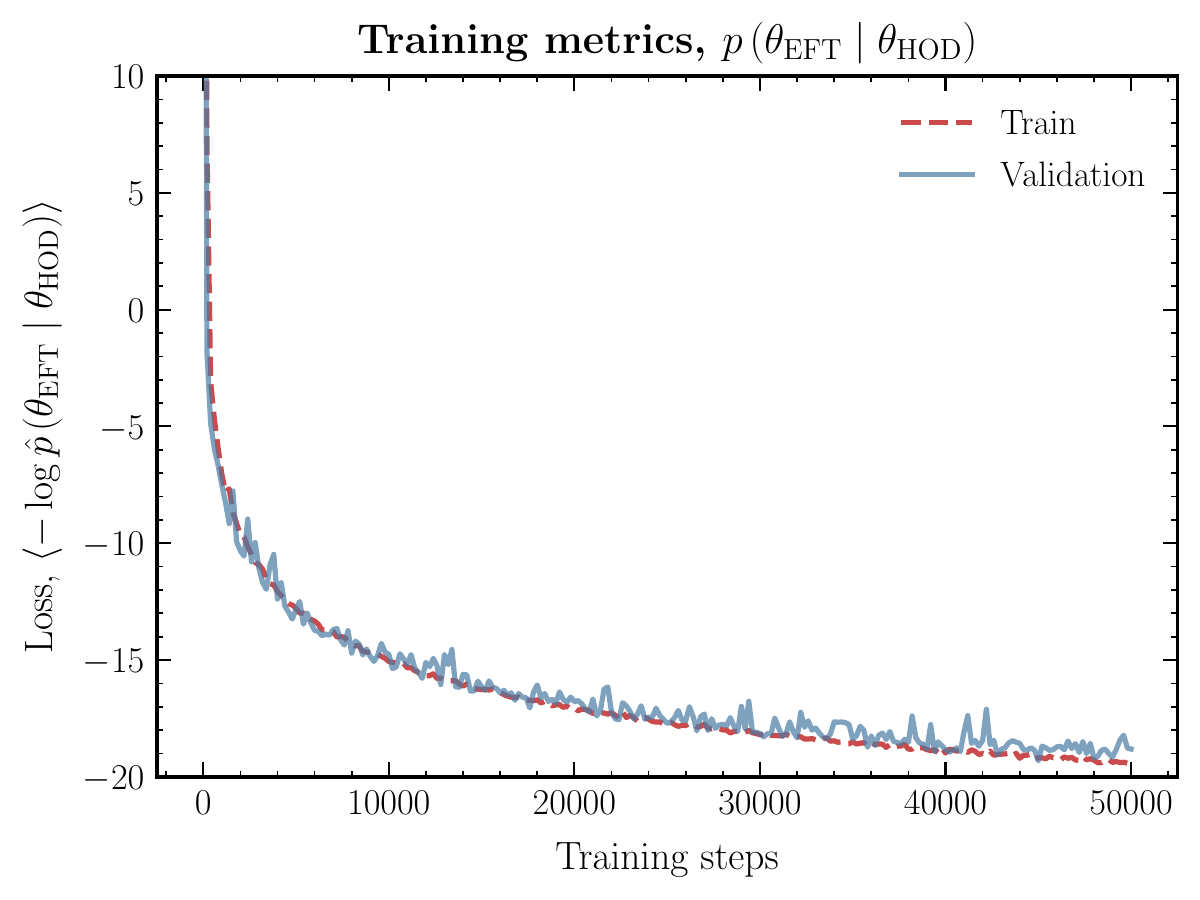}
   \caption{Training (dashed red line) and validation (blue) line losses over the course of training, for the marginal EFT parameter distribution (left) and HOD-conditional EFT parameter distribution (right).} 
\label{fig:flow_training}
\end{figure*}

Normalizing flows are implemented using the \texttt{nflows} library, with training and evaluation performed using \texttt{PyTorch}~\cite{paszke2019pytorch}. Masked Autoregressive Flows~\cite{papamakarios2017masked} are used, with 6 flow transformation, each parameterized using a 2-layer masked autoregressive neural network~\cite{germain2015made} with GELU activations. 10\% of the samples are held out for validation, and the flow is trained with batch size 128 for either 30,000 or 50,000 steps (for the EFT and HOD-conditional EFT distributions, respectively) using the Adam optimizer~\cite{kingma2014adam} with learning rate $3 \times 10^{-4}$. 

Since we have a finite number of sampled points for the HOD and EFT parameters, it is imperative to validate that features of our approximated distribution are not artificially sculpted due these points. Figure \ref{fig:flow_training} shows the training (dashed red) and validation (blue) losses over the course of training, with validation being done every 200 steps on a held out set of points. We can see that the loss asymptotes for both the marginal EFT parameter fit (left) and the HOD-conditional fit (right), with no signs of overfitting. 

\begin{figure*}[htbp]
    \centering
    \includegraphics[width=\textwidth]{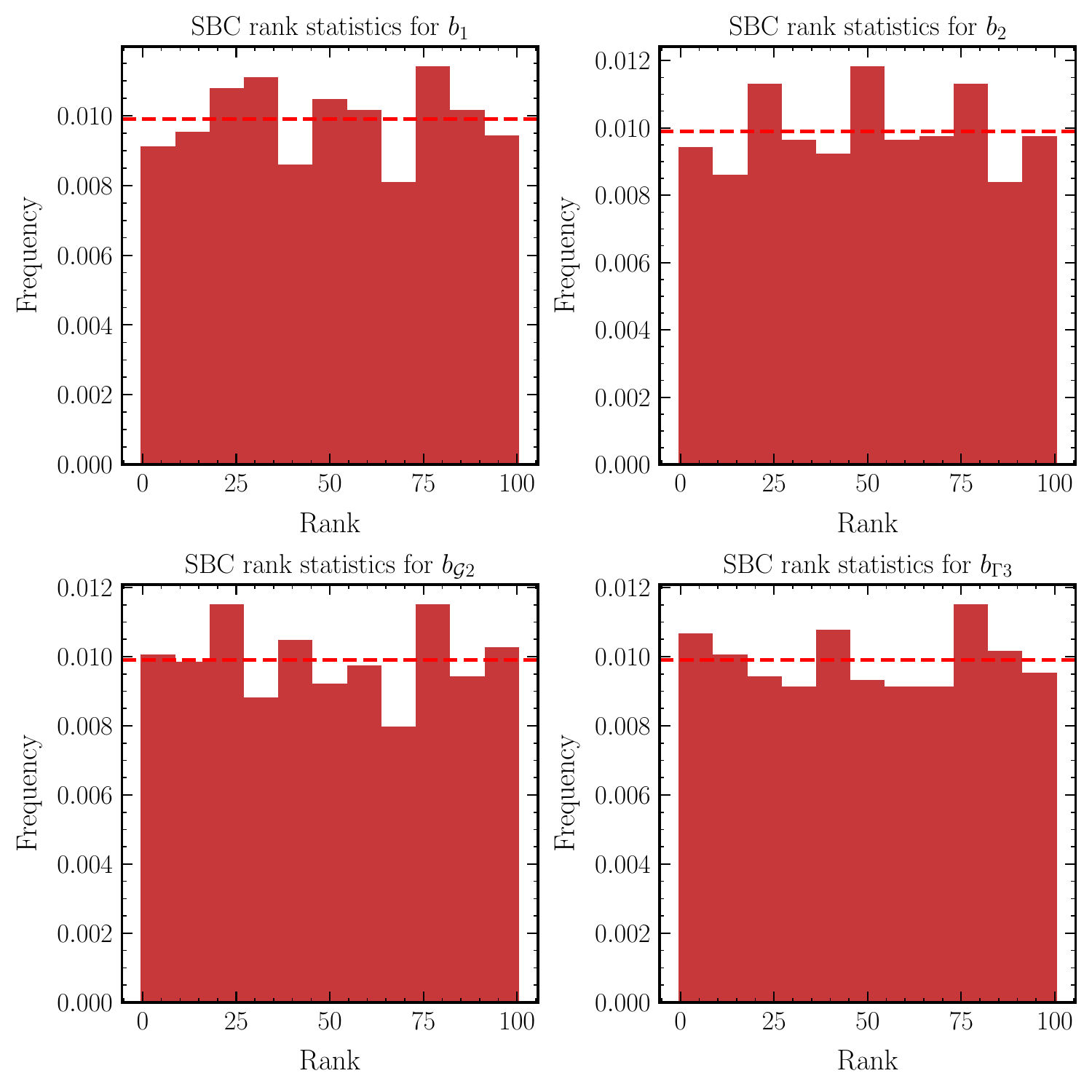}
    \caption{Simulation-based calibration rank statistics for EFT parameters. Each subplot shows the distribution of ranks for a different parameter: $b_1$ (top left), $b_2$ (top right), $b_{\mathcal{G}2}$ (bottom left), and $b_{\Gamma3}$ (bottom right). The horizontal dashed line represents the expected uniform distribution for perfect calibration.}
    \label{fig:sbc_ranks}
\end{figure*}

Finally, as an additional coverage 
test,
we perform a simulation-based
calibration analysis~\cite{2018arXiv180406788T} of our
normalizing flow model. 
To that end we generate 
100 samples from our trained normalizing flow model and compare them to validation samples.
In this process, we
calculate the rank of each validation sample among the set of 100 posterior samples drawn from the flow, and plot aggregate statistics, shown in Fig.~\ref{fig:sbc_ranks}. With perfect calibration, these ranks should be uniformly distributed. No significant deviation from uniformity can be observed, validating the calibration and convergence properties of our learned posterior.

\section{Test of UV-sensitivity}
\label{sec:uvsens}

In this section, we provide 
a convergence test for our 
transfer function measurements.
This analysis also provides 
an estimate of the residual 
UV-dependence of the 
bias parameters extracted 
with our field-level technique.

The application of the 
CIC window leads to an
implicit smoothing of 
all fields in our calculations. 
Note that this smoothing is quite soft
due to the particular shape of the CIC window. 
The cell size 
of the grid is a proxy 
of the filtering scale. 
In our baseline analysis, 
we use $256$ grid points $N_g$,
resulting in an effective 
smoothing scale $R_s \simeq \pi k^{-1}_{\rm Nyquist}= 6$ $\Mpch$.

\begin{figure*}[htbp]
    \centering
    \includegraphics[width=\textwidth]{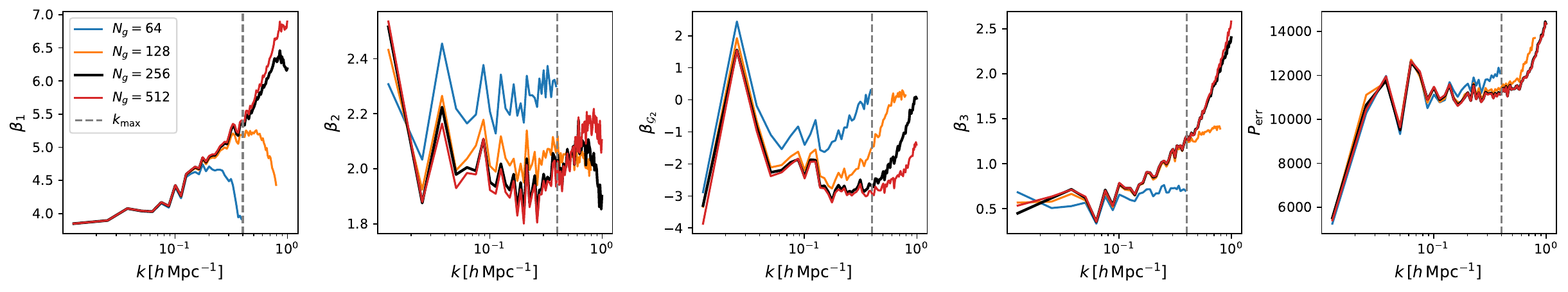}
    \caption{Dependence of the transfer functions on the Eulerian grid resolution, which acts as a cutoff in the field-level 
    calculation. We show results for the number of grid points $N_g=64,128,256,512$, which correspond 
    to effective smoothing scales $\pi k^{-1}_{\rm Nyquist}=24,12,6,3$ $\Mpch$, respectively.}
    \label{fig:uv}
\end{figure*}

In Fig.~\ref{fig:uv}
we display transfer functions 
for one of the HOD mocks 
in our catalog with 
four choices of the grid size, 
$N_g=64,128,256$ and $512$, 
which corresponds to the smoothing 
lengths $\pi k^{-1}_{\rm Nyquist} = 24,12,6,3$ $\Mpch$,
respectively. As expected, we see that 
using a relatively large cutoff $R_s$
affects the transfer functions 
quite significantly. In particular, 
the 
$R_s=8~\Mpch$ transfer functions 
for the non-linear bias parameters 
take different values 
in the $k\to 0$ limit. 
In contrast, the low-$k$ limit of $\beta_1$ is unaffected and yields the renormalized linear bias parameter $b_1$~\cite{Schmittfull:2018yuk}.

The appearance of the scale dependence
can be understood from 
perturbation theory arguments. 
For example,  at the one-loop order
$\beta_2$ reads, 
\be 
\beta_2(k) =\frac{b_2}{2}+\frac{\langle \tilde \delta_2^\perp 
\mathcal{G}_2^\perp \rangle}{\langle \tilde \delta_2^\perp 
\delta_2^\perp \rangle}\,.
\ee 
The rightmost term above vanishes
in the $k\to 0$ limit because 
the power spectrum of $\delta_2$
is constant on large scales. 
However,
this constant vanishes 
quickly
as one 
lowers the cutoff, i.e. for $R_s\simeq 10$ $\Mpch$ its value is reduced
by a factor of 3 compared to the 
$R_s\to 0$ limit. 
Hence, the scale dependence of the second term
starts playing a more important role
as one lowers the cutoff, 
and ultimately contaminates 
the low-$k$ limit of $\beta_2$.
On top of that, there is an additional scale dependence generated by higher-order
loop corrections. 

In our particular case, 
however, Fig.~\ref{fig:uv}
suggests that the cutoff-dependence of the transfer functions, especially their low-$k$ limit, is 
negligible for the 
momentum cut $\kmax=0.4~\hMpc$
used in the analysis. 


\bibliography{short.bib}

\begin{thebibliography}{124}%
\makeatletter
\providecommand \@ifxundefined [1]{%
 \@ifx{#1\undefined}
}%
\providecommand \@ifnum [1]{%
 \ifnum #1\expandafter \@firstoftwo
 \else \expandafter \@secondoftwo
 \fi
}%
\providecommand \@ifx [1]{%
 \ifx #1\expandafter \@firstoftwo
 \else \expandafter \@secondoftwo
 \fi
}%
\providecommand \natexlab [1]{#1}%
\providecommand \enquote  [1]{``#1''}%
\providecommand \bibnamefont  [1]{#1}%
\providecommand \bibfnamefont [1]{#1}%
\providecommand \citenamefont [1]{#1}%
\providecommand \href@noop [0]{\@secondoftwo}%
\providecommand \href [0]{\begingroup \@sanitize@url \@href}%
\providecommand \@href[1]{\@@startlink{#1}\@@href}%
\providecommand \@@href[1]{\endgroup#1\@@endlink}%
\providecommand \@sanitize@url [0]{\catcode `\\12\catcode `\$12\catcode
  `\&12\catcode `\#12\catcode `\^12\catcode `\_12\catcode `\%12\relax}%
\providecommand \@@startlink[1]{}%
\providecommand \@@endlink[0]{}%
\providecommand \url  [0]{\begingroup\@sanitize@url \@url }%
\providecommand \@url [1]{\endgroup\@href {#1}{\urlprefix }}%
\providecommand \urlprefix  [0]{URL }%
\providecommand \Eprint [0]{\href }%
\providecommand \doibase [0]{http://dx.doi.org/}%
\providecommand \selectlanguage [0]{\@gobble}%
\providecommand \bibinfo  [0]{\@secondoftwo}%
\providecommand \bibfield  [0]{\@secondoftwo}%
\providecommand \translation [1]{[#1]}%
\providecommand \BibitemOpen [0]{}%
\providecommand \bibitemStop [0]{}%
\providecommand \bibitemNoStop [0]{.\EOS\space}%
\providecommand \EOS [0]{\spacefactor3000\relax}%
\providecommand \BibitemShut  [1]{\csname bibitem#1\endcsname}%
\let\auto@bib@innerbib\@empty
\bibitem [{\citenamefont {Aghamousa}\ \emph {et~al.}(2016)\citenamefont
  {Aghamousa} \emph {et~al.}}]{Aghamousa:2016zmz}%
  \BibitemOpen
  \bibfield  {author} {\bibinfo {author} {\bibfnamefont {A.}~\bibnamefont
  {Aghamousa}} \emph {et~al.} (\bibinfo {collaboration} {DESI}),\ }\href@noop
  {} {\  (\bibinfo {year} {2016})},\ \Eprint {http://arxiv.org/abs/1611.00036}
  {arXiv:1611.00036 [astro-ph.IM]} \BibitemShut {NoStop}%
\bibitem [{\citenamefont {Laureijs}\ \emph {et~al.}(2011)\citenamefont
  {Laureijs} \emph {et~al.}}]{Laureijs:2011gra}%
  \BibitemOpen
  \bibfield  {author} {\bibinfo {author} {\bibfnamefont {R.}~\bibnamefont
  {Laureijs}} \emph {et~al.} (\bibinfo {collaboration} {EUCLID}),\ }\href@noop
  {} {\  (\bibinfo {year} {2011})},\ \Eprint {http://arxiv.org/abs/1110.3193}
  {arXiv:1110.3193 [astro-ph.CO]} \BibitemShut {NoStop}%
\bibitem [{\citenamefont {Ivezi\'c}\ \emph {et~al.}(2019)\citenamefont
  {Ivezi\'c} \emph {et~al.}}]{LSST:2008ijt}%
  \BibitemOpen
  \bibfield  {author} {\bibinfo {author} {\bibfnamefont {v.}~\bibnamefont
  {Ivezi\'c}} \emph {et~al.} (\bibinfo {collaboration} {LSST}),\ }\href
  {\doibase 10.3847/1538-4357/ab042c} {\bibfield  {journal} {\bibinfo
  {journal} {Astrophys. J.}\ }\textbf {\bibinfo {volume} {873}},\ \bibinfo
  {pages} {111} (\bibinfo {year} {2019})},\ \Eprint
  {http://arxiv.org/abs/0805.2366} {arXiv:0805.2366 [astro-ph]} \BibitemShut
  {NoStop}%
\bibitem [{\citenamefont {Akeson}\ \emph {et~al.}(2019)\citenamefont {Akeson}
  \emph {et~al.}}]{Akeson:2019biv}%
  \BibitemOpen
  \bibfield  {author} {\bibinfo {author} {\bibfnamefont {R.}~\bibnamefont
  {Akeson}} \emph {et~al.},\ }\href@noop {} {\  (\bibinfo {year} {2019})},\
  \Eprint {http://arxiv.org/abs/1902.05569} {arXiv:1902.05569 [astro-ph.IM]}
  \BibitemShut {NoStop}%
\bibitem [{\citenamefont {Baumann}\ \emph {et~al.}(2012)\citenamefont
  {Baumann}, \citenamefont {Nicolis}, \citenamefont {Senatore},\ and\
  \citenamefont {Zaldarriaga}}]{Baumann:2010tm}%
  \BibitemOpen
  \bibfield  {author} {\bibinfo {author} {\bibfnamefont {D.}~\bibnamefont
  {Baumann}}, \bibinfo {author} {\bibfnamefont {A.}~\bibnamefont {Nicolis}},
  \bibinfo {author} {\bibfnamefont {L.}~\bibnamefont {Senatore}}, \ and\
  \bibinfo {author} {\bibfnamefont {M.}~\bibnamefont {Zaldarriaga}},\ }\href
  {\doibase 10.1088/1475-7516/2012/07/051} {\bibfield  {journal} {\bibinfo
  {journal} {JCAP}\ }\textbf {\bibinfo {volume} {1207}},\ \bibinfo {pages}
  {051} (\bibinfo {year} {2012})},\ \Eprint {http://arxiv.org/abs/1004.2488}
  {arXiv:1004.2488 [astro-ph.CO]} \BibitemShut {NoStop}%
\bibitem [{\citenamefont {Carrasco}\ \emph {et~al.}(2012)\citenamefont
  {Carrasco}, \citenamefont {Hertzberg},\ and\ \citenamefont
  {Senatore}}]{Carrasco:2012cv}%
  \BibitemOpen
  \bibfield  {author} {\bibinfo {author} {\bibfnamefont {J.~J.~M.}\
  \bibnamefont {Carrasco}}, \bibinfo {author} {\bibfnamefont {M.~P.}\
  \bibnamefont {Hertzberg}}, \ and\ \bibinfo {author} {\bibfnamefont
  {L.}~\bibnamefont {Senatore}},\ }\href {\doibase 10.1007/JHEP09(2012)082}
  {\bibfield  {journal} {\bibinfo  {journal} {JHEP}\ }\textbf {\bibinfo
  {volume} {09}},\ \bibinfo {pages} {082} (\bibinfo {year} {2012})},\ \Eprint
  {http://arxiv.org/abs/1206.2926} {arXiv:1206.2926 [astro-ph.CO]} \BibitemShut
  {NoStop}%
\bibitem [{\citenamefont {Ivanov}(2022)}]{Ivanov:2022mrd}%
  \BibitemOpen
  \bibfield  {author} {\bibinfo {author} {\bibfnamefont {M.~M.}\ \bibnamefont
  {Ivanov}},\ }\href@noop {} {\  (\bibinfo {year} {2022})},\ \Eprint
  {http://arxiv.org/abs/2212.08488} {arXiv:2212.08488 [astro-ph.CO]}
  \BibitemShut {NoStop}%
\bibitem [{\citenamefont {Nishimichi}\ \emph {et~al.}(2020)\citenamefont
  {Nishimichi}, \citenamefont {D'Amico}, \citenamefont {Ivanov}, \citenamefont
  {Senatore}, \citenamefont {Simonovi\'c}, \citenamefont {Takada},
  \citenamefont {Zaldarriaga},\ and\ \citenamefont
  {Zhang}}]{Nishimichi:2020tvu}%
  \BibitemOpen
  \bibfield  {author} {\bibinfo {author} {\bibfnamefont {T.}~\bibnamefont
  {Nishimichi}}, \bibinfo {author} {\bibfnamefont {G.}~\bibnamefont {D'Amico}},
  \bibinfo {author} {\bibfnamefont {M.~M.}\ \bibnamefont {Ivanov}}, \bibinfo
  {author} {\bibfnamefont {L.}~\bibnamefont {Senatore}}, \bibinfo {author}
  {\bibfnamefont {M.}~\bibnamefont {Simonovi\'c}}, \bibinfo {author}
  {\bibfnamefont {M.}~\bibnamefont {Takada}}, \bibinfo {author} {\bibfnamefont
  {M.}~\bibnamefont {Zaldarriaga}}, \ and\ \bibinfo {author} {\bibfnamefont
  {P.}~\bibnamefont {Zhang}},\ }\href {\doibase 10.1103/PhysRevD.102.123541}
  {\bibfield  {journal} {\bibinfo  {journal} {Phys. Rev. D}\ }\textbf {\bibinfo
  {volume} {102}},\ \bibinfo {pages} {123541} (\bibinfo {year} {2020})},\
  \Eprint {http://arxiv.org/abs/2003.08277} {arXiv:2003.08277 [astro-ph.CO]}
  \BibitemShut {NoStop}%
\bibitem [{\citenamefont {Ivanov}(2021)}]{Ivanov:2021zmi}%
  \BibitemOpen
  \bibfield  {author} {\bibinfo {author} {\bibfnamefont {M.~M.}\ \bibnamefont
  {Ivanov}},\ }\href@noop {} {\  (\bibinfo {year} {2021})},\ \Eprint
  {http://arxiv.org/abs/2106.12580} {arXiv:2106.12580 [astro-ph.CO]}
  \BibitemShut {NoStop}%
\bibitem [{\citenamefont {Chudaykin}\ \emph {et~al.}(2020)\citenamefont
  {Chudaykin}, \citenamefont {Ivanov}, \citenamefont {Philcox},\ and\
  \citenamefont {Simonovi\'c}}]{Chudaykin:2020aoj}%
  \BibitemOpen
  \bibfield  {author} {\bibinfo {author} {\bibfnamefont {A.}~\bibnamefont
  {Chudaykin}}, \bibinfo {author} {\bibfnamefont {M.~M.}\ \bibnamefont
  {Ivanov}}, \bibinfo {author} {\bibfnamefont {O.~H.~E.}\ \bibnamefont
  {Philcox}}, \ and\ \bibinfo {author} {\bibfnamefont {M.}~\bibnamefont
  {Simonovi\'c}},\ }\href {\doibase 10.1103/PhysRevD.102.063533} {\bibfield
  {journal} {\bibinfo  {journal} {Phys. Rev. D}\ }\textbf {\bibinfo {volume}
  {102}},\ \bibinfo {pages} {063533} (\bibinfo {year} {2020})},\ \Eprint
  {http://arxiv.org/abs/2004.10607} {arXiv:2004.10607 [astro-ph.CO]}
  \BibitemShut {NoStop}%
\bibitem [{\citenamefont {Chen}\ \emph {et~al.}(2021)\citenamefont {Chen},
  \citenamefont {Vlah}, \citenamefont {Castorina},\ and\ \citenamefont
  {White}}]{Chen:2020zjt}%
  \BibitemOpen
  \bibfield  {author} {\bibinfo {author} {\bibfnamefont {S.-F.}\ \bibnamefont
  {Chen}}, \bibinfo {author} {\bibfnamefont {Z.}~\bibnamefont {Vlah}}, \bibinfo
  {author} {\bibfnamefont {E.}~\bibnamefont {Castorina}}, \ and\ \bibinfo
  {author} {\bibfnamefont {M.}~\bibnamefont {White}},\ }\href {\doibase
  10.1088/1475-7516/2021/03/100} {\bibfield  {journal} {\bibinfo  {journal}
  {JCAP}\ }\textbf {\bibinfo {volume} {03}},\ \bibinfo {pages} {100} (\bibinfo
  {year} {2021})},\ \Eprint {http://arxiv.org/abs/2012.04636} {arXiv:2012.04636
  [astro-ph.CO]} \BibitemShut {NoStop}%
\bibitem [{\citenamefont {D'Amico}\ \emph {et~al.}(2021)\citenamefont
  {D'Amico}, \citenamefont {Senatore},\ and\ \citenamefont
  {Zhang}}]{DAmico:2020kxu}%
  \BibitemOpen
  \bibfield  {author} {\bibinfo {author} {\bibfnamefont {G.}~\bibnamefont
  {D'Amico}}, \bibinfo {author} {\bibfnamefont {L.}~\bibnamefont {Senatore}}, \
  and\ \bibinfo {author} {\bibfnamefont {P.}~\bibnamefont {Zhang}},\ }\href
  {\doibase 10.1088/1475-7516/2021/01/006} {\bibfield  {journal} {\bibinfo
  {journal} {JCAP}\ }\textbf {\bibinfo {volume} {01}},\ \bibinfo {pages} {006}
  (\bibinfo {year} {2021})},\ \Eprint {http://arxiv.org/abs/2003.07956}
  {arXiv:2003.07956 [astro-ph.CO]} \BibitemShut {NoStop}%
\bibitem [{\citenamefont {Ivanov}\ \emph
  {et~al.}(2020{\natexlab{a}})\citenamefont {Ivanov}, \citenamefont
  {McDonough}, \citenamefont {Hill}, \citenamefont {Simonovi\'c}, \citenamefont
  {Toomey}, \citenamefont {Alexander},\ and\ \citenamefont
  {Zaldarriaga}}]{Ivanov:2020ril}%
  \BibitemOpen
  \bibfield  {author} {\bibinfo {author} {\bibfnamefont {M.~M.}\ \bibnamefont
  {Ivanov}}, \bibinfo {author} {\bibfnamefont {E.}~\bibnamefont {McDonough}},
  \bibinfo {author} {\bibfnamefont {J.~C.}\ \bibnamefont {Hill}}, \bibinfo
  {author} {\bibfnamefont {M.}~\bibnamefont {Simonovi\'c}}, \bibinfo {author}
  {\bibfnamefont {M.~W.}\ \bibnamefont {Toomey}}, \bibinfo {author}
  {\bibfnamefont {S.}~\bibnamefont {Alexander}}, \ and\ \bibinfo {author}
  {\bibfnamefont {M.}~\bibnamefont {Zaldarriaga}},\ }\href {\doibase
  10.1103/PhysRevD.102.103502} {\bibfield  {journal} {\bibinfo  {journal}
  {Phys. Rev. D}\ }\textbf {\bibinfo {volume} {102}},\ \bibinfo {pages}
  {103502} (\bibinfo {year} {2020}{\natexlab{a}})},\ \Eprint
  {http://arxiv.org/abs/2006.11235} {arXiv:2006.11235 [astro-ph.CO]}
  \BibitemShut {NoStop}%
\bibitem [{\citenamefont {Chudaykin}\ \emph
  {et~al.}(2021{\natexlab{a}})\citenamefont {Chudaykin}, \citenamefont
  {Dolgikh},\ and\ \citenamefont {Ivanov}}]{Chudaykin:2020ghx}%
  \BibitemOpen
  \bibfield  {author} {\bibinfo {author} {\bibfnamefont {A.}~\bibnamefont
  {Chudaykin}}, \bibinfo {author} {\bibfnamefont {K.}~\bibnamefont {Dolgikh}},
  \ and\ \bibinfo {author} {\bibfnamefont {M.~M.}\ \bibnamefont {Ivanov}},\
  }\href {\doibase 10.1103/PhysRevD.103.023507} {\bibfield  {journal} {\bibinfo
   {journal} {Phys. Rev. D}\ }\textbf {\bibinfo {volume} {103}},\ \bibinfo
  {pages} {023507} (\bibinfo {year} {2021}{\natexlab{a}})},\ \Eprint
  {http://arxiv.org/abs/2009.10106} {arXiv:2009.10106 [astro-ph.CO]}
  \BibitemShut {NoStop}%
\bibitem [{\citenamefont {Ivanov}\ \emph
  {et~al.}(2020{\natexlab{b}})\citenamefont {Ivanov}, \citenamefont
  {Simonovi\'c},\ and\ \citenamefont {Zaldarriaga}}]{Ivanov:2019pdj}%
  \BibitemOpen
  \bibfield  {author} {\bibinfo {author} {\bibfnamefont {M.~M.}\ \bibnamefont
  {Ivanov}}, \bibinfo {author} {\bibfnamefont {M.}~\bibnamefont {Simonovi\'c}},
  \ and\ \bibinfo {author} {\bibfnamefont {M.}~\bibnamefont {Zaldarriaga}},\
  }\href {\doibase 10.1088/1475-7516/2020/05/042} {\bibfield  {journal}
  {\bibinfo  {journal} {JCAP}\ }\textbf {\bibinfo {volume} {05}},\ \bibinfo
  {pages} {042} (\bibinfo {year} {2020}{\natexlab{b}})},\ \Eprint
  {http://arxiv.org/abs/1909.05277} {arXiv:1909.05277 [astro-ph.CO]}
  \BibitemShut {NoStop}%
\bibitem [{\citenamefont {D'Amico}\ \emph {et~al.}(2019)\citenamefont
  {D'Amico}, \citenamefont {Gleyzes}, \citenamefont {Kokron}, \citenamefont
  {Markovic}, \citenamefont {Senatore}, \citenamefont {Zhang}, \citenamefont
  {Beutler},\ and\ \citenamefont {Gil-Marín}}]{DAmico:2019fhj}%
  \BibitemOpen
  \bibfield  {author} {\bibinfo {author} {\bibfnamefont {G.}~\bibnamefont
  {D'Amico}}, \bibinfo {author} {\bibfnamefont {J.}~\bibnamefont {Gleyzes}},
  \bibinfo {author} {\bibfnamefont {N.}~\bibnamefont {Kokron}}, \bibinfo
  {author} {\bibfnamefont {D.}~\bibnamefont {Markovic}}, \bibinfo {author}
  {\bibfnamefont {L.}~\bibnamefont {Senatore}}, \bibinfo {author}
  {\bibfnamefont {P.}~\bibnamefont {Zhang}}, \bibinfo {author} {\bibfnamefont
  {F.}~\bibnamefont {Beutler}}, \ and\ \bibinfo {author} {\bibfnamefont
  {H.}~\bibnamefont {Gil-Marín}},\ }\href@noop {} {\  (\bibinfo {year}
  {2019})},\ \Eprint {http://arxiv.org/abs/1909.05271} {arXiv:1909.05271
  [astro-ph.CO]} \BibitemShut {NoStop}%
\bibitem [{\citenamefont {Philcox}\ and\ \citenamefont
  {Ivanov}(2022)}]{Philcox:2021kcw}%
  \BibitemOpen
  \bibfield  {author} {\bibinfo {author} {\bibfnamefont {O.~H.~E.}\
  \bibnamefont {Philcox}}\ and\ \bibinfo {author} {\bibfnamefont {M.~M.}\
  \bibnamefont {Ivanov}},\ }\href {\doibase 10.1103/PhysRevD.105.043517}
  {\bibfield  {journal} {\bibinfo  {journal} {Phys. Rev. D}\ }\textbf {\bibinfo
  {volume} {105}},\ \bibinfo {pages} {043517} (\bibinfo {year} {2022})},\
  \Eprint {http://arxiv.org/abs/2112.04515} {arXiv:2112.04515 [astro-ph.CO]}
  \BibitemShut {NoStop}%
\bibitem [{\citenamefont {Chen}\ \emph {et~al.}(2022)\citenamefont {Chen},
  \citenamefont {Vlah},\ and\ \citenamefont {White}}]{Chen:2021wdi}%
  \BibitemOpen
  \bibfield  {author} {\bibinfo {author} {\bibfnamefont {S.-F.}\ \bibnamefont
  {Chen}}, \bibinfo {author} {\bibfnamefont {Z.}~\bibnamefont {Vlah}}, \ and\
  \bibinfo {author} {\bibfnamefont {M.}~\bibnamefont {White}},\ }\href
  {\doibase 10.1088/1475-7516/2022/02/008} {\bibfield  {journal} {\bibinfo
  {journal} {JCAP}\ }\textbf {\bibinfo {volume} {02}},\ \bibinfo {pages} {008}
  (\bibinfo {year} {2022})},\ \Eprint {http://arxiv.org/abs/2110.05530}
  {arXiv:2110.05530 [astro-ph.CO]} \BibitemShut {NoStop}%
\bibitem [{\citenamefont {Alam}\ \emph
  {et~al.}(2017{\natexlab{a}})\citenamefont {Alam} \emph
  {et~al.}}]{Alam:2016hwk}%
  \BibitemOpen
  \bibfield  {author} {\bibinfo {author} {\bibfnamefont {S.}~\bibnamefont
  {Alam}} \emph {et~al.} (\bibinfo {collaboration} {BOSS}),\ }\href {\doibase
  10.1093/mnras/stx721} {\bibfield  {journal} {\bibinfo  {journal} {Mon. Not.
  Roy. Astron. Soc.}\ }\textbf {\bibinfo {volume} {470}},\ \bibinfo {pages}
  {2617} (\bibinfo {year} {2017}{\natexlab{a}})},\ \Eprint
  {http://arxiv.org/abs/1607.03155} {arXiv:1607.03155 [astro-ph.CO]}
  \BibitemShut {NoStop}%
\bibitem [{\citenamefont {Desjacques}\ \emph {et~al.}(2018)\citenamefont
  {Desjacques}, \citenamefont {Jeong},\ and\ \citenamefont
  {Schmidt}}]{Desjacques:2016bnm}%
  \BibitemOpen
  \bibfield  {author} {\bibinfo {author} {\bibfnamefont {V.}~\bibnamefont
  {Desjacques}}, \bibinfo {author} {\bibfnamefont {D.}~\bibnamefont {Jeong}}, \
  and\ \bibinfo {author} {\bibfnamefont {F.}~\bibnamefont {Schmidt}},\ }\href
  {\doibase 10.1016/j.physrep.2017.12.002} {\bibfield  {journal} {\bibinfo
  {journal} {Phys. Rept.}\ }\textbf {\bibinfo {volume} {733}},\ \bibinfo
  {pages} {1} (\bibinfo {year} {2018})},\ \Eprint
  {http://arxiv.org/abs/1611.09787} {arXiv:1611.09787 [astro-ph.CO]}
  \BibitemShut {NoStop}%
\bibitem [{\citenamefont {Wadekar}\ \emph {et~al.}(2020)\citenamefont
  {Wadekar}, \citenamefont {Ivanov},\ and\ \citenamefont
  {Scoccimarro}}]{Wadekar:2020hax}%
  \BibitemOpen
  \bibfield  {author} {\bibinfo {author} {\bibfnamefont {D.}~\bibnamefont
  {Wadekar}}, \bibinfo {author} {\bibfnamefont {M.~M.}\ \bibnamefont {Ivanov}},
  \ and\ \bibinfo {author} {\bibfnamefont {R.}~\bibnamefont {Scoccimarro}},\
  }\href {\doibase 10.1103/PhysRevD.102.123521} {\bibfield  {journal} {\bibinfo
   {journal} {Phys. Rev. D}\ }\textbf {\bibinfo {volume} {102}},\ \bibinfo
  {pages} {123521} (\bibinfo {year} {2020})},\ \Eprint
  {http://arxiv.org/abs/2009.00622} {arXiv:2009.00622 [astro-ph.CO]}
  \BibitemShut {NoStop}%
\bibitem [{\citenamefont {Cabass}\ \emph
  {et~al.}(2022{\natexlab{a}})\citenamefont {Cabass}, \citenamefont {Ivanov},
  \citenamefont {Philcox}, \citenamefont {Simonovic},\ and\ \citenamefont
  {Zaldarriaga}}]{Cabass:2022epm}%
  \BibitemOpen
  \bibfield  {author} {\bibinfo {author} {\bibfnamefont {G.}~\bibnamefont
  {Cabass}}, \bibinfo {author} {\bibfnamefont {M.~M.}\ \bibnamefont {Ivanov}},
  \bibinfo {author} {\bibfnamefont {O.~H.~E.}\ \bibnamefont {Philcox}},
  \bibinfo {author} {\bibfnamefont {M.}~\bibnamefont {Simonovic}}, \ and\
  \bibinfo {author} {\bibfnamefont {M.}~\bibnamefont {Zaldarriaga}},\
  }\href@noop {} {\  (\bibinfo {year} {2022}{\natexlab{a}})},\ \Eprint
  {http://arxiv.org/abs/2211.14899} {arXiv:2211.14899 [astro-ph.CO]}
  \BibitemShut {NoStop}%
\bibitem [{\citenamefont {Philcox}\ \emph {et~al.}(2022)\citenamefont
  {Philcox}, \citenamefont {Ivanov}, \citenamefont {Cabass}, \citenamefont
  {Simonovi\'c}, \citenamefont {Zaldarriaga},\ and\ \citenamefont
  {Nishimichi}}]{Philcox:2022frc}%
  \BibitemOpen
  \bibfield  {author} {\bibinfo {author} {\bibfnamefont {O.~H.~E.}\
  \bibnamefont {Philcox}}, \bibinfo {author} {\bibfnamefont {M.~M.}\
  \bibnamefont {Ivanov}}, \bibinfo {author} {\bibfnamefont {G.}~\bibnamefont
  {Cabass}}, \bibinfo {author} {\bibfnamefont {M.}~\bibnamefont {Simonovi\'c}},
  \bibinfo {author} {\bibfnamefont {M.}~\bibnamefont {Zaldarriaga}}, \ and\
  \bibinfo {author} {\bibfnamefont {T.}~\bibnamefont {Nishimichi}},\ }\href
  {\doibase 10.1103/PhysRevD.106.043530} {\bibfield  {journal} {\bibinfo
  {journal} {Phys. Rev. D}\ }\textbf {\bibinfo {volume} {106}},\ \bibinfo
  {pages} {043530} (\bibinfo {year} {2022})},\ \Eprint
  {http://arxiv.org/abs/2206.02800} {arXiv:2206.02800 [astro-ph.CO]}
  \BibitemShut {NoStop}%
\bibitem [{\citenamefont {Ivanov}\ \emph {et~al.}(2023)\citenamefont {Ivanov},
  \citenamefont {Philcox}, \citenamefont {Cabass}, \citenamefont {Nishimichi},
  \citenamefont {Simonovi\'c},\ and\ \citenamefont
  {Zaldarriaga}}]{Ivanov:2023qzb}%
  \BibitemOpen
  \bibfield  {author} {\bibinfo {author} {\bibfnamefont {M.~M.}\ \bibnamefont
  {Ivanov}}, \bibinfo {author} {\bibfnamefont {O.~H.~E.}\ \bibnamefont
  {Philcox}}, \bibinfo {author} {\bibfnamefont {G.}~\bibnamefont {Cabass}},
  \bibinfo {author} {\bibfnamefont {T.}~\bibnamefont {Nishimichi}}, \bibinfo
  {author} {\bibfnamefont {M.}~\bibnamefont {Simonovi\'c}}, \ and\ \bibinfo
  {author} {\bibfnamefont {M.}~\bibnamefont {Zaldarriaga}},\ }\href {\doibase
  10.1103/PhysRevD.107.083515} {\bibfield  {journal} {\bibinfo  {journal}
  {Phys. Rev. D}\ }\textbf {\bibinfo {volume} {107}},\ \bibinfo {pages}
  {083515} (\bibinfo {year} {2023})},\ \Eprint
  {http://arxiv.org/abs/2302.04414} {arXiv:2302.04414 [astro-ph.CO]}
  \BibitemShut {NoStop}%
\bibitem [{\citenamefont {Eggemeier}\ \emph {et~al.}(2021)\citenamefont
  {Eggemeier}, \citenamefont {Scoccimarro}, \citenamefont {Smith},
  \citenamefont {Crocce}, \citenamefont {Pezzotta},\ and\ \citenamefont
  {S\'anchez}}]{Eggemeier:2021cam}%
  \BibitemOpen
  \bibfield  {author} {\bibinfo {author} {\bibfnamefont {A.}~\bibnamefont
  {Eggemeier}}, \bibinfo {author} {\bibfnamefont {R.}~\bibnamefont
  {Scoccimarro}}, \bibinfo {author} {\bibfnamefont {R.~E.}\ \bibnamefont
  {Smith}}, \bibinfo {author} {\bibfnamefont {M.}~\bibnamefont {Crocce}},
  \bibinfo {author} {\bibfnamefont {A.}~\bibnamefont {Pezzotta}}, \ and\
  \bibinfo {author} {\bibfnamefont {A.~G.}\ \bibnamefont {S\'anchez}},\
  }\href@noop {} {\  (\bibinfo {year} {2021})},\ \Eprint
  {http://arxiv.org/abs/2102.06902} {arXiv:2102.06902 [astro-ph.CO]}
  \BibitemShut {NoStop}%
\bibitem [{\citenamefont {Beutler}\ \emph {et~al.}(2017)\citenamefont {Beutler}
  \emph {et~al.}}]{Beutler:2016arn}%
  \BibitemOpen
  \bibfield  {author} {\bibinfo {author} {\bibfnamefont {F.}~\bibnamefont
  {Beutler}} \emph {et~al.} (\bibinfo {collaboration} {BOSS}),\ }\href
  {\doibase 10.1093/mnras/stw3298} {\bibfield  {journal} {\bibinfo  {journal}
  {Mon. Not. Roy. Astron. Soc.}\ }\textbf {\bibinfo {volume} {466}},\ \bibinfo
  {pages} {2242} (\bibinfo {year} {2017})},\ \Eprint
  {http://arxiv.org/abs/1607.03150} {arXiv:1607.03150 [astro-ph.CO]}
  \BibitemShut {NoStop}%
\bibitem [{\citenamefont {Abidi}\ and\ \citenamefont
  {Baldauf}(2018)}]{Abidi:2018eyd}%
  \BibitemOpen
  \bibfield  {author} {\bibinfo {author} {\bibfnamefont {M.~M.}\ \bibnamefont
  {Abidi}}\ and\ \bibinfo {author} {\bibfnamefont {T.}~\bibnamefont
  {Baldauf}},\ }\href {\doibase 10.1088/1475-7516/2018/07/029} {\bibfield
  {journal} {\bibinfo  {journal} {JCAP}\ }\textbf {\bibinfo {volume} {1807}},\
  \bibinfo {pages} {029} (\bibinfo {year} {2018})},\ \Eprint
  {http://arxiv.org/abs/1802.07622} {arXiv:1802.07622 [astro-ph.CO]}
  \BibitemShut {NoStop}%
\bibitem [{\citenamefont {Ivanov}\ \emph
  {et~al.}(2022{\natexlab{a}})\citenamefont {Ivanov}, \citenamefont {Philcox},
  \citenamefont {Nishimichi}, \citenamefont {Simonovi\'c}, \citenamefont
  {Takada},\ and\ \citenamefont {Zaldarriaga}}]{Ivanov:2021kcd}%
  \BibitemOpen
  \bibfield  {author} {\bibinfo {author} {\bibfnamefont {M.~M.}\ \bibnamefont
  {Ivanov}}, \bibinfo {author} {\bibfnamefont {O.~H.~E.}\ \bibnamefont
  {Philcox}}, \bibinfo {author} {\bibfnamefont {T.}~\bibnamefont {Nishimichi}},
  \bibinfo {author} {\bibfnamefont {M.}~\bibnamefont {Simonovi\'c}}, \bibinfo
  {author} {\bibfnamefont {M.}~\bibnamefont {Takada}}, \ and\ \bibinfo {author}
  {\bibfnamefont {M.}~\bibnamefont {Zaldarriaga}},\ }\href {\doibase
  10.1103/PhysRevD.105.063512} {\bibfield  {journal} {\bibinfo  {journal}
  {Phys. Rev. D}\ }\textbf {\bibinfo {volume} {105}},\ \bibinfo {pages}
  {063512} (\bibinfo {year} {2022}{\natexlab{a}})},\ \Eprint
  {http://arxiv.org/abs/2110.10161} {arXiv:2110.10161 [astro-ph.CO]}
  \BibitemShut {NoStop}%
\bibitem [{\citenamefont {Barreira}\ \emph {et~al.}(2021)\citenamefont
  {Barreira}, \citenamefont {Lazeyras},\ and\ \citenamefont
  {Schmidt}}]{Barreira:2021ukk}%
  \BibitemOpen
  \bibfield  {author} {\bibinfo {author} {\bibfnamefont {A.}~\bibnamefont
  {Barreira}}, \bibinfo {author} {\bibfnamefont {T.}~\bibnamefont {Lazeyras}},
  \ and\ \bibinfo {author} {\bibfnamefont {F.}~\bibnamefont {Schmidt}},\
  }\href@noop {} {\  (\bibinfo {year} {2021})},\ \Eprint
  {http://arxiv.org/abs/2105.02876} {arXiv:2105.02876 [astro-ph.CO]}
  \BibitemShut {NoStop}%
\bibitem [{\citenamefont {Berlind}\ and\ \citenamefont
  {Weinberg}(2002)}]{Berlind:2001xk}%
  \BibitemOpen
  \bibfield  {author} {\bibinfo {author} {\bibfnamefont {A.~A.}\ \bibnamefont
  {Berlind}}\ and\ \bibinfo {author} {\bibfnamefont {D.~H.}\ \bibnamefont
  {Weinberg}},\ }\href {\doibase 10.1086/341469} {\bibfield  {journal}
  {\bibinfo  {journal} {Astrophys. J.}\ }\textbf {\bibinfo {volume} {575}},\
  \bibinfo {pages} {587} (\bibinfo {year} {2002})},\ \Eprint
  {http://arxiv.org/abs/astro-ph/0109001} {arXiv:astro-ph/0109001} \BibitemShut
  {NoStop}%
\bibitem [{\citenamefont {Zheng}\ \emph {et~al.}(2005)\citenamefont {Zheng},
  \citenamefont {Berlind}, \citenamefont {Weinberg}, \citenamefont {Benson},
  \citenamefont {Baugh}, \citenamefont {Cole}, \citenamefont {Dave},
  \citenamefont {Frenk}, \citenamefont {Katz},\ and\ \citenamefont
  {Lacey}}]{Zheng:2004id}%
  \BibitemOpen
  \bibfield  {author} {\bibinfo {author} {\bibfnamefont {Z.}~\bibnamefont
  {Zheng}}, \bibinfo {author} {\bibfnamefont {A.~A.}\ \bibnamefont {Berlind}},
  \bibinfo {author} {\bibfnamefont {D.~H.}\ \bibnamefont {Weinberg}}, \bibinfo
  {author} {\bibfnamefont {A.~J.}\ \bibnamefont {Benson}}, \bibinfo {author}
  {\bibfnamefont {C.~M.}\ \bibnamefont {Baugh}}, \bibinfo {author}
  {\bibfnamefont {S.}~\bibnamefont {Cole}}, \bibinfo {author} {\bibfnamefont
  {R.}~\bibnamefont {Dave}}, \bibinfo {author} {\bibfnamefont {C.~S.}\
  \bibnamefont {Frenk}}, \bibinfo {author} {\bibfnamefont {N.}~\bibnamefont
  {Katz}}, \ and\ \bibinfo {author} {\bibfnamefont {C.~G.}\ \bibnamefont
  {Lacey}},\ }\href {\doibase 10.1086/466510} {\bibfield  {journal} {\bibinfo
  {journal} {Astrophys. J.}\ }\textbf {\bibinfo {volume} {633}},\ \bibinfo
  {pages} {791} (\bibinfo {year} {2005})},\ \Eprint
  {http://arxiv.org/abs/astro-ph/0408564} {arXiv:astro-ph/0408564} \BibitemShut
  {NoStop}%
\bibitem [{\citenamefont {Zheng}\ \emph {et~al.}(2007)\citenamefont {Zheng},
  \citenamefont {Coil},\ and\ \citenamefont {Zehavi}}]{Zheng:2007zg}%
  \BibitemOpen
  \bibfield  {author} {\bibinfo {author} {\bibfnamefont {Z.}~\bibnamefont
  {Zheng}}, \bibinfo {author} {\bibfnamefont {A.~L.}\ \bibnamefont {Coil}}, \
  and\ \bibinfo {author} {\bibfnamefont {I.}~\bibnamefont {Zehavi}},\ }\href
  {\doibase 10.1086/521074} {\bibfield  {journal} {\bibinfo  {journal}
  {Astrophys. J.}\ }\textbf {\bibinfo {volume} {667}},\ \bibinfo {pages} {760}
  (\bibinfo {year} {2007})},\ \Eprint {http://arxiv.org/abs/astro-ph/0703457}
  {arXiv:astro-ph/0703457} \BibitemShut {NoStop}%
\bibitem [{\citenamefont {Wechsler}\ and\ \citenamefont
  {Tinker}(2018)}]{Wechsler:2018pic}%
  \BibitemOpen
  \bibfield  {author} {\bibinfo {author} {\bibfnamefont {R.~H.}\ \bibnamefont
  {Wechsler}}\ and\ \bibinfo {author} {\bibfnamefont {J.~L.}\ \bibnamefont
  {Tinker}},\ }\href {\doibase 10.1146/annurev-astro-081817-051756} {\bibfield
  {journal} {\bibinfo  {journal} {Ann. Rev. Astron. Astrophys.}\ }\textbf
  {\bibinfo {volume} {56}},\ \bibinfo {pages} {435} (\bibinfo {year} {2018})},\
  \Eprint {http://arxiv.org/abs/1804.03097} {arXiv:1804.03097 [astro-ph.GA]}
  \BibitemShut {NoStop}%
\bibitem [{\citenamefont {Kobayashi}\ \emph {et~al.}(2022)\citenamefont
  {Kobayashi}, \citenamefont {Nishimichi}, \citenamefont {Takada},\ and\
  \citenamefont {Miyatake}}]{Kobayashi:2021oud}%
  \BibitemOpen
  \bibfield  {author} {\bibinfo {author} {\bibfnamefont {Y.}~\bibnamefont
  {Kobayashi}}, \bibinfo {author} {\bibfnamefont {T.}~\bibnamefont
  {Nishimichi}}, \bibinfo {author} {\bibfnamefont {M.}~\bibnamefont {Takada}},
  \ and\ \bibinfo {author} {\bibfnamefont {H.}~\bibnamefont {Miyatake}},\
  }\href {\doibase 10.1103/PhysRevD.105.083517} {\bibfield  {journal} {\bibinfo
   {journal} {Phys. Rev. D}\ }\textbf {\bibinfo {volume} {105}},\ \bibinfo
  {pages} {083517} (\bibinfo {year} {2022})},\ \Eprint
  {http://arxiv.org/abs/2110.06969} {arXiv:2110.06969 [astro-ph.CO]}
  \BibitemShut {NoStop}%
\bibitem [{\citenamefont {Paillas}\ \emph {et~al.}(2023)\citenamefont {Paillas}
  \emph {et~al.}}]{Paillas:2023cpk}%
  \BibitemOpen
  \bibfield  {author} {\bibinfo {author} {\bibfnamefont {E.}~\bibnamefont
  {Paillas}} \emph {et~al.},\ }\href@noop {} {\  (\bibinfo {year} {2023})},\
  \Eprint {http://arxiv.org/abs/2309.16541} {arXiv:2309.16541 [astro-ph.CO]}
  \BibitemShut {NoStop}%
\bibitem [{\citenamefont {Cuesta-Lazaro}\ \emph {et~al.}(2023)\citenamefont
  {Cuesta-Lazaro} \emph {et~al.}}]{Cuesta-Lazaro:2023gbv}%
  \BibitemOpen
  \bibfield  {author} {\bibinfo {author} {\bibfnamefont {C.}~\bibnamefont
  {Cuesta-Lazaro}} \emph {et~al.},\ }\href@noop {} {\  (\bibinfo {year}
  {2023})},\ \Eprint {http://arxiv.org/abs/2309.16539} {arXiv:2309.16539
  [astro-ph.CO]} \BibitemShut {NoStop}%
\bibitem [{\citenamefont {Hahn}\ \emph {et~al.}(2023)\citenamefont {Hahn} \emph
  {et~al.}}]{Hahn:2023udg}%
  \BibitemOpen
  \bibfield  {author} {\bibinfo {author} {\bibfnamefont {C.}~\bibnamefont
  {Hahn}} \emph {et~al.},\ }\href@noop {} {\  (\bibinfo {year} {2023})},\
  \Eprint {http://arxiv.org/abs/2310.15246} {arXiv:2310.15246 [astro-ph.CO]}
  \BibitemShut {NoStop}%
\bibitem [{\citenamefont {Valogiannis}\ \emph {et~al.}(2023)\citenamefont
  {Valogiannis}, \citenamefont {Yuan},\ and\ \citenamefont
  {Dvorkin}}]{Valogiannis:2023mxf}%
  \BibitemOpen
  \bibfield  {author} {\bibinfo {author} {\bibfnamefont {G.}~\bibnamefont
  {Valogiannis}}, \bibinfo {author} {\bibfnamefont {S.}~\bibnamefont {Yuan}}, \
  and\ \bibinfo {author} {\bibfnamefont {C.}~\bibnamefont {Dvorkin}},\
  }\href@noop {} {\  (\bibinfo {year} {2023})},\ \Eprint
  {http://arxiv.org/abs/2310.16116} {arXiv:2310.16116 [astro-ph.CO]}
  \BibitemShut {NoStop}%
\bibitem [{\citenamefont {Barreira}\ \emph {et~al.}(2020)\citenamefont
  {Barreira}, \citenamefont {Cabass}, \citenamefont {Schmidt}, \citenamefont
  {Pillepich},\ and\ \citenamefont {Nelson}}]{Barreira:2020kvh}%
  \BibitemOpen
  \bibfield  {author} {\bibinfo {author} {\bibfnamefont {A.}~\bibnamefont
  {Barreira}}, \bibinfo {author} {\bibfnamefont {G.}~\bibnamefont {Cabass}},
  \bibinfo {author} {\bibfnamefont {F.}~\bibnamefont {Schmidt}}, \bibinfo
  {author} {\bibfnamefont {A.}~\bibnamefont {Pillepich}}, \ and\ \bibinfo
  {author} {\bibfnamefont {D.}~\bibnamefont {Nelson}},\ }\href {\doibase
  10.1088/1475-7516/2020/12/013} {\bibfield  {journal} {\bibinfo  {journal}
  {JCAP}\ }\textbf {\bibinfo {volume} {12}},\ \bibinfo {pages} {013} (\bibinfo
  {year} {2020})},\ \Eprint {http://arxiv.org/abs/2006.09368} {arXiv:2006.09368
  [astro-ph.CO]} \BibitemShut {NoStop}%
\bibitem [{\citenamefont {Kokron}\ \emph {et~al.}(2022)\citenamefont {Kokron},
  \citenamefont {DeRose}, \citenamefont {Chen}, \citenamefont {White},\ and\
  \citenamefont {Wechsler}}]{Kokron:2021faa}%
  \BibitemOpen
  \bibfield  {author} {\bibinfo {author} {\bibfnamefont {N.}~\bibnamefont
  {Kokron}}, \bibinfo {author} {\bibfnamefont {J.}~\bibnamefont {DeRose}},
  \bibinfo {author} {\bibfnamefont {S.-F.}\ \bibnamefont {Chen}}, \bibinfo
  {author} {\bibfnamefont {M.}~\bibnamefont {White}}, \ and\ \bibinfo {author}
  {\bibfnamefont {R.~H.}\ \bibnamefont {Wechsler}},\ }\href {\doibase
  10.1093/mnras/stac1420} {\bibfield  {journal} {\bibinfo  {journal} {Mon. Not.
  Roy. Astron. Soc.}\ }\textbf {\bibinfo {volume} {514}},\ \bibinfo {pages}
  {2198} (\bibinfo {year} {2022})},\ \Eprint {http://arxiv.org/abs/2112.00012}
  {arXiv:2112.00012 [astro-ph.CO]} \BibitemShut {NoStop}%
\bibitem [{\citenamefont {Zennaro}\ \emph {et~al.}(2022)\citenamefont
  {Zennaro}, \citenamefont {Angulo}, \citenamefont {Contreras}, \citenamefont
  {Pellejero-Ib\'a\~nez},\ and\ \citenamefont {Maion}}]{Zennaro:2021pbe}%
  \BibitemOpen
  \bibfield  {author} {\bibinfo {author} {\bibfnamefont {M.}~\bibnamefont
  {Zennaro}}, \bibinfo {author} {\bibfnamefont {R.~E.}\ \bibnamefont {Angulo}},
  \bibinfo {author} {\bibfnamefont {S.}~\bibnamefont {Contreras}}, \bibinfo
  {author} {\bibfnamefont {M.}~\bibnamefont {Pellejero-Ib\'a\~nez}}, \ and\
  \bibinfo {author} {\bibfnamefont {F.}~\bibnamefont {Maion}},\ }\href
  {\doibase 10.1093/mnras/stac1673} {\bibfield  {journal} {\bibinfo  {journal}
  {Mon. Not. Roy. Astron. Soc.}\ }\textbf {\bibinfo {volume} {514}},\ \bibinfo
  {pages} {5443} (\bibinfo {year} {2022})},\ \Eprint
  {http://arxiv.org/abs/2110.05408} {arXiv:2110.05408 [astro-ph.CO]}
  \BibitemShut {NoStop}%
\bibitem [{\citenamefont {Sullivan}\ \emph {et~al.}(2021)\citenamefont
  {Sullivan}, \citenamefont {Seljak},\ and\ \citenamefont
  {Singh}}]{Sullivan:2021sof}%
  \BibitemOpen
  \bibfield  {author} {\bibinfo {author} {\bibfnamefont {J.~M.}\ \bibnamefont
  {Sullivan}}, \bibinfo {author} {\bibfnamefont {U.}~\bibnamefont {Seljak}}, \
  and\ \bibinfo {author} {\bibfnamefont {S.}~\bibnamefont {Singh}},\ }\href
  {\doibase 10.1088/1475-7516/2021/11/026} {\bibfield  {journal} {\bibinfo
  {journal} {JCAP}\ }\textbf {\bibinfo {volume} {11}},\ \bibinfo {pages} {026}
  (\bibinfo {year} {2021})},\ \Eprint {http://arxiv.org/abs/2104.10676}
  {arXiv:2104.10676 [astro-ph.CO]} \BibitemShut {NoStop}%
\bibitem [{\citenamefont {Arkani-Hamed}\ \emph {et~al.}(2004)\citenamefont
  {Arkani-Hamed}, \citenamefont {Creminelli}, \citenamefont {Mukohyama},\ and\
  \citenamefont {Zaldarriaga}}]{Arkani-Hamed:2003juy}%
  \BibitemOpen
  \bibfield  {author} {\bibinfo {author} {\bibfnamefont {N.}~\bibnamefont
  {Arkani-Hamed}}, \bibinfo {author} {\bibfnamefont {P.}~\bibnamefont
  {Creminelli}}, \bibinfo {author} {\bibfnamefont {S.}~\bibnamefont
  {Mukohyama}}, \ and\ \bibinfo {author} {\bibfnamefont {M.}~\bibnamefont
  {Zaldarriaga}},\ }\href {\doibase 10.1088/1475-7516/2004/04/001} {\bibfield
  {journal} {\bibinfo  {journal} {JCAP}\ }\textbf {\bibinfo {volume} {04}},\
  \bibinfo {pages} {001} (\bibinfo {year} {2004})},\ \Eprint
  {http://arxiv.org/abs/hep-th/0312100} {arXiv:hep-th/0312100} \BibitemShut
  {NoStop}%
\bibitem [{\citenamefont {Alishahiha}\ \emph {et~al.}(2004)\citenamefont
  {Alishahiha}, \citenamefont {Silverstein},\ and\ \citenamefont
  {Tong}}]{Alishahiha:2004eh}%
  \BibitemOpen
  \bibfield  {author} {\bibinfo {author} {\bibfnamefont {M.}~\bibnamefont
  {Alishahiha}}, \bibinfo {author} {\bibfnamefont {E.}~\bibnamefont
  {Silverstein}}, \ and\ \bibinfo {author} {\bibfnamefont {D.}~\bibnamefont
  {Tong}},\ }\href {\doibase 10.1103/PhysRevD.70.123505} {\bibfield  {journal}
  {\bibinfo  {journal} {Phys. Rev. D}\ }\textbf {\bibinfo {volume} {70}},\
  \bibinfo {pages} {123505} (\bibinfo {year} {2004})},\ \Eprint
  {http://arxiv.org/abs/hep-th/0404084} {arXiv:hep-th/0404084} \BibitemShut
  {NoStop}%
\bibitem [{\citenamefont {Senatore}(2005)}]{Senatore:2004rj}%
  \BibitemOpen
  \bibfield  {author} {\bibinfo {author} {\bibfnamefont {L.}~\bibnamefont
  {Senatore}},\ }\href {\doibase 10.1103/PhysRevD.71.043512} {\bibfield
  {journal} {\bibinfo  {journal} {Phys. Rev. D}\ }\textbf {\bibinfo {volume}
  {71}},\ \bibinfo {pages} {043512} (\bibinfo {year} {2005})},\ \Eprint
  {http://arxiv.org/abs/astro-ph/0406187} {arXiv:astro-ph/0406187} \BibitemShut
  {NoStop}%
\bibitem [{\citenamefont {Chen}\ \emph {et~al.}(2007)\citenamefont {Chen},
  \citenamefont {Huang}, \citenamefont {Kachru},\ and\ \citenamefont
  {Shiu}}]{Chen:2006nt}%
  \BibitemOpen
  \bibfield  {author} {\bibinfo {author} {\bibfnamefont {X.}~\bibnamefont
  {Chen}}, \bibinfo {author} {\bibfnamefont {M.-x.}\ \bibnamefont {Huang}},
  \bibinfo {author} {\bibfnamefont {S.}~\bibnamefont {Kachru}}, \ and\ \bibinfo
  {author} {\bibfnamefont {G.}~\bibnamefont {Shiu}},\ }\href {\doibase
  10.1088/1475-7516/2007/01/002} {\bibfield  {journal} {\bibinfo  {journal}
  {JCAP}\ }\textbf {\bibinfo {volume} {01}},\ \bibinfo {pages} {002} (\bibinfo
  {year} {2007})},\ \Eprint {http://arxiv.org/abs/hep-th/0605045}
  {arXiv:hep-th/0605045} \BibitemShut {NoStop}%
\bibitem [{\citenamefont {Creminelli}\ \emph {et~al.}(2006)\citenamefont
  {Creminelli}, \citenamefont {Luty}, \citenamefont {Nicolis},\ and\
  \citenamefont {Senatore}}]{Creminelli:2006xe}%
  \BibitemOpen
  \bibfield  {author} {\bibinfo {author} {\bibfnamefont {P.}~\bibnamefont
  {Creminelli}}, \bibinfo {author} {\bibfnamefont {M.~A.}\ \bibnamefont
  {Luty}}, \bibinfo {author} {\bibfnamefont {A.}~\bibnamefont {Nicolis}}, \
  and\ \bibinfo {author} {\bibfnamefont {L.}~\bibnamefont {Senatore}},\ }\href
  {\doibase 10.1088/1126-6708/2006/12/080} {\bibfield  {journal} {\bibinfo
  {journal} {JHEP}\ }\textbf {\bibinfo {volume} {12}},\ \bibinfo {pages} {080}
  (\bibinfo {year} {2006})},\ \Eprint {http://arxiv.org/abs/hep-th/0606090}
  {arXiv:hep-th/0606090} \BibitemShut {NoStop}%
\bibitem [{\citenamefont {Cheung}\ \emph
  {et~al.}(2008{\natexlab{a}})\citenamefont {Cheung}, \citenamefont
  {Creminelli}, \citenamefont {Fitzpatrick}, \citenamefont {Kaplan},\ and\
  \citenamefont {Senatore}}]{Cheung:2007st}%
  \BibitemOpen
  \bibfield  {author} {\bibinfo {author} {\bibfnamefont {C.}~\bibnamefont
  {Cheung}}, \bibinfo {author} {\bibfnamefont {P.}~\bibnamefont {Creminelli}},
  \bibinfo {author} {\bibfnamefont {A.~L.}\ \bibnamefont {Fitzpatrick}},
  \bibinfo {author} {\bibfnamefont {J.}~\bibnamefont {Kaplan}}, \ and\ \bibinfo
  {author} {\bibfnamefont {L.}~\bibnamefont {Senatore}},\ }\href {\doibase
  10.1088/1126-6708/2008/03/014} {\bibfield  {journal} {\bibinfo  {journal}
  {JHEP}\ }\textbf {\bibinfo {volume} {03}},\ \bibinfo {pages} {014} (\bibinfo
  {year} {2008}{\natexlab{a}})},\ \Eprint {http://arxiv.org/abs/0709.0293}
  {arXiv:0709.0293 [hep-th]} \BibitemShut {NoStop}%
\bibitem [{\citenamefont {Cheung}\ \emph
  {et~al.}(2008{\natexlab{b}})\citenamefont {Cheung}, \citenamefont
  {Fitzpatrick}, \citenamefont {Kaplan},\ and\ \citenamefont
  {Senatore}}]{Cheung:2007sv}%
  \BibitemOpen
  \bibfield  {author} {\bibinfo {author} {\bibfnamefont {C.}~\bibnamefont
  {Cheung}}, \bibinfo {author} {\bibfnamefont {A.~L.}\ \bibnamefont
  {Fitzpatrick}}, \bibinfo {author} {\bibfnamefont {J.}~\bibnamefont {Kaplan}},
  \ and\ \bibinfo {author} {\bibfnamefont {L.}~\bibnamefont {Senatore}},\
  }\href {\doibase 10.1088/1475-7516/2008/02/021} {\bibfield  {journal}
  {\bibinfo  {journal} {JCAP}\ }\textbf {\bibinfo {volume} {02}},\ \bibinfo
  {pages} {021} (\bibinfo {year} {2008}{\natexlab{b}})},\ \Eprint
  {http://arxiv.org/abs/0709.0295} {arXiv:0709.0295 [hep-th]} \BibitemShut
  {NoStop}%
\bibitem [{\citenamefont {Senatore}\ \emph {et~al.}(2010)\citenamefont
  {Senatore}, \citenamefont {Smith},\ and\ \citenamefont
  {Zaldarriaga}}]{Senatore:2009gt}%
  \BibitemOpen
  \bibfield  {author} {\bibinfo {author} {\bibfnamefont {L.}~\bibnamefont
  {Senatore}}, \bibinfo {author} {\bibfnamefont {K.~M.}\ \bibnamefont {Smith}},
  \ and\ \bibinfo {author} {\bibfnamefont {M.}~\bibnamefont {Zaldarriaga}},\
  }\href {\doibase 10.1088/1475-7516/2010/01/028} {\bibfield  {journal}
  {\bibinfo  {journal} {JCAP}\ }\textbf {\bibinfo {volume} {01}},\ \bibinfo
  {pages} {028} (\bibinfo {year} {2010})},\ \Eprint
  {http://arxiv.org/abs/0905.3746} {arXiv:0905.3746 [astro-ph.CO]} \BibitemShut
  {NoStop}%
\bibitem [{\citenamefont {Cabass}\ \emph
  {et~al.}(2022{\natexlab{b}})\citenamefont {Cabass}, \citenamefont {Ivanov},
  \citenamefont {Philcox}, \citenamefont {Simonovi\'c},\ and\ \citenamefont
  {Zaldarriaga}}]{Cabass:2022wjy}%
  \BibitemOpen
  \bibfield  {author} {\bibinfo {author} {\bibfnamefont {G.}~\bibnamefont
  {Cabass}}, \bibinfo {author} {\bibfnamefont {M.~M.}\ \bibnamefont {Ivanov}},
  \bibinfo {author} {\bibfnamefont {O.~H.~E.}\ \bibnamefont {Philcox}},
  \bibinfo {author} {\bibfnamefont {M.}~\bibnamefont {Simonovi\'c}}, \ and\
  \bibinfo {author} {\bibfnamefont {M.}~\bibnamefont {Zaldarriaga}},\
  }\href@noop {} {\  (\bibinfo {year} {2022}{\natexlab{b}})},\ \Eprint
  {http://arxiv.org/abs/2201.07238} {arXiv:2201.07238 [astro-ph.CO]}
  \BibitemShut {NoStop}%
\bibitem [{\citenamefont {Cabass}\ \emph
  {et~al.}(2022{\natexlab{c}})\citenamefont {Cabass}, \citenamefont {Ivanov},
  \citenamefont {Philcox}, \citenamefont {Simonovi\'c},\ and\ \citenamefont
  {Zaldarriaga}}]{Cabass:2022ymb}%
  \BibitemOpen
  \bibfield  {author} {\bibinfo {author} {\bibfnamefont {G.}~\bibnamefont
  {Cabass}}, \bibinfo {author} {\bibfnamefont {M.~M.}\ \bibnamefont {Ivanov}},
  \bibinfo {author} {\bibfnamefont {O.~H.~E.}\ \bibnamefont {Philcox}},
  \bibinfo {author} {\bibfnamefont {M.}~\bibnamefont {Simonovi\'c}}, \ and\
  \bibinfo {author} {\bibfnamefont {M.}~\bibnamefont {Zaldarriaga}},\
  }\href@noop {} {\  (\bibinfo {year} {2022}{\natexlab{c}})},\ \Eprint
  {http://arxiv.org/abs/2204.01781} {arXiv:2204.01781 [astro-ph.CO]}
  \BibitemShut {NoStop}%
\bibitem [{\citenamefont {D'Amico}\ \emph
  {et~al.}(2022{\natexlab{a}})\citenamefont {D'Amico}, \citenamefont
  {Lewandowski}, \citenamefont {Senatore},\ and\ \citenamefont
  {Zhang}}]{DAmico:2022gki}%
  \BibitemOpen
  \bibfield  {author} {\bibinfo {author} {\bibfnamefont {G.}~\bibnamefont
  {D'Amico}}, \bibinfo {author} {\bibfnamefont {M.}~\bibnamefont
  {Lewandowski}}, \bibinfo {author} {\bibfnamefont {L.}~\bibnamefont
  {Senatore}}, \ and\ \bibinfo {author} {\bibfnamefont {P.}~\bibnamefont
  {Zhang}},\ }\href@noop {} {\  (\bibinfo {year} {2022}{\natexlab{a}})},\
  \Eprint {http://arxiv.org/abs/2201.11518} {arXiv:2201.11518 [astro-ph.CO]}
  \BibitemShut {NoStop}%
\bibitem [{\citenamefont {Chen}\ \emph {et~al.}(2024)\citenamefont {Chen},
  \citenamefont {Chakraborty},\ and\ \citenamefont {Dvorkin}}]{Chen:2024bdg}%
  \BibitemOpen
  \bibfield  {author} {\bibinfo {author} {\bibfnamefont {S.-F.}\ \bibnamefont
  {Chen}}, \bibinfo {author} {\bibfnamefont {P.}~\bibnamefont {Chakraborty}}, \
  and\ \bibinfo {author} {\bibfnamefont {C.}~\bibnamefont {Dvorkin}},\
  }\href@noop {} {\  (\bibinfo {year} {2024})},\ \Eprint
  {http://arxiv.org/abs/2401.13036} {arXiv:2401.13036 [astro-ph.CO]}
  \BibitemShut {NoStop}%
\bibitem [{\citenamefont {Castorina}\ \emph {et~al.}(2019)\citenamefont
  {Castorina} \emph {et~al.}}]{Castorina:2019wmr}%
  \BibitemOpen
  \bibfield  {author} {\bibinfo {author} {\bibfnamefont {E.}~\bibnamefont
  {Castorina}} \emph {et~al.},\ }\href {\doibase 10.1088/1475-7516/2019/09/010}
  {\bibfield  {journal} {\bibinfo  {journal} {JCAP}\ }\textbf {\bibinfo
  {volume} {09}},\ \bibinfo {pages} {010} (\bibinfo {year} {2019})},\ \Eprint
  {http://arxiv.org/abs/1904.08859} {arXiv:1904.08859 [astro-ph.CO]}
  \BibitemShut {NoStop}%
\bibitem [{\citenamefont {Moradinezhad~Dizgah}\ \emph
  {et~al.}(2021)\citenamefont {Moradinezhad~Dizgah}, \citenamefont {Biagetti},
  \citenamefont {Sefusatti}, \citenamefont {Desjacques},\ and\ \citenamefont
  {Nore\~na}}]{MoradinezhadDizgah:2020whw}%
  \BibitemOpen
  \bibfield  {author} {\bibinfo {author} {\bibfnamefont {A.}~\bibnamefont
  {Moradinezhad~Dizgah}}, \bibinfo {author} {\bibfnamefont {M.}~\bibnamefont
  {Biagetti}}, \bibinfo {author} {\bibfnamefont {E.}~\bibnamefont {Sefusatti}},
  \bibinfo {author} {\bibfnamefont {V.}~\bibnamefont {Desjacques}}, \ and\
  \bibinfo {author} {\bibfnamefont {J.}~\bibnamefont {Nore\~na}},\ }\href
  {\doibase 10.1088/1475-7516/2021/05/015} {\bibfield  {journal} {\bibinfo
  {journal} {JCAP}\ }\textbf {\bibinfo {volume} {05}},\ \bibinfo {pages} {015}
  (\bibinfo {year} {2021})},\ \Eprint {http://arxiv.org/abs/2010.14523}
  {arXiv:2010.14523 [astro-ph.CO]} \BibitemShut {NoStop}%
\bibitem [{\citenamefont {Lazeyras}\ \emph {et~al.}(2023)\citenamefont
  {Lazeyras}, \citenamefont {Barreira}, \citenamefont {Schmidt},\ and\
  \citenamefont {Desjacques}}]{Lazeyras:2022koc}%
  \BibitemOpen
  \bibfield  {author} {\bibinfo {author} {\bibfnamefont {T.}~\bibnamefont
  {Lazeyras}}, \bibinfo {author} {\bibfnamefont {A.}~\bibnamefont {Barreira}},
  \bibinfo {author} {\bibfnamefont {F.}~\bibnamefont {Schmidt}}, \ and\
  \bibinfo {author} {\bibfnamefont {V.}~\bibnamefont {Desjacques}},\ }\href
  {\doibase 10.1088/1475-7516/2023/01/023} {\bibfield  {journal} {\bibinfo
  {journal} {JCAP}\ }\textbf {\bibinfo {volume} {01}},\ \bibinfo {pages} {023}
  (\bibinfo {year} {2023})},\ \Eprint {http://arxiv.org/abs/2209.07251}
  {arXiv:2209.07251 [astro-ph.CO]} \BibitemShut {NoStop}%
\bibitem [{\citenamefont {Barreira}(2022)}]{Barreira:2022sey}%
  \BibitemOpen
  \bibfield  {author} {\bibinfo {author} {\bibfnamefont {A.}~\bibnamefont
  {Barreira}},\ }\href {\doibase 10.1088/1475-7516/2022/11/013} {\bibfield
  {journal} {\bibinfo  {journal} {JCAP}\ }\textbf {\bibinfo {volume} {11}},\
  \bibinfo {pages} {013} (\bibinfo {year} {2022})},\ \Eprint
  {http://arxiv.org/abs/2205.05673} {arXiv:2205.05673 [astro-ph.CO]}
  \BibitemShut {NoStop}%
\bibitem [{\citenamefont {Barreira}\ and\ \citenamefont
  {Krause}(2023)}]{Barreira:2023rxn}%
  \BibitemOpen
  \bibfield  {author} {\bibinfo {author} {\bibfnamefont {A.}~\bibnamefont
  {Barreira}}\ and\ \bibinfo {author} {\bibfnamefont {E.}~\bibnamefont
  {Krause}},\ }\href {\doibase 10.1088/1475-7516/2023/10/044} {\bibfield
  {journal} {\bibinfo  {journal} {JCAP}\ }\textbf {\bibinfo {volume} {10}},\
  \bibinfo {pages} {044} (\bibinfo {year} {2023})},\ \Eprint
  {http://arxiv.org/abs/2302.09066} {arXiv:2302.09066 [astro-ph.CO]}
  \BibitemShut {NoStop}%
\bibitem [{\citenamefont {Green}\ \emph {et~al.}(2023)\citenamefont {Green},
  \citenamefont {Guo}, \citenamefont {Han},\ and\ \citenamefont
  {Wallisch}}]{Green:2023uyz}%
  \BibitemOpen
  \bibfield  {author} {\bibinfo {author} {\bibfnamefont {D.}~\bibnamefont
  {Green}}, \bibinfo {author} {\bibfnamefont {Y.}~\bibnamefont {Guo}}, \bibinfo
  {author} {\bibfnamefont {J.}~\bibnamefont {Han}}, \ and\ \bibinfo {author}
  {\bibfnamefont {B.}~\bibnamefont {Wallisch}},\ }\href@noop {} {\  (\bibinfo
  {year} {2023})},\ \Eprint {http://arxiv.org/abs/2311.04882} {arXiv:2311.04882
  [astro-ph.CO]} \BibitemShut {NoStop}%
\bibitem [{\citenamefont {Saito}\ \emph {et~al.}(2014)\citenamefont {Saito},
  \citenamefont {Baldauf}, \citenamefont {Vlah}, \citenamefont {Seljak},
  \citenamefont {Okumura},\ and\ \citenamefont {McDonald}}]{Saito:2014qha}%
  \BibitemOpen
  \bibfield  {author} {\bibinfo {author} {\bibfnamefont {S.}~\bibnamefont
  {Saito}}, \bibinfo {author} {\bibfnamefont {T.}~\bibnamefont {Baldauf}},
  \bibinfo {author} {\bibfnamefont {Z.}~\bibnamefont {Vlah}}, \bibinfo {author}
  {\bibfnamefont {U.}~\bibnamefont {Seljak}}, \bibinfo {author} {\bibfnamefont
  {T.}~\bibnamefont {Okumura}}, \ and\ \bibinfo {author} {\bibfnamefont
  {P.}~\bibnamefont {McDonald}},\ }\href {\doibase 10.1103/PhysRevD.90.123522}
  {\bibfield  {journal} {\bibinfo  {journal} {Phys. Rev. D}\ }\textbf {\bibinfo
  {volume} {90}},\ \bibinfo {pages} {123522} (\bibinfo {year} {2014})},\
  \Eprint {http://arxiv.org/abs/1405.1447} {arXiv:1405.1447 [astro-ph.CO]}
  \BibitemShut {NoStop}%
\bibitem [{\citenamefont {Schmittfull}\ \emph {et~al.}(2015)\citenamefont
  {Schmittfull}, \citenamefont {Baldauf},\ and\ \citenamefont
  {Seljak}}]{Schmittfull:2014tca}%
  \BibitemOpen
  \bibfield  {author} {\bibinfo {author} {\bibfnamefont {M.}~\bibnamefont
  {Schmittfull}}, \bibinfo {author} {\bibfnamefont {T.}~\bibnamefont
  {Baldauf}}, \ and\ \bibinfo {author} {\bibfnamefont {U.}~\bibnamefont
  {Seljak}},\ }\href {\doibase 10.1103/PhysRevD.91.043530} {\bibfield
  {journal} {\bibinfo  {journal} {Phys. Rev. D}\ }\textbf {\bibinfo {volume}
  {91}},\ \bibinfo {pages} {043530} (\bibinfo {year} {2015})},\ \Eprint
  {http://arxiv.org/abs/1411.6595} {arXiv:1411.6595 [astro-ph.CO]} \BibitemShut
  {NoStop}%
\bibitem [{\citenamefont {Lazeyras}\ and\ \citenamefont
  {Schmidt}(2018)}]{Lazeyras:2017hxw}%
  \BibitemOpen
  \bibfield  {author} {\bibinfo {author} {\bibfnamefont {T.}~\bibnamefont
  {Lazeyras}}\ and\ \bibinfo {author} {\bibfnamefont {F.}~\bibnamefont
  {Schmidt}},\ }\href {\doibase 10.1088/1475-7516/2018/09/008} {\bibfield
  {journal} {\bibinfo  {journal} {JCAP}\ }\textbf {\bibinfo {volume} {1809}},\
  \bibinfo {pages} {008} (\bibinfo {year} {2018})},\ \Eprint
  {http://arxiv.org/abs/1712.07531} {arXiv:1712.07531 [astro-ph.CO]}
  \BibitemShut {NoStop}%
\bibitem [{\citenamefont {Schmidt}\ \emph {et~al.}(2019)\citenamefont
  {Schmidt}, \citenamefont {Elsner}, \citenamefont {Jasche}, \citenamefont
  {Nguyen},\ and\ \citenamefont {Lavaux}}]{Schmidt:2018bkr}%
  \BibitemOpen
  \bibfield  {author} {\bibinfo {author} {\bibfnamefont {F.}~\bibnamefont
  {Schmidt}}, \bibinfo {author} {\bibfnamefont {F.}~\bibnamefont {Elsner}},
  \bibinfo {author} {\bibfnamefont {J.}~\bibnamefont {Jasche}}, \bibinfo
  {author} {\bibfnamefont {N.~M.}\ \bibnamefont {Nguyen}}, \ and\ \bibinfo
  {author} {\bibfnamefont {G.}~\bibnamefont {Lavaux}},\ }\href {\doibase
  10.1088/1475-7516/2019/01/042} {\bibfield  {journal} {\bibinfo  {journal}
  {JCAP}\ }\textbf {\bibinfo {volume} {01}},\ \bibinfo {pages} {042} (\bibinfo
  {year} {2019})},\ \Eprint {http://arxiv.org/abs/1808.02002} {arXiv:1808.02002
  [astro-ph.CO]} \BibitemShut {NoStop}%
\bibitem [{\citenamefont {Schmittfull}\ \emph {et~al.}(2019)\citenamefont
  {Schmittfull}, \citenamefont {Simonović}, \citenamefont {Assassi},\ and\
  \citenamefont {Zaldarriaga}}]{Schmittfull:2018yuk}%
  \BibitemOpen
  \bibfield  {author} {\bibinfo {author} {\bibfnamefont {M.}~\bibnamefont
  {Schmittfull}}, \bibinfo {author} {\bibfnamefont {M.}~\bibnamefont
  {Simonović}}, \bibinfo {author} {\bibfnamefont {V.}~\bibnamefont {Assassi}},
  \ and\ \bibinfo {author} {\bibfnamefont {M.}~\bibnamefont {Zaldarriaga}},\
  }\href {\doibase 10.1103/PhysRevD.100.043514} {\bibfield  {journal} {\bibinfo
   {journal} {Phys.\ Rev.\ D}\ }\textbf {\bibinfo {volume} {100}},\ \bibinfo
  {pages} {043514} (\bibinfo {year} {2019})},\ \Eprint
  {http://arxiv.org/abs/1811.10640} {arXiv:1811.10640 [astro-ph.CO]}
  \BibitemShut {NoStop}%
\bibitem [{\citenamefont {Elsner}\ \emph {et~al.}(2020)\citenamefont {Elsner},
  \citenamefont {Schmidt}, \citenamefont {Jasche}, \citenamefont {Lavaux},\
  and\ \citenamefont {Nguyen}}]{Elsner:2019rql}%
  \BibitemOpen
  \bibfield  {author} {\bibinfo {author} {\bibfnamefont {F.}~\bibnamefont
  {Elsner}}, \bibinfo {author} {\bibfnamefont {F.}~\bibnamefont {Schmidt}},
  \bibinfo {author} {\bibfnamefont {J.}~\bibnamefont {Jasche}}, \bibinfo
  {author} {\bibfnamefont {G.}~\bibnamefont {Lavaux}}, \ and\ \bibinfo {author}
  {\bibfnamefont {N.-M.}\ \bibnamefont {Nguyen}},\ }\href {\doibase
  10.1088/1475-7516/2020/01/029} {\bibfield  {journal} {\bibinfo  {journal}
  {JCAP}\ }\textbf {\bibinfo {volume} {01}},\ \bibinfo {pages} {029} (\bibinfo
  {year} {2020})},\ \Eprint {http://arxiv.org/abs/1906.07143} {arXiv:1906.07143
  [astro-ph.CO]} \BibitemShut {NoStop}%
\bibitem [{\citenamefont {Cabass}\ and\ \citenamefont
  {Schmidt}(2020)}]{Cabass:2019lqx}%
  \BibitemOpen
  \bibfield  {author} {\bibinfo {author} {\bibfnamefont {G.}~\bibnamefont
  {Cabass}}\ and\ \bibinfo {author} {\bibfnamefont {F.}~\bibnamefont
  {Schmidt}},\ }\href {\doibase 10.1088/1475-7516/2020/04/042} {\bibfield
  {journal} {\bibinfo  {journal} {JCAP}\ }\textbf {\bibinfo {volume} {04}},\
  \bibinfo {pages} {042} (\bibinfo {year} {2020})},\ \Eprint
  {http://arxiv.org/abs/1909.04022} {arXiv:1909.04022 [astro-ph.CO]}
  \BibitemShut {NoStop}%
\bibitem [{\citenamefont {Modi}\ \emph {et~al.}(2020)\citenamefont {Modi},
  \citenamefont {Chen},\ and\ \citenamefont {White}}]{Modi:2019qbt}%
  \BibitemOpen
  \bibfield  {author} {\bibinfo {author} {\bibfnamefont {C.}~\bibnamefont
  {Modi}}, \bibinfo {author} {\bibfnamefont {S.-F.}\ \bibnamefont {Chen}}, \
  and\ \bibinfo {author} {\bibfnamefont {M.}~\bibnamefont {White}},\ }\href
  {\doibase 10.1093/mnras/staa251} {\bibfield  {journal} {\bibinfo  {journal}
  {Mon. Not. Roy. Astron. Soc.}\ }\textbf {\bibinfo {volume} {492}},\ \bibinfo
  {pages} {5754} (\bibinfo {year} {2020})},\ \Eprint
  {http://arxiv.org/abs/1910.07097} {arXiv:1910.07097 [astro-ph.CO]}
  \BibitemShut {NoStop}%
\bibitem [{\citenamefont {Schmidt}(2021)}]{Schmidt:2020tao}%
  \BibitemOpen
  \bibfield  {author} {\bibinfo {author} {\bibfnamefont {F.}~\bibnamefont
  {Schmidt}},\ }\href {\doibase 10.1088/1475-7516/2021/04/032} {\bibfield
  {journal} {\bibinfo  {journal} {JCAP}\ }\textbf {\bibinfo {volume} {04}},\
  \bibinfo {pages} {032} (\bibinfo {year} {2021})},\ \Eprint
  {http://arxiv.org/abs/2009.14176} {arXiv:2009.14176 [astro-ph.CO]}
  \BibitemShut {NoStop}%
\bibitem [{\citenamefont {Schmidt}\ \emph {et~al.}(2020)\citenamefont
  {Schmidt}, \citenamefont {Cabass}, \citenamefont {Jasche},\ and\
  \citenamefont {Lavaux}}]{Schmidt:2020viy}%
  \BibitemOpen
  \bibfield  {author} {\bibinfo {author} {\bibfnamefont {F.}~\bibnamefont
  {Schmidt}}, \bibinfo {author} {\bibfnamefont {G.}~\bibnamefont {Cabass}},
  \bibinfo {author} {\bibfnamefont {J.}~\bibnamefont {Jasche}}, \ and\ \bibinfo
  {author} {\bibfnamefont {G.}~\bibnamefont {Lavaux}},\ }\href {\doibase
  10.1088/1475-7516/2020/11/008} {\bibfield  {journal} {\bibinfo  {journal}
  {JCAP}\ }\textbf {\bibinfo {volume} {11}},\ \bibinfo {pages} {008} (\bibinfo
  {year} {2020})},\ \Eprint {http://arxiv.org/abs/2004.06707} {arXiv:2004.06707
  [astro-ph.CO]} \BibitemShut {NoStop}%
\bibitem [{\citenamefont {Schmittfull}\ \emph {et~al.}(2021)\citenamefont
  {Schmittfull}, \citenamefont {Simonovi\'c}, \citenamefont {Ivanov},
  \citenamefont {Philcox},\ and\ \citenamefont
  {Zaldarriaga}}]{Schmittfull:2020trd}%
  \BibitemOpen
  \bibfield  {author} {\bibinfo {author} {\bibfnamefont {M.}~\bibnamefont
  {Schmittfull}}, \bibinfo {author} {\bibfnamefont {M.}~\bibnamefont
  {Simonovi\'c}}, \bibinfo {author} {\bibfnamefont {M.~M.}\ \bibnamefont
  {Ivanov}}, \bibinfo {author} {\bibfnamefont {O.~H.~E.}\ \bibnamefont
  {Philcox}}, \ and\ \bibinfo {author} {\bibfnamefont {M.}~\bibnamefont
  {Zaldarriaga}},\ }\href {\doibase 10.1088/1475-7516/2021/05/059} {\bibfield
  {journal} {\bibinfo  {journal} {JCAP}\ }\textbf {\bibinfo {volume} {05}},\
  \bibinfo {pages} {059} (\bibinfo {year} {2021})},\ \Eprint
  {http://arxiv.org/abs/2012.03334} {arXiv:2012.03334 [astro-ph.CO]}
  \BibitemShut {NoStop}%
\bibitem [{\citenamefont {Lazeyras}\ \emph {et~al.}(2021)\citenamefont
  {Lazeyras}, \citenamefont {Barreira},\ and\ \citenamefont
  {Schmidt}}]{Lazeyras:2021dar}%
  \BibitemOpen
  \bibfield  {author} {\bibinfo {author} {\bibfnamefont {T.}~\bibnamefont
  {Lazeyras}}, \bibinfo {author} {\bibfnamefont {A.}~\bibnamefont {Barreira}},
  \ and\ \bibinfo {author} {\bibfnamefont {F.}~\bibnamefont {Schmidt}},\ }\href
  {\doibase 10.1088/1475-7516/2021/10/063} {\bibfield  {journal} {\bibinfo
  {journal} {JCAP}\ }\textbf {\bibinfo {volume} {10}},\ \bibinfo {pages} {063}
  (\bibinfo {year} {2021})},\ \Eprint {http://arxiv.org/abs/2106.14713}
  {arXiv:2106.14713 [astro-ph.CO]} \BibitemShut {NoStop}%
\bibitem [{\citenamefont {Stadler}\ \emph {et~al.}(2023)\citenamefont
  {Stadler}, \citenamefont {Schmidt},\ and\ \citenamefont
  {Reinecke}}]{Stadler:2023hea}%
  \BibitemOpen
  \bibfield  {author} {\bibinfo {author} {\bibfnamefont {J.}~\bibnamefont
  {Stadler}}, \bibinfo {author} {\bibfnamefont {F.}~\bibnamefont {Schmidt}}, \
  and\ \bibinfo {author} {\bibfnamefont {M.}~\bibnamefont {Reinecke}},\ }\href
  {\doibase 10.1088/1475-7516/2023/10/069} {\bibfield  {journal} {\bibinfo
  {journal} {JCAP}\ }\textbf {\bibinfo {volume} {10}},\ \bibinfo {pages} {069}
  (\bibinfo {year} {2023})},\ \Eprint {http://arxiv.org/abs/2303.09876}
  {arXiv:2303.09876 [astro-ph.CO]} \BibitemShut {NoStop}%
\bibitem [{\citenamefont {Rubira}\ and\ \citenamefont
  {Schmidt}(2024)}]{Rubira:2023vzw}%
  \BibitemOpen
  \bibfield  {author} {\bibinfo {author} {\bibfnamefont {H.}~\bibnamefont
  {Rubira}}\ and\ \bibinfo {author} {\bibfnamefont {F.}~\bibnamefont
  {Schmidt}},\ }\href {\doibase 10.1088/1475-7516/2024/01/031} {\bibfield
  {journal} {\bibinfo  {journal} {JCAP}\ }\textbf {\bibinfo {volume} {01}},\
  \bibinfo {pages} {031} (\bibinfo {year} {2024})},\ \Eprint
  {http://arxiv.org/abs/2307.15031} {arXiv:2307.15031 [astro-ph.CO]}
  \BibitemShut {NoStop}%
\bibitem [{\citenamefont {Nguyen}\ \emph {et~al.}(2024)\citenamefont {Nguyen},
  \citenamefont {Schmidt}, \citenamefont {Tucci}, \citenamefont {Reinecke},\
  and\ \citenamefont {Kosti\'c}}]{Nguyen:2024yth}%
  \BibitemOpen
  \bibfield  {author} {\bibinfo {author} {\bibfnamefont {N.-M.}\ \bibnamefont
  {Nguyen}}, \bibinfo {author} {\bibfnamefont {F.}~\bibnamefont {Schmidt}},
  \bibinfo {author} {\bibfnamefont {B.}~\bibnamefont {Tucci}}, \bibinfo
  {author} {\bibfnamefont {M.}~\bibnamefont {Reinecke}}, \ and\ \bibinfo
  {author} {\bibfnamefont {A.}~\bibnamefont {Kosti\'c}},\ }\href@noop {} {\
  (\bibinfo {year} {2024})},\ \Eprint {http://arxiv.org/abs/2403.03220}
  {arXiv:2403.03220 [astro-ph.CO]} \BibitemShut {NoStop}%
\bibitem [{\citenamefont {Maksimova}\ \emph {et~al.}(2021)\citenamefont
  {Maksimova}, \citenamefont {Garrison}, \citenamefont {Eisenstein},
  \citenamefont {Hadzhiyska}, \citenamefont {Bose},\ and\ \citenamefont
  {Satterthwaite}}]{Maksimova:2021ynf}%
  \BibitemOpen
  \bibfield  {author} {\bibinfo {author} {\bibfnamefont {N.~A.}\ \bibnamefont
  {Maksimova}}, \bibinfo {author} {\bibfnamefont {L.~H.}\ \bibnamefont
  {Garrison}}, \bibinfo {author} {\bibfnamefont {D.~J.}\ \bibnamefont
  {Eisenstein}}, \bibinfo {author} {\bibfnamefont {B.}~\bibnamefont
  {Hadzhiyska}}, \bibinfo {author} {\bibfnamefont {S.}~\bibnamefont {Bose}}, \
  and\ \bibinfo {author} {\bibfnamefont {T.~P.}\ \bibnamefont
  {Satterthwaite}},\ }\href {\doibase 10.1093/mnras/stab2484} {\bibfield
  {journal} {\bibinfo  {journal} {Mon. Not. Roy. Astron. Soc.}\ }\textbf
  {\bibinfo {volume} {508}},\ \bibinfo {pages} {4017} (\bibinfo {year}
  {2021})},\ \Eprint {http://arxiv.org/abs/2110.11398} {arXiv:2110.11398
  [astro-ph.CO]} \BibitemShut {NoStop}%
\bibitem [{\citenamefont {Aghanim}\ \emph {et~al.}(2018)\citenamefont {Aghanim}
  \emph {et~al.}}]{Aghanim:2018eyx}%
  \BibitemOpen
  \bibfield  {author} {\bibinfo {author} {\bibfnamefont {N.}~\bibnamefont
  {Aghanim}} \emph {et~al.} (\bibinfo {collaboration} {Planck}),\ }\href@noop
  {} {\  (\bibinfo {year} {2018})},\ \Eprint {http://arxiv.org/abs/1807.06209}
  {arXiv:1807.06209 [astro-ph.CO]} \BibitemShut {NoStop}%
\bibitem [{\citenamefont {Yuan}\ \emph {et~al.}(2022)\citenamefont {Yuan},
  \citenamefont {Garrison}, \citenamefont {Hadzhiyska}, \citenamefont {Bose},\
  and\ \citenamefont {Eisenstein}}]{Yuan:2021izi}%
  \BibitemOpen
  \bibfield  {author} {\bibinfo {author} {\bibfnamefont {S.}~\bibnamefont
  {Yuan}}, \bibinfo {author} {\bibfnamefont {L.~H.}\ \bibnamefont {Garrison}},
  \bibinfo {author} {\bibfnamefont {B.}~\bibnamefont {Hadzhiyska}}, \bibinfo
  {author} {\bibfnamefont {S.}~\bibnamefont {Bose}}, \ and\ \bibinfo {author}
  {\bibfnamefont {D.~J.}\ \bibnamefont {Eisenstein}},\ }\href {\doibase
  10.1093/mnras/stab3355} {\bibfield  {journal} {\bibinfo  {journal} {Mon. Not.
  Roy. Astron. Soc.}\ }\textbf {\bibinfo {volume} {510}},\ \bibinfo {pages}
  {3301} (\bibinfo {year} {2022})},\ \Eprint {http://arxiv.org/abs/2110.11412}
  {arXiv:2110.11412 [astro-ph.CO]} \BibitemShut {NoStop}%
\bibitem [{\citenamefont {{Hadzhiyska}}\ \emph {et~al.}(2020)\citenamefont
  {{Hadzhiyska}}, \citenamefont {{Bose}}, \citenamefont {{Eisenstein}},
  \citenamefont {{Hernquist}},\ and\ \citenamefont
  {{Spergel}}}]{2020MNRAS.493.5506H}%
  \BibitemOpen
  \bibfield  {author} {\bibinfo {author} {\bibfnamefont {B.}~\bibnamefont
  {{Hadzhiyska}}}, \bibinfo {author} {\bibfnamefont {S.}~\bibnamefont
  {{Bose}}}, \bibinfo {author} {\bibfnamefont {D.}~\bibnamefont
  {{Eisenstein}}}, \bibinfo {author} {\bibfnamefont {L.}~\bibnamefont
  {{Hernquist}}}, \ and\ \bibinfo {author} {\bibfnamefont {D.~N.}\ \bibnamefont
  {{Spergel}}},\ }\href {\doibase 10.1093/mnras/staa623} {\bibfield  {journal}
  {\bibinfo  {journal} {\mnras}\ }\textbf {\bibinfo {volume} {493}},\ \bibinfo
  {pages} {5506} (\bibinfo {year} {2020})},\ \Eprint
  {http://arxiv.org/abs/1911.02610} {arXiv:1911.02610 [astro-ph.CO]}
  \BibitemShut {NoStop}%
\bibitem [{\citenamefont {{Xu}}\ \emph {et~al.}(2021)\citenamefont {{Xu}},
  \citenamefont {{Zehavi}},\ and\ \citenamefont
  {{Contreras}}}]{2021MNRAS.502.3242X}%
  \BibitemOpen
  \bibfield  {author} {\bibinfo {author} {\bibfnamefont {X.}~\bibnamefont
  {{Xu}}}, \bibinfo {author} {\bibfnamefont {I.}~\bibnamefont {{Zehavi}}}, \
  and\ \bibinfo {author} {\bibfnamefont {S.}~\bibnamefont {{Contreras}}},\
  }\href {\doibase 10.1093/mnras/stab100} {\bibfield  {journal} {\bibinfo
  {journal} {\mnras}\ }\textbf {\bibinfo {volume} {502}},\ \bibinfo {pages}
  {3242} (\bibinfo {year} {2021})},\ \Eprint {http://arxiv.org/abs/2007.05545}
  {arXiv:2007.05545 [astro-ph.GA]} \BibitemShut {NoStop}%
\bibitem [{\citenamefont {Assassi}\ \emph {et~al.}(2014)\citenamefont
  {Assassi}, \citenamefont {Baumann}, \citenamefont {Green},\ and\
  \citenamefont {Zaldarriaga}}]{Assassi:2014fva}%
  \BibitemOpen
  \bibfield  {author} {\bibinfo {author} {\bibfnamefont {V.}~\bibnamefont
  {Assassi}}, \bibinfo {author} {\bibfnamefont {D.}~\bibnamefont {Baumann}},
  \bibinfo {author} {\bibfnamefont {D.}~\bibnamefont {Green}}, \ and\ \bibinfo
  {author} {\bibfnamefont {M.}~\bibnamefont {Zaldarriaga}},\ }\href {\doibase
  10.1088/1475-7516/2014/08/056} {\bibfield  {journal} {\bibinfo  {journal}
  {JCAP}\ }\textbf {\bibinfo {volume} {1408}},\ \bibinfo {pages} {056}
  (\bibinfo {year} {2014})},\ \Eprint {http://arxiv.org/abs/1402.5916}
  {arXiv:1402.5916 [astro-ph.CO]} \BibitemShut {NoStop}%
\bibitem [{\citenamefont {Bernardeau}\ \emph {et~al.}(2002)\citenamefont
  {Bernardeau}, \citenamefont {Colombi}, \citenamefont {Gaztanaga},\ and\
  \citenamefont {Scoccimarro}}]{Bernardeau:2001qr}%
  \BibitemOpen
  \bibfield  {author} {\bibinfo {author} {\bibfnamefont {F.}~\bibnamefont
  {Bernardeau}}, \bibinfo {author} {\bibfnamefont {S.}~\bibnamefont {Colombi}},
  \bibinfo {author} {\bibfnamefont {E.}~\bibnamefont {Gaztanaga}}, \ and\
  \bibinfo {author} {\bibfnamefont {R.}~\bibnamefont {Scoccimarro}},\ }\href
  {\doibase 10.1016/S0370-1573(02)00135-7} {\bibfield  {journal} {\bibinfo
  {journal} {Phys. Rept.}\ }\textbf {\bibinfo {volume} {367}},\ \bibinfo
  {pages} {1} (\bibinfo {year} {2002})},\ \Eprint
  {http://arxiv.org/abs/astro-ph/0112551} {arXiv:astro-ph/0112551 [astro-ph]}
  \BibitemShut {NoStop}%
\bibitem [{\citenamefont {Zeldovich}(1970)}]{Zeldovich:1969sb}%
  \BibitemOpen
  \bibfield  {author} {\bibinfo {author} {\bibfnamefont {Y.~B.}\ \bibnamefont
  {Zeldovich}},\ }\href@noop {} {\bibfield  {journal} {\bibinfo  {journal}
  {Astron. Astrophys.}\ }\textbf {\bibinfo {volume} {5}},\ \bibinfo {pages}
  {84} (\bibinfo {year} {1970})}\BibitemShut {NoStop}%
\bibitem [{\citenamefont {Obuljen}\ \emph {et~al.}(2023)\citenamefont
  {Obuljen}, \citenamefont {Simonovi\'c}, \citenamefont {Schneider},\ and\
  \citenamefont {Feldmann}}]{Obuljen:2022cjo}%
  \BibitemOpen
  \bibfield  {author} {\bibinfo {author} {\bibfnamefont {A.}~\bibnamefont
  {Obuljen}}, \bibinfo {author} {\bibfnamefont {M.}~\bibnamefont
  {Simonovi\'c}}, \bibinfo {author} {\bibfnamefont {A.}~\bibnamefont
  {Schneider}}, \ and\ \bibinfo {author} {\bibfnamefont {R.}~\bibnamefont
  {Feldmann}},\ }\href {\doibase 10.1103/PhysRevD.108.083528} {\bibfield
  {journal} {\bibinfo  {journal} {Phys. Rev. D}\ }\textbf {\bibinfo {volume}
  {108}},\ \bibinfo {pages} {083528} (\bibinfo {year} {2023})},\ \Eprint
  {http://arxiv.org/abs/2207.12398} {arXiv:2207.12398 [astro-ph.CO]}
  \BibitemShut {NoStop}%
\bibitem [{\citenamefont {Senatore}\ and\ \citenamefont
  {Zaldarriaga}(2015)}]{Senatore:2014via}%
  \BibitemOpen
  \bibfield  {author} {\bibinfo {author} {\bibfnamefont {L.}~\bibnamefont
  {Senatore}}\ and\ \bibinfo {author} {\bibfnamefont {M.}~\bibnamefont
  {Zaldarriaga}},\ }\href {\doibase 10.1088/1475-7516/2015/02/013} {\bibfield
  {journal} {\bibinfo  {journal} {JCAP}\ }\textbf {\bibinfo {volume} {1502}},\
  \bibinfo {pages} {013} (\bibinfo {year} {2015})},\ \Eprint
  {http://arxiv.org/abs/1404.5954} {arXiv:1404.5954 [astro-ph.CO]} \BibitemShut
  {NoStop}%
\bibitem [{\citenamefont {Baldauf}\ \emph {et~al.}(2015)\citenamefont
  {Baldauf}, \citenamefont {Mirbabayi}, \citenamefont {Simonović},\ and\
  \citenamefont {Zaldarriaga}}]{Baldauf:2015xfa}%
  \BibitemOpen
  \bibfield  {author} {\bibinfo {author} {\bibfnamefont {T.}~\bibnamefont
  {Baldauf}}, \bibinfo {author} {\bibfnamefont {M.}~\bibnamefont {Mirbabayi}},
  \bibinfo {author} {\bibfnamefont {M.}~\bibnamefont {Simonović}}, \ and\
  \bibinfo {author} {\bibfnamefont {M.}~\bibnamefont {Zaldarriaga}},\ }\href
  {\doibase 10.1103/PhysRevD.92.043514} {\bibfield  {journal} {\bibinfo
  {journal} {Phys. Rev.}\ }\textbf {\bibinfo {volume} {D92}},\ \bibinfo {pages}
  {043514} (\bibinfo {year} {2015})},\ \Eprint
  {http://arxiv.org/abs/1504.04366} {arXiv:1504.04366 [astro-ph.CO]}
  \BibitemShut {NoStop}%
\bibitem [{\citenamefont {Blas}\ \emph
  {et~al.}(2016{\natexlab{a}})\citenamefont {Blas}, \citenamefont {Garny},
  \citenamefont {Ivanov},\ and\ \citenamefont {Sibiryakov}}]{Blas:2015qsi}%
  \BibitemOpen
  \bibfield  {author} {\bibinfo {author} {\bibfnamefont {D.}~\bibnamefont
  {Blas}}, \bibinfo {author} {\bibfnamefont {M.}~\bibnamefont {Garny}},
  \bibinfo {author} {\bibfnamefont {M.~M.}\ \bibnamefont {Ivanov}}, \ and\
  \bibinfo {author} {\bibfnamefont {S.}~\bibnamefont {Sibiryakov}},\ }\href
  {\doibase 10.1088/1475-7516/2016/07/052} {\bibfield  {journal} {\bibinfo
  {journal} {JCAP}\ }\textbf {\bibinfo {volume} {1607}},\ \bibinfo {pages}
  {052} (\bibinfo {year} {2016}{\natexlab{a}})},\ \Eprint
  {http://arxiv.org/abs/1512.05807} {arXiv:1512.05807 [astro-ph.CO]}
  \BibitemShut {NoStop}%
\bibitem [{\citenamefont {Blas}\ \emph
  {et~al.}(2016{\natexlab{b}})\citenamefont {Blas}, \citenamefont {Garny},
  \citenamefont {Ivanov},\ and\ \citenamefont {Sibiryakov}}]{Blas:2016sfa}%
  \BibitemOpen
  \bibfield  {author} {\bibinfo {author} {\bibfnamefont {D.}~\bibnamefont
  {Blas}}, \bibinfo {author} {\bibfnamefont {M.}~\bibnamefont {Garny}},
  \bibinfo {author} {\bibfnamefont {M.~M.}\ \bibnamefont {Ivanov}}, \ and\
  \bibinfo {author} {\bibfnamefont {S.}~\bibnamefont {Sibiryakov}},\ }\href
  {\doibase 10.1088/1475-7516/2016/07/028} {\bibfield  {journal} {\bibinfo
  {journal} {JCAP}\ }\textbf {\bibinfo {volume} {1607}},\ \bibinfo {pages}
  {028} (\bibinfo {year} {2016}{\natexlab{b}})},\ \Eprint
  {http://arxiv.org/abs/1605.02149} {arXiv:1605.02149 [astro-ph.CO]}
  \BibitemShut {NoStop}%
\bibitem [{\citenamefont {Ivanov}\ and\ \citenamefont
  {Sibiryakov}(2018)}]{Ivanov:2018gjr}%
  \BibitemOpen
  \bibfield  {author} {\bibinfo {author} {\bibfnamefont {M.~M.}\ \bibnamefont
  {Ivanov}}\ and\ \bibinfo {author} {\bibfnamefont {S.}~\bibnamefont
  {Sibiryakov}},\ }\href {\doibase 10.1088/1475-7516/2018/07/053} {\bibfield
  {journal} {\bibinfo  {journal} {JCAP}\ }\textbf {\bibinfo {volume} {1807}},\
  \bibinfo {pages} {053} (\bibinfo {year} {2018})},\ \Eprint
  {http://arxiv.org/abs/1804.05080} {arXiv:1804.05080 [astro-ph.CO]}
  \BibitemShut {NoStop}%
\bibitem [{\citenamefont {Eggemeier}\ \emph {et~al.}(2018)\citenamefont
  {Eggemeier}, \citenamefont {Scoccimarro},\ and\ \citenamefont
  {Smith}}]{Eggemeier:2018qae}%
  \BibitemOpen
  \bibfield  {author} {\bibinfo {author} {\bibfnamefont {A.}~\bibnamefont
  {Eggemeier}}, \bibinfo {author} {\bibfnamefont {R.}~\bibnamefont
  {Scoccimarro}}, \ and\ \bibinfo {author} {\bibfnamefont {R.~E.}\ \bibnamefont
  {Smith}},\ }\href@noop {} {\  (\bibinfo {year} {2018})},\ \Eprint
  {http://arxiv.org/abs/1812.03208} {arXiv:1812.03208 [astro-ph.CO]}
  \BibitemShut {NoStop}%
\bibitem [{\citenamefont {D'Amico}\ \emph
  {et~al.}(2022{\natexlab{b}})\citenamefont {D'Amico}, \citenamefont {Donath},
  \citenamefont {Lewandowski}, \citenamefont {Senatore},\ and\ \citenamefont
  {Zhang}}]{DAmico:2022ukl}%
  \BibitemOpen
  \bibfield  {author} {\bibinfo {author} {\bibfnamefont {G.}~\bibnamefont
  {D'Amico}}, \bibinfo {author} {\bibfnamefont {Y.}~\bibnamefont {Donath}},
  \bibinfo {author} {\bibfnamefont {M.}~\bibnamefont {Lewandowski}}, \bibinfo
  {author} {\bibfnamefont {L.}~\bibnamefont {Senatore}}, \ and\ \bibinfo
  {author} {\bibfnamefont {P.}~\bibnamefont {Zhang}},\ }\href@noop {} {\
  (\bibinfo {year} {2022}{\natexlab{b}})},\ \Eprint
  {http://arxiv.org/abs/2211.17130} {arXiv:2211.17130 [astro-ph.CO]}
  \BibitemShut {NoStop}%
\bibitem [{\citenamefont {Lazeyras}\ \emph {et~al.}(2016)\citenamefont
  {Lazeyras}, \citenamefont {Wagner}, \citenamefont {Baldauf},\ and\
  \citenamefont {Schmidt}}]{Lazeyras:2015lgp}%
  \BibitemOpen
  \bibfield  {author} {\bibinfo {author} {\bibfnamefont {T.}~\bibnamefont
  {Lazeyras}}, \bibinfo {author} {\bibfnamefont {C.}~\bibnamefont {Wagner}},
  \bibinfo {author} {\bibfnamefont {T.}~\bibnamefont {Baldauf}}, \ and\
  \bibinfo {author} {\bibfnamefont {F.}~\bibnamefont {Schmidt}},\ }\href
  {\doibase 10.1088/1475-7516/2016/02/018} {\bibfield  {journal} {\bibinfo
  {journal} {JCAP}\ }\textbf {\bibinfo {volume} {1602}},\ \bibinfo {pages}
  {018} (\bibinfo {year} {2016})},\ \Eprint {http://arxiv.org/abs/1511.01096}
  {arXiv:1511.01096 [astro-ph.CO]} \BibitemShut {NoStop}%
\bibitem [{\citenamefont {Chudaykin}\ \emph
  {et~al.}(2021{\natexlab{b}})\citenamefont {Chudaykin}, \citenamefont
  {Ivanov},\ and\ \citenamefont {Simonovi\'c}}]{Chudaykin:2020hbf}%
  \BibitemOpen
  \bibfield  {author} {\bibinfo {author} {\bibfnamefont {A.}~\bibnamefont
  {Chudaykin}}, \bibinfo {author} {\bibfnamefont {M.~M.}\ \bibnamefont
  {Ivanov}}, \ and\ \bibinfo {author} {\bibfnamefont {M.}~\bibnamefont
  {Simonovi\'c}},\ }\href {\doibase 10.1103/PhysRevD.103.043525} {\bibfield
  {journal} {\bibinfo  {journal} {Phys. Rev. D}\ }\textbf {\bibinfo {volume}
  {103}},\ \bibinfo {pages} {043525} (\bibinfo {year} {2021}{\natexlab{b}})},\
  \Eprint {http://arxiv.org/abs/2009.10724} {arXiv:2009.10724 [astro-ph.CO]}
  \BibitemShut {NoStop}%
\bibitem [{\citenamefont {Ivanov}\ \emph
  {et~al.}(2022{\natexlab{b}})\citenamefont {Ivanov}, \citenamefont {Philcox},
  \citenamefont {Simonovi\'c}, \citenamefont {Zaldarriaga}, \citenamefont
  {Nischimichi},\ and\ \citenamefont {Takada}}]{Ivanov:2021fbu}%
  \BibitemOpen
  \bibfield  {author} {\bibinfo {author} {\bibfnamefont {M.~M.}\ \bibnamefont
  {Ivanov}}, \bibinfo {author} {\bibfnamefont {O.~H.~E.}\ \bibnamefont
  {Philcox}}, \bibinfo {author} {\bibfnamefont {M.}~\bibnamefont
  {Simonovi\'c}}, \bibinfo {author} {\bibfnamefont {M.}~\bibnamefont
  {Zaldarriaga}}, \bibinfo {author} {\bibfnamefont {T.}~\bibnamefont
  {Nischimichi}}, \ and\ \bibinfo {author} {\bibfnamefont {M.}~\bibnamefont
  {Takada}},\ }\href {\doibase 10.1103/PhysRevD.105.043531} {\bibfield
  {journal} {\bibinfo  {journal} {Phys. Rev. D}\ }\textbf {\bibinfo {volume}
  {105}},\ \bibinfo {pages} {043531} (\bibinfo {year} {2022}{\natexlab{b}})},\
  \Eprint {http://arxiv.org/abs/2110.00006} {arXiv:2110.00006 [astro-ph.CO]}
  \BibitemShut {NoStop}%
\bibitem [{\citenamefont {Rezende}\ and\ \citenamefont
  {Mohamed}(2015)}]{rezende2015variational}%
  \BibitemOpen
  \bibfield  {author} {\bibinfo {author} {\bibfnamefont {D.}~\bibnamefont
  {Rezende}}\ and\ \bibinfo {author} {\bibfnamefont {S.}~\bibnamefont
  {Mohamed}},\ }in\ \href@noop {} {\emph {\bibinfo {booktitle} {International
  conference on machine learning}}}\ (\bibinfo {organization} {PMLR},\ \bibinfo
  {year} {2015})\ pp.\ \bibinfo {pages} {1530--1538}\BibitemShut {NoStop}%
\bibitem [{\citenamefont {Paszke}\ \emph {et~al.}(2019)\citenamefont {Paszke},
  \citenamefont {Gross}, \citenamefont {Massa}, \citenamefont {Lerer},
  \citenamefont {Bradbury}, \citenamefont {Chanan}, \citenamefont {Killeen},
  \citenamefont {Lin}, \citenamefont {Gimelshein}, \citenamefont {Antiga} \emph
  {et~al.}}]{paszke2019pytorch}%
  \BibitemOpen
  \bibfield  {author} {\bibinfo {author} {\bibfnamefont {A.}~\bibnamefont
  {Paszke}}, \bibinfo {author} {\bibfnamefont {S.}~\bibnamefont {Gross}},
  \bibinfo {author} {\bibfnamefont {F.}~\bibnamefont {Massa}}, \bibinfo
  {author} {\bibfnamefont {A.}~\bibnamefont {Lerer}}, \bibinfo {author}
  {\bibfnamefont {J.}~\bibnamefont {Bradbury}}, \bibinfo {author}
  {\bibfnamefont {G.}~\bibnamefont {Chanan}}, \bibinfo {author} {\bibfnamefont
  {T.}~\bibnamefont {Killeen}}, \bibinfo {author} {\bibfnamefont
  {Z.}~\bibnamefont {Lin}}, \bibinfo {author} {\bibfnamefont {N.}~\bibnamefont
  {Gimelshein}}, \bibinfo {author} {\bibfnamefont {L.}~\bibnamefont {Antiga}},
  \emph {et~al.},\ }\href@noop {} {\bibfield  {journal} {\bibinfo  {journal}
  {Advances in neural information processing systems}\ }\textbf {\bibinfo
  {volume} {32}} (\bibinfo {year} {2019})}\BibitemShut {NoStop}%
\bibitem [{\citenamefont {Casas-Miranda}\ \emph {et~al.}(2002)\citenamefont
  {Casas-Miranda}, \citenamefont {Mo}, \citenamefont {Sheth},\ and\
  \citenamefont {Boerner}}]{Casas-Miranda:2001dwz}%
  \BibitemOpen
  \bibfield  {author} {\bibinfo {author} {\bibfnamefont {R.}~\bibnamefont
  {Casas-Miranda}}, \bibinfo {author} {\bibfnamefont {H.~J.}\ \bibnamefont
  {Mo}}, \bibinfo {author} {\bibfnamefont {R.~K.}\ \bibnamefont {Sheth}}, \
  and\ \bibinfo {author} {\bibfnamefont {G.}~\bibnamefont {Boerner}},\ }\href
  {\doibase 10.1046/j.1365-8711.2002.05378.x} {\bibfield  {journal} {\bibinfo
  {journal} {Mon. Not. Roy. Astron. Soc.}\ }\textbf {\bibinfo {volume} {333}},\
  \bibinfo {pages} {730} (\bibinfo {year} {2002})},\ \Eprint
  {http://arxiv.org/abs/astro-ph/0105008} {arXiv:astro-ph/0105008} \BibitemShut
  {NoStop}%
\bibitem [{\citenamefont {Cooray}\ and\ \citenamefont
  {Sheth}(2002)}]{Cooray:2002dia}%
  \BibitemOpen
  \bibfield  {author} {\bibinfo {author} {\bibfnamefont {A.}~\bibnamefont
  {Cooray}}\ and\ \bibinfo {author} {\bibfnamefont {R.~K.}\ \bibnamefont
  {Sheth}},\ }\href {\doibase 10.1016/S0370-1573(02)00276-4} {\bibfield
  {journal} {\bibinfo  {journal} {Phys. Rept.}\ }\textbf {\bibinfo {volume}
  {372}},\ \bibinfo {pages} {1} (\bibinfo {year} {2002})},\ \Eprint
  {http://arxiv.org/abs/astro-ph/0206508} {arXiv:astro-ph/0206508} \BibitemShut
  {NoStop}%
\bibitem [{\citenamefont {Baldauf}\ \emph {et~al.}(2013)\citenamefont
  {Baldauf}, \citenamefont {Seljak}, \citenamefont {Smith}, \citenamefont
  {Hamaus},\ and\ \citenamefont {Desjacques}}]{Baldauf:2013hka}%
  \BibitemOpen
  \bibfield  {author} {\bibinfo {author} {\bibfnamefont {T.}~\bibnamefont
  {Baldauf}}, \bibinfo {author} {\bibfnamefont {U.}~\bibnamefont {Seljak}},
  \bibinfo {author} {\bibfnamefont {R.~E.}\ \bibnamefont {Smith}}, \bibinfo
  {author} {\bibfnamefont {N.}~\bibnamefont {Hamaus}}, \ and\ \bibinfo {author}
  {\bibfnamefont {V.}~\bibnamefont {Desjacques}},\ }\href {\doibase
  10.1103/PhysRevD.88.083507} {\bibfield  {journal} {\bibinfo  {journal} {Phys.
  Rev. D}\ }\textbf {\bibinfo {volume} {88}},\ \bibinfo {pages} {083507}
  (\bibinfo {year} {2013})},\ \Eprint {http://arxiv.org/abs/1305.2917}
  {arXiv:1305.2917 [astro-ph.CO]} \BibitemShut {NoStop}%
\bibitem [{\citenamefont {Baldauf}\ \emph {et~al.}(2016)\citenamefont
  {Baldauf}, \citenamefont {Codis}, \citenamefont {Desjacques},\ and\
  \citenamefont {Pichon}}]{Baldauf:2015fbu}%
  \BibitemOpen
  \bibfield  {author} {\bibinfo {author} {\bibfnamefont {T.}~\bibnamefont
  {Baldauf}}, \bibinfo {author} {\bibfnamefont {S.}~\bibnamefont {Codis}},
  \bibinfo {author} {\bibfnamefont {V.}~\bibnamefont {Desjacques}}, \ and\
  \bibinfo {author} {\bibfnamefont {C.}~\bibnamefont {Pichon}},\ }\href
  {\doibase 10.1093/mnras/stv2973} {\bibfield  {journal} {\bibinfo  {journal}
  {Mon. Not. Roy. Astron. Soc.}\ }\textbf {\bibinfo {volume} {456}},\ \bibinfo
  {pages} {3985} (\bibinfo {year} {2016})},\ \Eprint
  {http://arxiv.org/abs/1510.09204} {arXiv:1510.09204 [astro-ph.CO]}
  \BibitemShut {NoStop}%
\bibitem [{\citenamefont {Hearin}\ \emph {et~al.}(2016)\citenamefont {Hearin},
  \citenamefont {Zentner}, \citenamefont {van~den Bosch}, \citenamefont
  {Campbell},\ and\ \citenamefont {Tollerud}}]{Hearin:2015jnf}%
  \BibitemOpen
  \bibfield  {author} {\bibinfo {author} {\bibfnamefont {A.~P.}\ \bibnamefont
  {Hearin}}, \bibinfo {author} {\bibfnamefont {A.~R.}\ \bibnamefont {Zentner}},
  \bibinfo {author} {\bibfnamefont {F.~C.}\ \bibnamefont {van~den Bosch}},
  \bibinfo {author} {\bibfnamefont {D.}~\bibnamefont {Campbell}}, \ and\
  \bibinfo {author} {\bibfnamefont {E.}~\bibnamefont {Tollerud}},\ }\href
  {\doibase 10.1093/mnras/stw840} {\bibfield  {journal} {\bibinfo  {journal}
  {Mon. Not. Roy. Astron. Soc.}\ }\textbf {\bibinfo {volume} {460}},\ \bibinfo
  {pages} {2552} (\bibinfo {year} {2016})},\ \Eprint
  {http://arxiv.org/abs/1512.03050} {arXiv:1512.03050 [astro-ph.CO]}
  \BibitemShut {NoStop}%
\bibitem [{\citenamefont {Akrami}\ \emph {et~al.}(2020)\citenamefont {Akrami}
  \emph {et~al.}}]{Planck:2019kim}%
  \BibitemOpen
  \bibfield  {author} {\bibinfo {author} {\bibfnamefont {Y.}~\bibnamefont
  {Akrami}} \emph {et~al.} (\bibinfo {collaboration} {Planck}),\ }\href
  {\doibase 10.1051/0004-6361/201935891} {\bibfield  {journal} {\bibinfo
  {journal} {Astron. Astrophys.}\ }\textbf {\bibinfo {volume} {641}},\ \bibinfo
  {pages} {A9} (\bibinfo {year} {2020})},\ \Eprint
  {http://arxiv.org/abs/1905.05697} {arXiv:1905.05697 [astro-ph.CO]}
  \BibitemShut {NoStop}%
\bibitem [{\citenamefont {Alam}\ \emph
  {et~al.}(2017{\natexlab{b}})\citenamefont {Alam} \emph
  {et~al.}}]{BOSS:2016wmc}%
  \BibitemOpen
  \bibfield  {author} {\bibinfo {author} {\bibfnamefont {S.}~\bibnamefont
  {Alam}} \emph {et~al.} (\bibinfo {collaboration} {BOSS}),\ }\href {\doibase
  10.1093/mnras/stx721} {\bibfield  {journal} {\bibinfo  {journal} {Mon. Not.
  Roy. Astron. Soc.}\ }\textbf {\bibinfo {volume} {470}},\ \bibinfo {pages}
  {2617} (\bibinfo {year} {2017}{\natexlab{b}})},\ \Eprint
  {http://arxiv.org/abs/1607.03155} {arXiv:1607.03155 [astro-ph.CO]}
  \BibitemShut {NoStop}%
\bibitem [{\citenamefont {Philcox}\ \emph {et~al.}(2020)\citenamefont
  {Philcox}, \citenamefont {Ivanov}, \citenamefont {Simonovi\'c},\ and\
  \citenamefont {Zaldarriaga}}]{Philcox:2020vvt}%
  \BibitemOpen
  \bibfield  {author} {\bibinfo {author} {\bibfnamefont {O.~H.~E.}\
  \bibnamefont {Philcox}}, \bibinfo {author} {\bibfnamefont {M.~M.}\
  \bibnamefont {Ivanov}}, \bibinfo {author} {\bibfnamefont {M.}~\bibnamefont
  {Simonovi\'c}}, \ and\ \bibinfo {author} {\bibfnamefont {M.}~\bibnamefont
  {Zaldarriaga}},\ }\href {\doibase 10.1088/1475-7516/2020/05/032} {\bibfield
  {journal} {\bibinfo  {journal} {JCAP}\ }\textbf {\bibinfo {volume} {05}},\
  \bibinfo {pages} {032} (\bibinfo {year} {2020})},\ \Eprint
  {http://arxiv.org/abs/2002.04035} {arXiv:2002.04035 [astro-ph.CO]}
  \BibitemShut {NoStop}%
\bibitem [{\citenamefont {Kitaura}\ \emph {et~al.}(2016)\citenamefont {Kitaura}
  \emph {et~al.}}]{Kitaura:2015uqa}%
  \BibitemOpen
  \bibfield  {author} {\bibinfo {author} {\bibfnamefont {F.-S.}\ \bibnamefont
  {Kitaura}} \emph {et~al.},\ }\href {\doibase 10.1093/mnras/stv2826}
  {\bibfield  {journal} {\bibinfo  {journal} {Mon. Not. Roy. Astron. Soc.}\
  }\textbf {\bibinfo {volume} {456}},\ \bibinfo {pages} {4156} (\bibinfo {year}
  {2016})},\ \Eprint {http://arxiv.org/abs/1509.06400} {arXiv:1509.06400
  [astro-ph.CO]} \BibitemShut {NoStop}%
\bibitem [{\citenamefont {Rodríguez-Torres}\ \emph {et~al.}(2016)\citenamefont
  {Rodríguez-Torres} \emph {et~al.}}]{Rodriguez-Torres:2015vqa}%
  \BibitemOpen
  \bibfield  {author} {\bibinfo {author} {\bibfnamefont {S.~A.}\ \bibnamefont
  {Rodríguez-Torres}} \emph {et~al.},\ }\href {\doibase 10.1093/mnras/stw1014}
  {\bibfield  {journal} {\bibinfo  {journal} {Mon. Not. Roy. Astron. Soc.}\
  }\textbf {\bibinfo {volume} {460}},\ \bibinfo {pages} {1173} (\bibinfo {year}
  {2016})},\ \Eprint {http://arxiv.org/abs/1509.06404} {arXiv:1509.06404
  [astro-ph.CO]} \BibitemShut {NoStop}%
\bibitem [{\citenamefont {Ivanov}(2024)}]{Ivanov:2023yla}%
  \BibitemOpen
  \bibfield  {author} {\bibinfo {author} {\bibfnamefont {M.~M.}\ \bibnamefont
  {Ivanov}},\ }\href {\doibase 10.1103/PhysRevD.109.023507} {\bibfield
  {journal} {\bibinfo  {journal} {Phys. Rev. D}\ }\textbf {\bibinfo {volume}
  {109}},\ \bibinfo {pages} {023507} (\bibinfo {year} {2024})},\ \Eprint
  {http://arxiv.org/abs/2309.10133} {arXiv:2309.10133 [astro-ph.CO]}
  \BibitemShut {NoStop}%
\bibitem [{\citenamefont {Hahn}\ \emph {et~al.}(2017)\citenamefont {Hahn},
  \citenamefont {Scoccimarro}, \citenamefont {Blanton}, \citenamefont
  {Tinker},\ and\ \citenamefont {Rodr\'\i{}guez-Torres}}]{Hahn:2016kiy}%
  \BibitemOpen
  \bibfield  {author} {\bibinfo {author} {\bibfnamefont {C.}~\bibnamefont
  {Hahn}}, \bibinfo {author} {\bibfnamefont {R.}~\bibnamefont {Scoccimarro}},
  \bibinfo {author} {\bibfnamefont {M.~R.}\ \bibnamefont {Blanton}}, \bibinfo
  {author} {\bibfnamefont {J.~L.}\ \bibnamefont {Tinker}}, \ and\ \bibinfo
  {author} {\bibfnamefont {S.~A.}\ \bibnamefont {Rodr\'\i{}guez-Torres}},\
  }\href {\doibase 10.1093/mnras/stx185} {\bibfield  {journal} {\bibinfo
  {journal} {Mon. Not. Roy. Astron. Soc.}\ }\textbf {\bibinfo {volume} {467}},\
  \bibinfo {pages} {1940} (\bibinfo {year} {2017})},\ \Eprint
  {http://arxiv.org/abs/1609.01714} {arXiv:1609.01714 [astro-ph.CO]}
  \BibitemShut {NoStop}%
\bibitem [{\citenamefont {Ivanov}\ \emph
  {et~al.}(2020{\natexlab{c}})\citenamefont {Ivanov}, \citenamefont
  {Simonovi\'c},\ and\ \citenamefont {Zaldarriaga}}]{Ivanov:2019hqk}%
  \BibitemOpen
  \bibfield  {author} {\bibinfo {author} {\bibfnamefont {M.~M.}\ \bibnamefont
  {Ivanov}}, \bibinfo {author} {\bibfnamefont {M.}~\bibnamefont {Simonovi\'c}},
  \ and\ \bibinfo {author} {\bibfnamefont {M.}~\bibnamefont {Zaldarriaga}},\
  }\href {\doibase 10.1103/PhysRevD.101.083504} {\bibfield  {journal} {\bibinfo
   {journal} {Phys. Rev. D}\ }\textbf {\bibinfo {volume} {101}},\ \bibinfo
  {pages} {083504} (\bibinfo {year} {2020}{\natexlab{c}})},\ \Eprint
  {http://arxiv.org/abs/1912.08208} {arXiv:1912.08208 [astro-ph.CO]}
  \BibitemShut {NoStop}%
\bibitem [{\citenamefont {Mirbabayi}\ \emph {et~al.}(2015)\citenamefont
  {Mirbabayi}, \citenamefont {Schmidt},\ and\ \citenamefont
  {Zaldarriaga}}]{Mirbabayi:2014zca}%
  \BibitemOpen
  \bibfield  {author} {\bibinfo {author} {\bibfnamefont {M.}~\bibnamefont
  {Mirbabayi}}, \bibinfo {author} {\bibfnamefont {F.}~\bibnamefont {Schmidt}},
  \ and\ \bibinfo {author} {\bibfnamefont {M.}~\bibnamefont {Zaldarriaga}},\
  }\href {\doibase 10.1088/1475-7516/2015/07/030} {\bibfield  {journal}
  {\bibinfo  {journal} {JCAP}\ }\textbf {\bibinfo {volume} {1507}},\ \bibinfo
  {pages} {030} (\bibinfo {year} {2015})},\ \Eprint
  {http://arxiv.org/abs/1412.5169} {arXiv:1412.5169 [astro-ph.CO]} \BibitemShut
  {NoStop}%
\bibitem [{\citenamefont {Philcox}(2021{\natexlab{a}})}]{Philcox:2020vbm}%
  \BibitemOpen
  \bibfield  {author} {\bibinfo {author} {\bibfnamefont {O.~H.~E.}\
  \bibnamefont {Philcox}},\ }\href {\doibase 10.1103/PhysRevD.103.103504}
  {\bibfield  {journal} {\bibinfo  {journal} {Phys. Rev. D}\ }\textbf {\bibinfo
  {volume} {103}},\ \bibinfo {pages} {103504} (\bibinfo {year}
  {2021}{\natexlab{a}})},\ \Eprint {http://arxiv.org/abs/2012.09389}
  {arXiv:2012.09389 [astro-ph.CO]} \BibitemShut {NoStop}%
\bibitem [{\citenamefont {Philcox}(2021{\natexlab{b}})}]{Philcox:2021ukg}%
  \BibitemOpen
  \bibfield  {author} {\bibinfo {author} {\bibfnamefont {O.~H.~E.}\
  \bibnamefont {Philcox}},\ }\href@noop {} {\  (\bibinfo {year}
  {2021}{\natexlab{b}})},\ \Eprint {http://arxiv.org/abs/2107.06287}
  {arXiv:2107.06287 [astro-ph.CO]} \BibitemShut {NoStop}%
\bibitem [{\citenamefont {Ata}\ \emph {et~al.}(2018)\citenamefont {Ata} \emph
  {et~al.}}]{Ata:2017dya}%
  \BibitemOpen
  \bibfield  {author} {\bibinfo {author} {\bibfnamefont {M.}~\bibnamefont
  {Ata}} \emph {et~al.},\ }\href {\doibase 10.1093/mnras/stx2630} {\bibfield
  {journal} {\bibinfo  {journal} {Mon. Not. Roy. Astron. Soc.}\ }\textbf
  {\bibinfo {volume} {473}},\ \bibinfo {pages} {4773} (\bibinfo {year}
  {2018})},\ \Eprint {http://arxiv.org/abs/1705.06373} {arXiv:1705.06373
  [astro-ph.CO]} \BibitemShut {NoStop}%
\bibitem [{\citenamefont {Hou}\ \emph {et~al.}(2020)\citenamefont {Hou} \emph
  {et~al.}}]{Hou:2020rse}%
  \BibitemOpen
  \bibfield  {author} {\bibinfo {author} {\bibfnamefont {J.}~\bibnamefont
  {Hou}} \emph {et~al.},\ }\href {\doibase 10.1093/mnras/staa3234} {\bibfield
  {journal} {\bibinfo  {journal} {Mon. Not. Roy. Astron. Soc.}\ }\textbf
  {\bibinfo {volume} {500}},\ \bibinfo {pages} {1201} (\bibinfo {year}
  {2020})},\ \Eprint {http://arxiv.org/abs/2007.08998} {arXiv:2007.08998
  [astro-ph.CO]} \BibitemShut {NoStop}%
\bibitem [{\citenamefont {Neveux}\ \emph {et~al.}(2020)\citenamefont {Neveux}
  \emph {et~al.}}]{Neveux:2020voa}%
  \BibitemOpen
  \bibfield  {author} {\bibinfo {author} {\bibfnamefont {R.}~\bibnamefont
  {Neveux}} \emph {et~al.},\ }\href {\doibase 10.1093/mnras/staa2780}
  {\bibfield  {journal} {\bibinfo  {journal} {Mon. Not. Roy. Astron. Soc.}\
  }\textbf {\bibinfo {volume} {499}},\ \bibinfo {pages} {210} (\bibinfo {year}
  {2020})},\ \Eprint {http://arxiv.org/abs/2007.08999} {arXiv:2007.08999
  [astro-ph.CO]} \BibitemShut {NoStop}%
\bibitem [{\citenamefont {Chudaykin}\ and\ \citenamefont
  {Ivanov}(2022)}]{Chudaykin:2022nru}%
  \BibitemOpen
  \bibfield  {author} {\bibinfo {author} {\bibfnamefont {A.}~\bibnamefont
  {Chudaykin}}\ and\ \bibinfo {author} {\bibfnamefont {M.~M.}\ \bibnamefont
  {Ivanov}},\ }\href@noop {} {\  (\bibinfo {year} {2022})},\ \Eprint
  {http://arxiv.org/abs/2210.17044} {arXiv:2210.17044 [astro-ph.CO]}
  \BibitemShut {NoStop}%
\bibitem [{\citenamefont {de~Mattia}\ \emph {et~al.}(2021)\citenamefont
  {de~Mattia} \emph {et~al.}}]{deMattia:2020fkb}%
  \BibitemOpen
  \bibfield  {author} {\bibinfo {author} {\bibfnamefont {A.}~\bibnamefont
  {de~Mattia}} \emph {et~al.},\ }\href {\doibase 10.1093/mnras/staa3891}
  {\bibfield  {journal} {\bibinfo  {journal} {Mon. Not. Roy. Astron. Soc.}\
  }\textbf {\bibinfo {volume} {501}},\ \bibinfo {pages} {5616} (\bibinfo {year}
  {2021})},\ \Eprint {http://arxiv.org/abs/2007.09008} {arXiv:2007.09008
  [astro-ph.CO]} \BibitemShut {NoStop}%
\bibitem [{\citenamefont {Modi}\ and\ \citenamefont
  {Philcox}(2023)}]{Modi:2023drt}%
  \BibitemOpen
  \bibfield  {author} {\bibinfo {author} {\bibfnamefont {C.}~\bibnamefont
  {Modi}}\ and\ \bibinfo {author} {\bibfnamefont {O.~H.~E.}\ \bibnamefont
  {Philcox}},\ }\href@noop {} {\  (\bibinfo {year} {2023})},\ \Eprint
  {http://arxiv.org/abs/2309.10270} {arXiv:2309.10270 [astro-ph.CO]}
  \BibitemShut {NoStop}%
\bibitem [{\citenamefont {Audren}\ \emph {et~al.}(2013)\citenamefont {Audren},
  \citenamefont {Lesgourgues}, \citenamefont {Benabed},\ and\ \citenamefont
  {Prunet}}]{Audren:2012wb}%
  \BibitemOpen
  \bibfield  {author} {\bibinfo {author} {\bibfnamefont {B.}~\bibnamefont
  {Audren}}, \bibinfo {author} {\bibfnamefont {J.}~\bibnamefont {Lesgourgues}},
  \bibinfo {author} {\bibfnamefont {K.}~\bibnamefont {Benabed}}, \ and\
  \bibinfo {author} {\bibfnamefont {S.}~\bibnamefont {Prunet}},\ }\href
  {\doibase 10.1088/1475-7516/2013/02/001} {\bibfield  {journal} {\bibinfo
  {journal} {JCAP}\ }\textbf {\bibinfo {volume} {1302}},\ \bibinfo {pages}
  {001} (\bibinfo {year} {2013})},\ \Eprint {http://arxiv.org/abs/1210.7183}
  {arXiv:1210.7183 [astro-ph.CO]} \BibitemShut {NoStop}%
\bibitem [{\citenamefont {Brinckmann}\ and\ \citenamefont
  {Lesgourgues}(2019)}]{Brinckmann:2018cvx}%
  \BibitemOpen
  \bibfield  {author} {\bibinfo {author} {\bibfnamefont {T.}~\bibnamefont
  {Brinckmann}}\ and\ \bibinfo {author} {\bibfnamefont {J.}~\bibnamefont
  {Lesgourgues}},\ }\href {\doibase 10.1016/j.dark.2018.100260} {\bibfield
  {journal} {\bibinfo  {journal} {Phys. Dark Univ.}\ }\textbf {\bibinfo
  {volume} {24}},\ \bibinfo {pages} {100260} (\bibinfo {year} {2019})},\
  \Eprint {http://arxiv.org/abs/1804.07261} {arXiv:1804.07261 [astro-ph.CO]}
  \BibitemShut {NoStop}%
\bibitem [{\citenamefont {Papamakarios}\ \emph {et~al.}(2017)\citenamefont
  {Papamakarios}, \citenamefont {Pavlakou},\ and\ \citenamefont
  {Murray}}]{papamakarios2017masked}%
  \BibitemOpen
  \bibfield  {author} {\bibinfo {author} {\bibfnamefont {G.}~\bibnamefont
  {Papamakarios}}, \bibinfo {author} {\bibfnamefont {T.}~\bibnamefont
  {Pavlakou}}, \ and\ \bibinfo {author} {\bibfnamefont {I.}~\bibnamefont
  {Murray}},\ }\href@noop {} {\bibfield  {journal} {\bibinfo  {journal}
  {Advances in neural information processing systems}\ }\textbf {\bibinfo
  {volume} {30}} (\bibinfo {year} {2017})}\BibitemShut {NoStop}%
\bibitem [{\citenamefont {Germain}\ \emph {et~al.}(2015)\citenamefont
  {Germain}, \citenamefont {Gregor}, \citenamefont {Murray},\ and\
  \citenamefont {Larochelle}}]{germain2015made}%
  \BibitemOpen
  \bibfield  {author} {\bibinfo {author} {\bibfnamefont {M.}~\bibnamefont
  {Germain}}, \bibinfo {author} {\bibfnamefont {K.}~\bibnamefont {Gregor}},
  \bibinfo {author} {\bibfnamefont {I.}~\bibnamefont {Murray}}, \ and\ \bibinfo
  {author} {\bibfnamefont {H.}~\bibnamefont {Larochelle}},\ }in\ \href@noop {}
  {\emph {\bibinfo {booktitle} {International conference on machine
  learning}}}\ (\bibinfo {organization} {PMLR},\ \bibinfo {year} {2015})\ pp.\
  \bibinfo {pages} {881--889}\BibitemShut {NoStop}%
\bibitem [{\citenamefont {Kingma}\ and\ \citenamefont
  {Ba}(2014)}]{kingma2014adam}%
  \BibitemOpen
  \bibfield  {author} {\bibinfo {author} {\bibfnamefont {D.~P.}\ \bibnamefont
  {Kingma}}\ and\ \bibinfo {author} {\bibfnamefont {J.}~\bibnamefont {Ba}},\
  }\href@noop {} {\bibfield  {journal} {\bibinfo  {journal} {arXiv preprint
  arXiv:1412.6980}\ } (\bibinfo {year} {2014})}\BibitemShut {NoStop}%
\bibitem [{\citenamefont {{Talts}}\ \emph {et~al.}(2018)\citenamefont
  {{Talts}}, \citenamefont {{Betancourt}}, \citenamefont {{Simpson}},
  \citenamefont {{Vehtari}},\ and\ \citenamefont
  {{Gelman}}}]{2018arXiv180406788T}%
  \BibitemOpen
  \bibfield  {author} {\bibinfo {author} {\bibfnamefont {S.}~\bibnamefont
  {{Talts}}}, \bibinfo {author} {\bibfnamefont {M.}~\bibnamefont
  {{Betancourt}}}, \bibinfo {author} {\bibfnamefont {D.}~\bibnamefont
  {{Simpson}}}, \bibinfo {author} {\bibfnamefont {A.}~\bibnamefont
  {{Vehtari}}}, \ and\ \bibinfo {author} {\bibfnamefont {A.}~\bibnamefont
  {{Gelman}}},\ }\href {\doibase 10.48550/arXiv.1804.06788} {\bibfield
  {journal} {\bibinfo  {journal} {arXiv e-prints}\ ,\ \bibinfo {eid}
  {arXiv:1804.06788}} (\bibinfo {year} {2018})},\ \Eprint
  {http://arxiv.org/abs/1804.06788} {arXiv:1804.06788 [stat.ME]} \BibitemShut
  {NoStop}%
\end{thebibliography}%

\end{document}